\pdfoutput=0
\documentclass[11pt]{article}
\usepackage{axodraw}
\usepackage{epsfig}
\usepackage{amsfonts}
\usepackage{amsmath}
\usepackage{bm,bbm}
\usepackage{cite}
\usepackage[hidelinks]{hyperref}
\allowdisplaybreaks[1]

\input pix.sty

\newcommand{\la}[1]{\label{#1}}
\newcommand{\be}{\begin{equation}}
\newcommand{\ee}{\end{equation}}
\newcommand{\ba}{\begin{eqnarray}}
\newcommand{\ea}{\end{eqnarray}}
\newcommand{\rmi}[1]{{\mbox{\scriptsize #1}}}
\newcommand{\fig}{Fig.~}
\newcommand{\figs}{Figs.~}
\newcommand{\eq}{Eq.~}
\newcommand{\eqs}{Eqs.~}
\newcommand{\se}{Sec.~}

\newcommand{\nr}[1]{(\ref{#1})}
\newcommand{\tr}{{\rm Tr\,}}
\newcommand{\nn}{\nonumber \\}
\newcommand{\fr}[2]{{\frac{#1}{#2}\,}}
\newcommand{\msbar}{{\overline{\mbox{\rm MS}}}}

\renewcommand{\vec}[1]{{\bf #1}}

\newcommand{\tfr}[2]{{\textstyle \frac{#1}{#2}\,}}

\newcommand{\kallen}{\sqrt{\lambda}}

\renewcommand{\eq}{eq.~}
\renewcommand{\eqs}{eqs.~}
\renewcommand{\se}{sec.~}

\renewcommand{\fig}{fig.~}
\renewcommand{\figs}{figs.~}
\newcommand{\PT}{\mathbbm{P}^\rmii{T}}
\newcommand{\PE}{\mathbbm{P}^\rmii{E}}
\newcommand{\PL}{\mathbbm{P}^\rmii{L}}
\newcommand{\aL}{a^{ }_\rmii{L}}
\newcommand{\aR}{a^{ }_\rmii{R}}

\newcommand{\F}{\mathcal{F}}
\renewcommand{\G}{\mathcal{G}}
\renewcommand{\H}{\mathcal{H}}

\newcommand{\K}{\mathcal{K}}
\renewcommand{\P}{\mathcal{P}}
\newcommand{\Q}{\mathcal{Q}}
\newcommand{\R}{\mathcal{R}}
\renewcommand{\S}{\mathcal{S}}

\newcommand{\T}{\rmii{$T$}}
\newcommand{\W}{\rmii{$W$}}
\newcommand{\iZ}{\rmiii{\it Z}}
\newcommand{\iW}{\rmiii{\it W}}
\newcommand{\Z}{\rmii{$Z$}}

\newcommand{\Nc}{N_{\rm c}}

\newcommand{\Tc}{T_{\rm c}}
\newcommand{\nG}{n_\rmii{G}}
\newcommand{\nS}{n_\rmii{S}}

\newcommand{\rmO}{{\mathcal{O}}}
\newcommand{\bmu}{\bar\mu}

\def\lsi{\raise0.3ex\hbox{$<$\kern-0.75em\raise-1.1ex\hbox{$\sim$}}}
\def\gsi{\raise0.3ex\hbox{$>$\kern-0.75em\raise-1.1ex\hbox{$\sim$}}}

\newcommand{\gsim}{\mathop{\gsi}}

\newcommand{\nB}{f_\rmii{B}} 
\newcommand{\niB}{f_\rmiii{B}} 
\newcommand{\rmii}[1]{{\mbox{\tiny\rm{#1}}}}
\newcommand{\rmiii}[1]{{\mbox{\tiny{$\scriptstyle{\rm#1}$}}}}
\newcommand{\re}{\mathop{\mbox{Re}}}
\newcommand{\im}{\mathop{\mbox{Im}}}

\newcommand{\Tint}[1]{{\hbox{$\sum$}\!\!\!\!\!\!\!\int\,}_{\!\!\!\!\raise-0.9ex\hbox{$\scriptstyle{#1}$}}}
\newcommand{\Tinti}[1]{{{\Sigma}\!\!\!\!\raise0.3ex\hbox{$\int$}_\rmii{${#1}$}}}

\newcommand{\bi}{\begin{itemize}}
\newcommand{\ei}{\end{itemize}}
\newcommand{\hide}[1]{ }
\newcommand{\blind}[1]{\fbox{$ ? $}} 
\newcommand{\bsl}[1]{\,\slash\!\!\!\!{#1}\,}

\newcommand{\deltabar}{\raise-0.02em\hbox{$\bar{}$}\hspace*{-0.8mm}{\delta}}
\newcommand{\ddeltabar}{\raise-0.18em\hbox{$\bar{}$}\hspace*{-0.8mm}{\delta}}

\newcommand{\aS}{\varphi} 

\newcommand{\vrel}{v^{ }_\rmi{rel}}

\def\TAsc(#1,#2)(#3,#4,#5)%
{\SetWidth{2.0}\CArc(#1,#2)(#3,#4,#5)\SetWidth{1.0}}
\def\Lwidth{3}

\def\TAgl(#1,#2)(#3,#4,#5){\SetWidth{2.0}\PhotonArc(#1,#2)(#3,#4,#5){\Lwidth}%
{6.283 #3 mul 360 div #4 #5 sub #4 #5 sub mul sqrt mul Tdensity mul}%
\SetWidth{1.0}}
\def\TLgl(#1,#2)(#3,#4){\SetWidth{2.0}\Photon(#1,#2)(#3,#4){\Lwidth}
{#1 #3 sub #1 #3 sub mul #2 #4 sub #2 #4 sub mul add sqrt Tdensity mul}%
\SetWidth{1.0}}

\renewcommand{\picc}[1]{\;\parbox[c]{60pt}{\begin{picture}(60,30)(0,-10)
\SetWidth{1.0}\SetScale{1.3} #1 \end{picture}}\; }
\def\Lwidth{1.3}

\newcommand{\picu}[1]{\;\parbox[c]{80pt}{\begin{picture}(80,30)(-15,-5)
\SetWidth{1.0}\SetScale{0.65} #1 \end{picture}}\; }

\def\AmplHa{\picu{%
 \CArc(40,15)(20,180,360)%
 \Asc(40,15)(20,0,180)%
 \Lsc(20,15)(60,15)%
 \COval(20,15)(2,2)(0){Black}{Black}%
 \Line(18,16.5)(0,16.5)%
 \Line(18,13.5)(0,13.5)
 \COval(60,15)(2,2)(0){Black}{Black}%
 \Line(62,16.5)(80,16.5)%
 \Line(62,13.5)(80,13.5)
}}
\def\AmplHb{\picu{%
 \CArc(40,15)(20,180,360)%
 \Asc(40,15)(20,0,180)%
 \Asc(20,35)(20,270,360)%
 \COval(20,15)(2,2)(0){Black}{Black}%
 \Line(18,16.5)(0,16.5)%
 \Line(18,13.5)(0,13.5)
 \COval(60,15)(2,2)(0){Black}{Black}%
 \Line(62,16.5)(80,16.5)%
 \Line(62,13.5)(80,13.5)
}}
\def\AmplHc{\picu{%
 \CArc(40,15)(20,180,360)%
 \Asc(40,15)(20,0,180)%
 \Asc(20,-5)(20,0,90)%
 \COval(40,-5)(2,2)(0){Black}{Black}%
 \COval(20,15)(2,2)(0){Black}{Black}%
 \Line(18,16.5)(0,16.5)%
 \Line(18,13.5)(0,13.5)
 \COval(60,15)(2,2)(0){Black}{Black}%
 \Line(62,16.5)(80,16.5)%
 \Line(62,13.5)(80,13.5)
}}
\def\AmplHd{\picu{%
 \CArc(40,15)(20,180,360)%
 \Asc(40,15)(20,0,180)%
 \Lsc(40,-5)(40,35)%
 \COval(40,-5)(2,2)(0){Black}{Black}%
 \COval(20,15)(2,2)(0){Black}{Black}%
 \Line(18,16.5)(0,16.5)%
 \Line(18,13.5)(0,13.5)
 \COval(60,15)(2,2)(0){Black}{Black}%
 \Line(62,16.5)(80,16.5)%
 \Line(62,13.5)(80,13.5)
}}
\def\AmplHe{\picu{%
 \CArc(40,15)(20,180,270)%
 \Asc(40,15)(20,270,360)%
 \Asc(40,15)(20,90,180)%
 \CArc(40,15)(20,0,90)%
 \Line(40,-5)(40,35)%
 \COval(40,35)(2,2)(0){Black}{Black}%
 \COval(40,-5)(2,2)(0){Black}{Black}%
 \COval(20,15)(2,2)(0){Black}{Black}%
 \Line(18,16.5)(0,16.5)%
 \Line(18,13.5)(0,13.5)
 \COval(60,15)(2,2)(0){Black}{Black}%
 \Line(62,16.5)(80,16.5)%
 \Line(62,13.5)(80,13.5)
}}
\def\AmplHf{\picu{%
 \CArc(40,15)(20,180,360)%
 \Asc(40,15)(20,0,180)%
 \Asc(40,42)(20,221,319)%
 \COval(20,15)(2,2)(0){Black}{Black}%
 \Line(18,16.5)(0,16.5)%
 \Line(18,13.5)(0,13.5)
 \COval(60,15)(2,2)(0){Black}{Black}%
 \Line(62,16.5)(80,16.5)%
 \Line(62,13.5)(80,13.5)
}}
\def\AmplHg{\picu{%
 \CArc(40,15)(20,180,360)%
 \Asc(40,15)(20,0,180)%
 \Asc(40,-12)(20,41,139)%
 \COval(25,2)(2,2)(0){Black}{Black}%
 \COval(55,2)(2,2)(0){Black}{Black}%
 \COval(20,15)(2,2)(0){Black}{Black}%
 \Line(18,16.5)(0,16.5)%
 \Line(18,13.5)(0,13.5)
 \COval(60,15)(2,2)(0){Black}{Black}%
 \Line(62,16.5)(80,16.5)%
 \Line(62,13.5)(80,13.5)
}}
\def\AmplHi{\picu{%
 \CArc(40,15)(20,180,360)%
 \Asc(40,15)(20,0,180)%
 \Agl(40,42)(20,221,319)%
 \COval(20,15)(2,2)(0){Black}{Black}%
 \Line(18,16.5)(0,16.5)%
 \Line(18,13.5)(0,13.5)
 \COval(60,15)(2,2)(0){Black}{Black}%
 \Line(62,16.5)(80,16.5)%
 \Line(62,13.5)(80,13.5)
}}
\def\AmplHj{\picu{%
 \CArc(40,15)(20,180,360)%
 \Asc(40,15)(20,0,41)%
 \Agl(40,15)(20,41,139)%
 \Asc(40,15)(20,139,180)%
 \Agl(40,42)(20,221,319)%
 \COval(20,15)(2,2)(0){Black}{Black}%
 \Line(18,16.5)(0,16.5)%
 \Line(18,13.5)(0,13.5)
 \COval(60,15)(2,2)(0){Black}{Black}%
 \Line(62,16.5)(80,16.5)%
 \Line(62,13.5)(80,13.5)
}}
\def\AmplHk{\picu{%
 \CArc(40,15)(20,180,360)%
 \Asc(40,15)(20,0,41)%
 \Agh(40,15)(20,41,139)%
 \Asc(40,15)(20,139,180)%
 \Agh(40,42)(20,221,319)%
 \COval(20,15)(2,2)(0){Black}{Black}%
 \Line(18,16.5)(0,16.5)%
 \Line(18,13.5)(0,13.5)
 \COval(60,15)(2,2)(0){Black}{Black}%
 \Line(62,16.5)(80,16.5)%
 \Line(62,13.5)(80,13.5)
}}
\def\AmplHl{\picu{%
 \CArc(40,15)(20,180,360)%
 \Asc(40,15)(20,0,41)%
 \Aqu(40,15)(20,41,139)%
 \Asc(40,15)(20,139,180)%
 \Aqu(40,42)(20,221,319)%
 \COval(20,15)(2,2)(0){Black}{Black}%
 \Line(18,16.5)(0,16.5)%
 \Line(18,13.5)(0,13.5)
 \COval(60,15)(2,2)(0){Black}{Black}%
 \Line(62,16.5)(80,16.5)%
 \Line(62,13.5)(80,13.5)
}}

\makeatletter \@addtoreset{equation}{section} \makeatother
\renewcommand{\theequation}{\arabic{section}.\arabic{equation}}
\makeatletter
\renewcommand\section{\@startsection {section}{1}{\z@}%
                                   {-5.5ex \@plus -1ex \@minus -.2ex}
                                   {2.3ex \@plus.2ex}%
                                   {\normalfont\large\bfseries}}
\renewcommand\subsection{\@startsection{subsection}{2}{\z@}%
                                     {-3.25ex\@plus -1ex \@minus -.2ex}%
                                     {1.5ex \@plus .2ex}%
                                     {\normalfont\normalsize\bfseries}}
\renewcommand\thesection {\@arabic\c@section}
\renewcommand\thesubsection   {\thesection.\@arabic\c@subsection}
\renewcommand{\@seccntformat}[1]{%
\csname the#1\endcsname.\hspace{1.0em}}
\makeatother

\begin{document}

\flushbottom

\begin{titlepage}

\begin{flushright}
August 2025
\end{flushright}
\begin{centering}
\vfill

{\Large{\bf
 Computing singlet scalar freeze-out \\[2mm]
 ~with plasmon and plasmino states
}} 

\vspace{0.8cm}

S.~Biondini$\hspace*{0.3mm}^\rmi{a}$,
M.~Eriksson$\hspace*{0.3mm}^\rmi{b}$,
M.~Laine$\hspace*{0.3mm}^\rmi{b}$

\vspace{0.8cm}

$^\rmi{a}$%
{\em 
Institute of Physics, University of Freiburg, \\
Hermann-Herder-Stra{\ss}e 3, 79014 Freiburg, Germany \\}

\vspace{0.3cm}

$^\rmi{b}$%
{\em
AEC, 
Institute for Theoretical Physics, 
University of Bern, \\ 
Sidlerstrasse 5, CH-3012 Bern, Switzerland \\}

\vspace*{0.8cm}

\mbox{\bf Abstract}
 
\end{centering}

\vspace*{0.3cm}
 
\noindent
The final-state particles from cosmological dark matter co-annihilation are 
expected to equilibrate. As dictated by Hard Thermal Loop resummation, 
the spectrum of equilibrated quasiparticles  
is richer than in vacuum, with a massless gauge field possessing three 
independent polarization states (``plasmons''), and a massless fermion 
developing a novel branch (``plasmino''). Furthermore, once the Higgs 
phenomenon sets in, vacuum and thermal mass corrections interfere. 
We collect together the corresponding poles and residues for 
the Standard Model around its crossover temperature. Choosing its 
singlet scalar extension for illustration, we subsequently demonstrate, 
both numerically and via power counting, and in accordance with general 
theoretical expectations, how in the freeze-out of TeV-scale dark matter, 
these effects remain well hidden in the inclusive annihilation cross section. 
In particular, the dominant (longitudinal) gauge channel is shown to be 
practically temperature-independent. Cosmological constraints on TeV-scale 
singlet scalars are reconfirmed. 

\vspace*{0.5cm}

\noindent
{\fontsize{9pt}{11pt}\selectfont
{\em Emails}: 
{\tt simone.biondini@physik.uni-freiburg.de},
{\tt magdalena.eriksson@unibe.ch}, 
{\tt laine@itp.unibe.ch}
}

\vfill


\end{titlepage}

\tableofcontents



%
\section{Introduction}

The relic density of cold dark matter, 
obtained by fitting Planck data
to the standard cosmological model, is given by 
$\Omega^{ }_\rmi{dm} h^2_{ } = 0.120\pm 0.001$~\cite{planck6}.
This precise measurement imposes a strong constraint on the nature
of dark matter and on the mechanism by which it is formed. 
In the freeze-out scenario~\cite{anc1,old0}, dark 
matter is at first in equilibrium with a thermal plasma. 
As the universe cools down, the dark matter density should 
become Boltzmann-suppressed. 
Dynamically, the density can decrease 
via $2\to 2$ co-annihilation processes. 
However, 
once the co-annihilation rate becomes smaller than the Hubble rate, 
the decrease terminates, and 
dark matter falls out of equilibrium. 
By combining tools from quantum field 
theory and non-equilibrium statistical physics, the corresponding
relic density 
$\Omega^{ }_\rmi{dm} h^2_{ }$ can be computed theoretically, 
as a function of the parameters of a given model. In principle, 
we should aim at a similar 1\% precision in these computations
as on the observational side.

Of course, as long as the nature of dark matter is 
unknown, the majority of the literature concentrates
on model building rather than on precision computations. That said, 
it is also appreciated that in some cases, the
non-equilibrium physics responsible for the freeze-out process
becomes non-trivial~\cite{old1,old2}. In concrete terms, we may 
wonder about the correct functional form of the kinetic equations
to be used, as well as about the radiative and thermal 
corrections that affect the ``matching coefficients'' 
that parametrize the kinetic equations. 

A unifying theoretical framework for computing 
radiative and thermal corrections to the matching coefficients
(at least in a class of theories) 
has been formulated only rather recently. A key insight is that
to understand the structure of the matching coefficients, we should
place ourselves on the side of the dark matter, viewing
the Standard Model plasma as a heat bath, whose microscopic 
details do not need to be resolved. Then the matching coefficients 
can be related to ``spectral functions''
of the portal operators by which dark matter decays. 
The structure of the spectral functions can, in turn, be 
analyzed through the framework of
the operator product expansion, 
as was first shown in the context of QCD~\cite{sch}.  
This way, thermal corrections can be shown to be powerlike, 
with the powers dictated by the dimensions of gauge-invariant
condensates. In the cosmological context, this type of computations
were first carried out for non-relativistic 
leptogenesis~\cite{salvio,nonrel,tum}, 
where thermal corrections go like $T^2_{ }$, and subsequently 
also for specific dark matter scenarios~\cite{mb,bis}. 

However, wonderful as a unifying framework is, one can still
have second thoughts. One is that 
the low-temperature expansion in powers of $T^2_{ }$
is asymptotic, and in fact shows poor convergence~\cite{relat}. 
A second is that whereas the cosmological freeze-out process is
fully ``inclusive'', if we instead think about dark matter
direct or indirect detection, or its production at colliders,  
we are forced to consider specific ``exclusive'' 
Standard Model channels.\footnote{%
 In an {\em inclusive} process, only the 
 initial state is specified, and all possible final states are included. 
 In an {\em exclusive} process, the final-state particles
 and their kinematics are specified as well.  
 } 
Would it not be nice to understand the freeze-out dynamics in 
the same language?

If we do adopt an exclusive viewpoint, the problem rapidly becomes
complicated. Concretely, for weakly interacting TeV-scale dark matter, 
freeze-out happens close to the temperature of the 
electroweak crossover. Then masses induced by the Higgs mechanism 
($\sim gv$) are of the same order as thermal mass corrections
($\sim gT$). In thermal field theory, these are known as 
{\em soft scales}. If the physics that we are considering is 
sensitive to the soft scales, perturbative computations need 
to be ``resummed'', through the framework of Hard Thermal Loop
effective theories~\cite{ht1,ht2,ht3,ht4}. Often, a further challenge
is that it may not be transparent in advance which scales really matter. 
Then we may be tempted to resum ``just in case''. The goal of the
present investigation is related to such an approach: we would like
to demonstrate, as pedagogically as possible, that such an exclusive
viewpoint, including resummations, is also viable, and in the end
only leads to very small thermal corrections, even if intermediate
steps get substantially modified compared with a vacuum-like
computation. 

Given that our main focus is on a formal side, we have decided
to adopt one of the simplest marginally
viable dark matter scenarios for illustration, 
obtained through the addition of a gauge-singlet scalar field to the Standard 
Model~\cite{singlet1,singlet2,singlet3,cas,singlet4,singlet5,singlet6}. 
This model has been employed for illustration purposes before. One example
is if the singlet scalar mass happens to be about 1/2 of the Higgs mass. 
Then dark matter can annihilate via a resonant mechanism with extremely 
small couplings~\cite{drake,abe,kk2,kinetic,pm}, 
which are not experimentally excluded~\cite{exp1,exp2}. 
Another possibility could be if the mass and couplings are pushed to 
become large. Then the dark matter particles become strongly
interacting, with associated effects like a 
Sommerfeld enhancement of their 
co-annihilation cross section~\cite{sommerfeld}
(though this is likely screened by the non-zero Higgs mass).
A further non-trivial scenario originates if the singlet scalar
converts the electroweak phase transition into 
a first-order one (cf., e.g.,\ refs.~\cite{fopt1,fopt2}).

In the present paper, we inspect a region of the parameter
space in which {\em no} exotic phenomena are expected, 
and a standard analysis
should suffice. Notably, we consider singlet scalar vacuum 
masses $m^{ }_{\varphi,\rmii{phys}} \sim$~a~few~TeV and couplings
$\kappa \sim 1$, which were mildly favoured by global 
fits some time ago~\cite{singlet5,singlet6}.\footnote{%
 It is appropriate to remark that, given the theoretical nature of our
 work, we have {\em not} applied the latest rather stringent
 constraints from the observational side. 
 A recent look at the strongly interacting regime of the singlet scalar 
 model, in light of 
 direct detection updates, can be found in ref.~\cite{eh}.
 } 
The question we pose is whether a resummed computation, 
with its new channels, could 
yield larger thermal effects than generally expected, 
particularly if the freeze-out takes place 
close to the electroweak crossover?
Though the results yield no surprises, 
the details do reveal a rich and partly unexplored structure, 
which might find use in other contexts as well. 

Our paper is organized as follows. Given the fact that our
computation leads us to some technical detail, we start by
outlining the approach on a general level, 
in \se\ref{se:formulation}. The computation itself can then
be factorized into two main parts. First, in \se\ref{se:feynman}, 
we establish the Feynman rules for the Hard Thermal Loop resummed
Standard Model, and use them to compute a particular 2-point function
of the dark matter field. At this stage, many cancellations 
can be verified, notably that of gauge
independence, providing for an important crosscheck.  
In the second part, in \se\ref{se:cuts},
we extract the ``cut'', or 
imaginary part, of the 2-point correlator, which is proportional
to the annihilation cross section. At this stage, 
tedious phase-space integrals are met, but given that cancellations had already
taken place, the set that needs to be tackled is the minimal one. 
The phase-space constraints permit us to set up a power counting, 
establishing the importance of various contributions. 
The results are inserted into a practical dark matter
freeze-out computation in \se\ref{se:pheno}, 
before we conclude in \se\ref{se:concl}. Further details
related to \se\ref{se:cuts} are relegated into four appendices.

%
\section{Formulation of the problem}
\la{se:formulation}

If we want to compute thermally averaged dark matter annihilation 
cross sections beyond leading order in couplings, 
or if we want to ``re-sum'' 
effects from arbitrarily high loop orders, the conventional 
Boltzmann equations do {\em not} offer for a robust
starting point. The particles appearing in them 
are vacuum-like asymptotic scattering states. However, within a medium, 
through Hard Thermal Loop (HTL) corrections~\cite{ht1,ht2,ht3,ht4}, 
particles develop thermal masses and thermal interaction rates. 
Moreover, the number of physical polarization states, defined 
through the on-shell energies that can be found for a given momentum, 
increases from that in vacuum, opening up new channels. 
In order to account
for these effects, we need to replace the Boltzmann equations through 
a quantum field theoretic framework. 

A relatively simple way to establish a well-defined
starting point is to relate the annihilation cross
section appearing in a Boltzmann equation to an {\em equilibration rate}. 
The benefit of equilibration rates is that they have an unambiguous
physical meaning within the linear response regime, and can
therefore be computed with any formalism, including
thermal quantum field theory. 
Specifically, as we are interested in inelastic reactions changing
particle number, we should be concerned with a chemical equilibration
rate~\cite{chem}. 

The chemical equilibration rate can be viewed as a subpart
of what has been called a maximal interaction rate~\cite{resonance}. 
For the maximal interaction rate, a simple recipe can be formulated. 
Suppose that we write the dark matter field $\varphi$ as 
$\varphi\to \widetilde\varphi + \varphi$, 
where 
$\widetilde\varphi$ is 
treated as a kinetically
equilibrated mode, 
and 
$\varphi$ as an out-of-equilibrium mode. 
The terms linear in $\varphi$ in the action can 
be expressed as 
\be
 S \;\supset\; \int_X \varphi(X) \mathcal{O}(X) 
  \; = \; \Tint{K} \varphi(K) \mathcal{O}(-K)
 \;, \la{S} 
\ee
where we employ 
the imaginary-time formalism, denoting the corresponding
four-momentum by $K = (\omega^{ }_n,\vec{k})$ and the sum-integral
by $\Tinti{K} \equiv T \sum_{\omega_n}\int_\vec{k}$. 
Let us then compute the self-energy of the $\varphi$ particle, 
\be
 \Pi^{ }_{K;\varphi} 
 \; \equiv \; 
 \Bigl\langle\, 
   \mathcal{O}(-K) \; \Tinti{Q} \, \mathcal{O}(Q)
 \,\Bigr\rangle
 \;. \la{Pi_def}
\ee
Once the frequency
is analytically continued to a Minkowskian one, 
$\omega^{ }_n \to - i [\omega + i 0^+_{ }]$, the Euclidean
self-energy turns into a retarded self-energy. Its imaginary part (``cut'')
defines the maximal interaction rate, 
\be
 \Gamma^\rmi{max}_{\K;\varphi} 
 \; \equiv \; \frac{\im \Pi^{ }_{\K;\varphi}}{\omega}
 \;, \quad
 \K \; \equiv \; (\omega,\vec{k})
 \;, \quad
 \omega \; \equiv \; \sqrt{k^2_{ } + m^2_\varphi}
 \;, \quad
 k \; \equiv \; |\vec{k}|
 \;, \la{Gamma_def}
\ee
where the momentum subscript $\K$ indicates that
analytic continuation has been carried out. 

We note in passing that in 
ref.~\cite{dbx}, a rate was obtained just like
in \eq\nr{Gamma_def}, and its physical relevance 
was rigorously established
in a context in which a single $\varphi$-field is weakly coupled
to the Standard Model. Here, our physical context is different, in that 
originally the coupling of $\varphi$ to Standard Model fields
could be quadratic. The linear 
appearance of \eq\nr{S} has only emerged after the shift
$\varphi\to \varphi + \widetilde\varphi$.

Once 
$
 \Gamma^\rmi{max}_{\K;\varphi}
$
has been computed, 
we would like to obtain the corresponding annihilation cross section. 
As a first step, we need to select the contribution of 
the inelastic processes to the rate, writing 
\be
 \Gamma^\rmi{max}_{\K;\varphi} 
 \; \equiv \; 
 \Gamma^\rmi{inel}_{\K;\varphi}
 + 
 \Gamma^\rmi{elas}_{\K;\varphi}
 \;. \la{inel_elas}
\ee
The inelastic processes are those in which, 
in addition to $\varphi$, 
at least one $\widetilde\varphi$ is pulled to the initial state. 
In total,  
$
 \Gamma^\rmi{inel}_{\K;\varphi}
$
can be represented as a sum of $i$ 
dark matter particles in the initial state, 
$
  \Gamma^\rmi{inel}_{\K;\varphi}
 = 
 \sum_{i\ge 2}
 \Gamma^\rmi{inel(i)}_{\K;\varphi}
$.
In the non-relativistic regime, $i=2$ gives the dominant
contribution, up to corrections suppressed 
by $e^{-m_\varphi^{ }/T}_{ } \ll 1$. Subsequently, we can average
over momenta, whereby the contribution of the elastic 
processes drops out. Assuming kinetic equilibrium, so that
the momenta of the dark matter particles follow
the Bose distribution, denoted by $\nB^{ }$, 
this yields
\be
 \langle \sigma \vrel \rangle 
 \;\equiv\; 
  \frac{
   \int_\vec{k}  \Gamma^\rmi{inel($2$)}_{\K;\varphi} \, \nB^{ }(\omega) 
      }{n^{2}_\rmi{eq}}   
 \;, \la{sigmav}
\ee
where the number density has been expressed as 
\be
  n^{ }_\rmi{eq} \equiv \int_\vec{k} \nB^{ }(\omega)
  \;. 
\ee
Even though we have given these as definitions, it can be 
verified that in the regime of validity of Boltzmann equations, 
\eq\nr{sigmav} agrees with the corresponding averaged cross section 
(cf.\ \se\ref{se:cuts}). 
Then the rate equation also has the familiar form~\cite{clas2}, 
\be
 \bigl( \partial^{ }_t + 3 H \bigr) n^{ }_\aS 
 \approx - 
 \langle \sigma \vrel \rangle 
 \bigl( n^2_\aS - n^2_\rmi{eq} \bigr)
 \;. \la{lw}
\ee

%
\section{Feynman rules, contractions, and gauge invariance}
\la{se:feynman}

%
\subsection{Feynman rules for the HTL-resummed Standard Model}
\la{ss:notation}

We would now like to compute the 
self-energy in \eq\nr{Pi_def}. For this, we need 
the Feynman rules for the Standard Model extended by the singlet
scalar field. Specifically, we assume the singlet scalar to be heavy, 
with a mass of a few~TeV. Then its decoupling takes place at
temperatures close to those of the electroweak crossover, 
$T\simeq 160$~GeV~\cite{dono}. The Standard Model gauge bosons, 
scalars, and fermions are ``soft'' in this regime, with vacuum
masses in general smaller than the ``hard'' thermal scale, $\sim \pi T$.
Then the Feynman rules need to be worked out in the context
of a HTL effective theory.\footnote{%
 In general, the HTL theory involves both propagator and vertex corrections.
 However, the vertices of a scalar particle do not get corrected, and since
 in our case the singlet scalar communicates with the Standard Model through
 the Higgs field, we do not need to worry about the vertex corrections. 
 }
Previously, 
soft real-time physics in the same temperature
regime has been considered, for instance,  
for low-scale leptogenesis~\cite{miura,broken} 
or for the equilibration rate
of the Standard Model Higgs field~\cite{htl_old,higgs_width}.

The theory considered is defined by the Lagrangian
\be
  \mathcal{L} \; = \; \mathcal{L}^{ }_\rmii{\it SM}
  + \, \biggl\{
     \frac{1}{2} \partial\hspace*{0.4mm}^{\mu}_{ }\varphi
  \, \partial^{ }_\mu\varphi
  - \, \biggl[
                 \frac{1}{2} \, m_{\aS 0}^2\, \varphi^2
               + \frac{1}{2} \, \kappa\,  \varphi^2 H^\dagger H
               + \frac{1}{4} \, \lambda^{ }_\aS\, \varphi^4
    \,\biggr]\,  \biggr\} 
 \;, \la{L}
\ee
where $H$ is the Standard Model Higgs doublet. After electroweak
symmetry breaking, the Higgs doublet can be written as 
\be
 H = \left( 
 \begin{array}{c}
   G^{ }_\W \\ 
   \tfr{1}{\sqrt{2}} \bigl[ v + h + i G^{ }_\Z \bigr]
 \end{array} 
 \right)
 \;, 
\ee
where the normalization is such that the expectation value
in vacuum reads
$v|^{ }_{T=0}\simeq 246$~GeV. The scalar singlet is assumed
to be heavy in the sense that 
$m^{2}_{\varphi 0} \gg \kappa v^2_{ }|^{ }_{T=0}$.

At finite temperature, the value of $v$ gets reduced. 
However, $v$ is a delicate quantity, first of all
because it is gauge dependent, and second, because it is sensitive 
to soft ($\sim gT$)~\cite{pba} and even ultrasoft 
($\sim g^2_{ }T/\pi$)~\cite{linde} scales. 
The latter issue can be rephrased by noting that the Higgs
mass parameter, $m_h^2$, receives a correction $\sim (g^2_{ }T/\pi)^2_{ }$, 
where part of the coefficient is non-perturbative. 
In order to handle this issue, which in the end is not important 
for our theoretical considerations, even though it plays a visible
numerical role, we make use of a lattice determination of the  
pseudocritical temperature $\Tc^{ }\simeq 160$~GeV
of the Standard Model crossover~\cite{dono}. With its help, 
we parametrize
$
 v \equiv 
 v|^{ }_\T \simeq v|^{ }_{T=0} \re\sqrt{1 - T^2/\Tc^2}
$. 
For the benefit of readers wishing to apply our formalism to BSM
frameworks for which no lattice input is available, we remark that
our main results remain practically 
unchanged even if an approximate perturbative
evaluation of $v|^{ }_\T$ is employed. 
In terms of diagrams,
the evolution of $v$ corresponds to effects originating
from tadpole diagrams, i.e.\ those that would contribute
to $\langle h \rangle$. As their effects have been 
accounted for through $v$, we have $\langle h \rangle\simeq 0$, 
and tadpole diagrams are 
omitted from our actual computation. 

As for the singlet scalar,  
in a thermal environment its mass becomes
\be
 m_\varphi^2
 \; \approx \;
 m^{2}_{\varphi 0}
 + \kappa\,\biggl( \frac{ v^2_{ }}{2} + \frac{ T^2_{ }}{6}\biggr)
 \;. \la{m_varphi}
\ee
Here the $T^2_{ }$-part is strictly speaking correct only for
$\pi T \gg m_h^{ }$, but we do not need to worry about this, since 
in the opposite limit the thermal correction 
is subdominant to the $v^2_{ }$-part.
The physical mass at $T=0$ is 
\be
  m_{\varphi,\rmii{phys}}^2
 \;
 \underset{T\,=\,0}{
 \overset{\rmii{\nr{m_varphi}}}{\approx}}
 \;
 m^{2}_{\varphi 0}
 + \frac{\kappa\,  v^2_{ }}{2} \biggr|^{ }_{T=0}
  \;. \la{m_phys}
\ee

For the electroweak sector, we work in a general $R^{ }_\xi$ gauge, 
with the gauge parameter denoted by $\xi$.
Apart from the Higgs field ($h$, illustrated with dashed lines) and 
the heavy singlet ($\widetilde\varphi$, illustrated with solid lines),
the particles playing a role in the loops are the charged and
neutral Goldstone modes ($G^{ }_\W$ and $G^{ }_\Z$, 
illustrated with dashed lines), 
the charged and
neutral ghosts ($c^{ }_\W$, $c^{ }_\Z$, 
illustrated with dotted lines), 
the gauge bosons ($W$, $Z$, 
illustrated with wiggly lines), 
and the fermions ($t$, $b$, illustrated with arrowed lines).
The corresponding tree-level physical masses 
in the thermal ground state
are 
$
 m_\W^2 = g_2^2 v^2_{ }/4
$, 
$
 m_\Z^2 = (g_1^2 + g_2^2) v^2_{ }/4
$, 
$
 m_t^2 = h_t^2 v^2_{ }/2
$, 
$
 m_b^2 = h_b^2 v^2_{ }/2
$
where $g^{ }_1$ and $g^{ }_2$ are the U(1) and SU(2) gauge couplings, 
respectively, and
$
 h^{ }_t
$
and 
$
 h^{ }_b
$ 
are the top and bottom Yukawa couplings. 
The masses of the Goldstones ($m^2_{G_{\W,\Z}}$), longitudinal
gauge field polarizations (${m'}^{2}_{\!\!\W,\Z}$), 
and ghosts (${m'}^{2}_{\!\!\W,\Z}$) agree, 
$
 m^2_{G_\W} \equiv {m'}^2_{\!\!\W} \equiv \xi m^2_\W
$
and
$
 m^2_{G_\Z} \equiv {m'}^2_{\!\!\Z} \equiv \xi m^2_\Z
$.
For the Higgs excitation, we adopt a different notation according
to whether the Higgs mechanism is active or not, 
\ba
 m_h^2
 &
 \overset{}{\simeq} 
 &
 \frac{m^2_{h,\rmii{phys}}}{2} 
  \biggl(\,
 \frac{T^2_{ }}{\Tc^2} - 1
 \,\biggr)
 + 3 \lambda^{ }_h v^2_{ }
 \; 
 \underset{ }{
 \overset{T\, <\, \Tc^{ }}{=}}
 \; 
 2\lambda^{ }_h v^2_{ }
 \; 
 = 
 \;  
 m^2_{h,\rmii{phys}}
 \, 
 \biggl(\,
 1 - \frac{T^2_{ }}{\Tc^2}
 \,\biggr)
 \;, \la{m_h} \hspace*{6mm} \\[2mm]
 m_\phi^2 
 & 
 \overset{T\, >\, \Tc^{ }}{\simeq} 
 & 
 \frac{ m^2_{h,\rmii{phys}} }{2}
 \, 
 \biggl(\,
 \frac{T^2_{ }}{\Tc^2} - 1
 \,\biggr)
 \;, \la{m_phi}
\ea
where $v$ is the thermally modified Higgs expectation value
and $m^{ }_{h,\rmii{phys}} \approx 125$~GeV. 
To be clear, we remark that \eq\nr{m_phi} is directly the (resummed) 
mass parameter appearing in the effective Lagrangian, 
whereas in the low-temperature result of \eq\nr{m_h}, the value of $v$ 
needs to be chosen so that we find ourselves in the Higgs minimum.

The propagators of the HTL-resummed theory 
are simple for the spin-$0$ particles, 
\ba
 &&
 \Delta^{-1}_{P;h} 
 \;\equiv\; 
 \frac{1}{P^2_{ } + m^2_h}
 \;, \quad
 \Delta^{-1}_{P;\widetilde\varphi} 
 \;\equiv\; 
 \frac{1}{P^2_{ } + m^2_{\varphi}}
 \;, \la{def_Delta} \\[2mm]
 &&
 \Delta^{-1}_{P;G_{\W,\Z}} 
 \;\equiv\; 
 \frac{1}{P^2_{ } + {m'}^2_{\!\!\W,\Z} }
 \;, \quad
 \Delta^{-1}_{P;c^{ }_{\W,\Z}} 
 \;\equiv\; 
 \frac{1}{P^2_{ } + {m'}^2_{\!\!\W,\Z} } 
 \;. 
\ea
For the gauge sector, we need the transverse (T), 
longitudinal (L), and
electric (E)  projectors, 
\be
 \PT_{\mu\nu} 
 \; \equiv \; 
 \delta^{ }_{\mu i}\delta^{ }_{\nu j}
 \biggl(
  \delta^{ }_{ij} - \frac{p^{ }_i p^{ }_j}{p^2} 
 \biggr)
 \;, \quad
 \PL_{\mu\nu} 
  \; \equiv \; 
 \frac{ P^{ }_\mu P^{ }_\nu }{  P^2_{ } }
 \;, \quad
 \PE_{\mu\nu}
 \; \equiv \; 
 \delta^{ }_{\mu\nu}
  - \PT_{\mu\nu}
  - \PL_{\mu\nu}
 \;, \la{projectors}
\ee 
where 
$
  p \; \equiv \; |\vec{p}|
$.
The corresponding HTL self-energies will ultimately
be employed after analytic continuation, so it is helpful to list 
those expressions already, 
\ba 
 \Pi^\rmii{T$j$}_{(-i(p^0_{ } + i 0^+_{ }),\vec{p})} & = & 
 \frac{m_\rmii{E$j$}^2}{2} 
 \biggl\{ 
   \frac{(p^0_{ })^2}{{p}^2} + 
   \frac{p^0_{ }}{2p}
   \biggl[
     1 -  \frac{(p^0_{ })^2}{{p}^2}
   \biggr] 
   \ln\frac{p^0_{ }  + p + i 0^+_{ }}{p^0_{ } - p + i 0^+_{ }}
 \biggr\} 
 \;, \la{PiT} \\
 \widehat\Pi^\rmii{E$j$}_{(-i(p^0_{ } + i 0^+_{ }),\vec{p})} & = & 
 \frac{ m_\rmii{E$j$}^2 }{{p}^2}
   \biggl[
     1 -     
     \frac{p^0_{ }}{2p}
   \ln\frac{p^0_{ } + p + i 0^+_{ }}{p^0_{ } - p + i 0^+_{ }}
   \biggr] 
 \;, \la{hatPiE}
\ea
where $j\in\{1,2\}$, and the thermal masses read
\be
 m_\rmii{E1}^2
 =
 \biggl( \fr{\nS}6 + \frac{5 \nG}{9} \biggr) g_1^2 T^2
 + \rmO(g_1^4)
 \;, \quad
 m_\rmii{E2}^2
 = 
 \biggl(  \fr23 + \fr{\nS}6 + \frac{\nG}{3} \biggr) g_2^2 T^2_{ }
 + \rmO(g_2^4)
 \;, \la{mE2}
\ee
where
$\nS^{ } \equiv 1$ is the number of Higgs doublets and 
$\nG^{ } \equiv 3$ the number of fermion generations.
Furthermore, for the electric part, we have written 
the original self-energy in a rescaled form, 
\be
 \Pi^\rmii{E}_{P}
  \; \equiv \; 
 P^2_{ } \,
 \widehat\Pi^\rmii{E}_{P} 
 \;. 
\ee

With the given self-energies, propagators can be worked out. 
It is helpful to 
recombine the structures appearing in them so that only a single 
pole appears in each part. In the charged gauge boson sector, this leads to 
\ba
 [\Delta^{-1}_{P;W}]^\rmii{ }_{\mu\nu}
 & = &  
 [\Delta^{-1}_{P;W}]^\rmi{phys}_{\mu\nu}
 \; - \;  
 \frac{P^{ }_\mu P^{ }_\nu}{m_\W^2 (P^2 + {m'}_{\!\!\W}^2)}
 \;, \la{htl_prop_w_pre} \\[2mm]
 [\Delta^{-1}_{P;W}]^\rmi{phys}_{\mu\nu}
 & \equiv &  
  \PT_{\mu\nu}\, \G^\rmii{T}_{P} 
 + 
  \bigl( \delta^{ }_{\mu\nu} - \PT_{\mu\nu} \bigr)\, \G^\rmii{E}_{P}
 + 
  \frac{P^{ }_\mu P^{ }_\nu}{m^2_\W}\,
  \, \widehat{\G}\hspace*{0.4mm}{}^\rmii{E}_{P}
  \;, \la{htl_prop_w} \\[2mm]
 \G^\rmii{T}_{P} & \equiv & 
 \frac{ 1 }{P^2 + \Pi^\rmii{T2}_{P}  + m_\W^2 } 
 \;, \la{G_T}
 \\[2mm]
 \G^\rmii{E}_{P} & \equiv & 
 \frac{1}{P^2 (1 + \widehat \Pi^\rmii{E2}_{P}) + m_\W^2}
 \;, \la{G_E}
 \\[2mm]
 \widehat{\G}\hspace*{0.3mm}{}^\rmii{E}_{P}
 & \equiv & 
 \frac{  1 + \widehat \Pi^\rmii{E2}_{P} }
      {P^2 (1 + \widehat \Pi^\rmii{E2}_{P}) + m_\W^2}
 \;. \la{hatG_E} 
\ea

The situation is more complicated in the neutral sector. 
We employ sign conventions in which the covariant derivative
acting on the Higgs doublet reads
\be
 D^{ }_\mu H \; = \; 
 \biggl( 
   \partial^{ }_\mu 
  + \frac{i g^{ }_1 B^{ }_\mu}{2}
  - i g^{ }_2\, T^a_{ } A^a_\mu{ }
 \biggr) H
 \;, \quad
 T^a_{ } \equiv \frac{\sigma^a_{ }}{2}
 \;, 
\ee
where $\sigma^a_{ }$ are the Pauli matrices. 
The vacuum mixing angles are defined as 
\be
 s \; \equiv \; \frac{g^{ }_1}{\sqrt{g_1^2 + g_2^2}} 
 \;, \quad 
 c \; \equiv \; \frac{g^{ }_2}{\sqrt{g_1^2 + g_2^2}} 
 \;.
\ee
In the basis of the original (Lagrangian) gauge fields, the quadratic
part of the HTL effect action contains
\ba
 S & \supset & \Tint{P}
 \frac{1}{2}
 \biggl(
  \begin{array}{c} 
    A^3_{\mu;-P}
      \\ 
    B^{ }_{\mu;-P}
  \end{array}
 \biggr)^T_{ }
  \biggl(
  \begin{array}{cc} 
     M^{ }_\rmiii{\it AA} & M^{ }_\rmiii{\it AB} 
     \\
     M^{ }_\rmiii{\it AB} & M^{ }_\rmiii{\it BB}  
  \end{array}
 \biggr)
  \biggl(
  \begin{array}{c} 
    A^3_{\mu;P}
     \\ 
    B^{ }_{\mu;P}
  \end{array}
 \biggr)
 \;, \la{S_2}
 \\[2mm]
 M^{ }_\rmii{$AA$} & = & 
     \delta^{ }_{\mu\nu} ( P^2_{ } + c^2_{ } m^2_\Z )
    - P^{ }_\mu P^{ }_\nu + \frac{P^{ }_\mu P^{ }_\nu}{\xi}
    + \mathbbm{P}^\rmii{T}_{\mu\nu} \Pi^\rmii{T2}_{P}
    + \mathbbm{P}^\rmii{E}_{\mu\nu} \Pi^\rmii{E2}_{P}
 \;, \\[2mm]
 M^{ }_\rmii{$AB$} & = & 
     \delta^{ }_{\mu\nu} c\, s\, m^2_\Z 
 \;, \\[2mm]
 M^{ }_\rmii{$BB$} & = & 
     \delta^{ }_{\mu\nu} ( P^2_{ } + s^2_{ } m^2_\Z )
    - P^{ }_\mu P^{ }_\nu + \frac{P^{ }_\mu P^{ }_\nu}{\xi}
    + \mathbbm{P}^\rmii{T}_{\mu\nu} \Pi^\rmii{T1}_{P}
    + \mathbbm{P}^\rmii{E}_{\mu\nu} \Pi^\rmii{E1}_{P}
 \;.
\ea
With the help of projectors from \eq\nr{projectors}, 
we can split \eq\nr{S_2} into three independent parts. Each of them
can be inverted separately.
Subsequently, we can rotate into the basis of the vacuum-like fields, 
\be
  \biggl(
  \begin{array}{c} 
    A^3_{\mu;P}
     \\ 
    B^{ }_{\mu;P}
  \end{array}
 \biggr)
 = 
   \biggl(
  \begin{array}{rr} 
    c & - s
     \\ 
    s &   c 
  \end{array}
 \biggr)
  \biggl(
  \begin{array}{c} 
    Z^{ }_{\mu;P}
     \\ 
    Q^{ }_{\mu;P}
  \end{array}
 \biggr)
 \;. 
\ee
We only need the $ZZ$ component
from the propagator matrix.
Furthermore, the terms proportional to $P^{ }_\mu P^{ }_\nu$
from $\mathbbm{P}^\rmii{E}_{\mu\nu}$ 
(cf.\ \eq\nr{projectors})
can be combined with those
from $\mathbbm{P}^\rmii{L}_{\mu\nu}$, in order to remove 
spurious poles. This way, we obtain an expression analogous
to that in \eqs\nr{htl_prop_w_pre} and \nr{htl_prop_w}, 
\ba
 [\Delta^{-1}_{P;Z}]^\rmii{ }_{\mu\nu}
 & = &  
 [\Delta^{-1}_{P;Z}]^\rmi{phys}_{\mu\nu}
 - 
 \frac{P^{ }_\mu P^{ }_\nu}{m_\Z^2 (P^2 + {m'}_{\!\!\Z}^2)}
 \;, \la{htl_prop_z_pre} \\[2mm]
 [\Delta^{-1}_{P;Z}]^\rmi{phys}_{\mu\nu}
 & = & 
  \PT_{\mu\nu}\, \F^\rmii{T}_{P} 
 + 
  \bigl( \delta^{ }_{\mu\nu} - \PT_{\mu\nu} \bigr)\, \F^\rmii{E}_{P}
 + 
  \frac{P^{ }_\mu P^{ }_\nu}{m^2_\Z}\,
  \, \widehat{\F}^\rmii{E}_{P}
 \;, \la{htl_prop_z_mid}
 \\[2mm] 
 \F^\rmii{T}_{P} & \equiv & 
 \frac{P^2_{ } + c^2_{ }\Pi^\rmiii{T1}_{P} + s^2_{ }\Pi^\rmiii{T2}_{P}}
 {
 (P^2_{ } + \Pi^\rmiii{T1}_{P} )
 (P^2_{ } + \Pi^\rmiii{T2}_{P} )
 + m^2_\Z
 (P^2_{ } + c^2_{ }\Pi^\rmiii{T1}_{P} + s^2_{ }\Pi^\rmiii{T2}_{P})
 }
 \;, \la{F_T}
 \\[2mm]
 \F^\rmii{E}_{P} & \equiv & 
 \frac{1 + c^2_{ }\widehat\Pi^\rmiii{E1}_{P} 
         + s^2_{ }\widehat\Pi^\rmiii{E2}_{P}}
 {
 P^2_{ }(1 + \widehat\Pi^\rmiii{E1}_{P} )
 (1 + \widehat\Pi^\rmiii{E2}_{P} )
 + m^2_\Z
 (1 + c^2_{ }\widehat\Pi^\rmiii{E1}_{P}
    + s^2_{ }\widehat\Pi^\rmiii{E2}_{P})
 }
 \;, \la{F_E}
 \\[2mm]
 \widehat{\F}^\rmii{E}_{P}
 & \equiv & 
 \frac{ 1 - m^2_\Z \F^\rmii{E}_{P} }{P^2_{ }} 
 \; = \; 
 \frac{(1 + \widehat\Pi^\rmiii{E1}_{P} )
 (1 + \widehat\Pi^\rmiii{E2}_{P} )
 }
 {
 P^2_{ }(1 + \widehat\Pi^\rmiii{E1}_{P} )
 (1 + \widehat\Pi^\rmiii{E2}_{P} )
 + m^2_\Z
 (1 + c^2_{ }\widehat\Pi^\rmiii{E1}_{P}
    + s^2_{ }\widehat\Pi^\rmiii{E2}_{P})
 }
 \;. \hspace*{5mm} \la{hatF_E}
\ea
If we set $s^2_{ }\to 0$, $c^2_{ }\to 1$, the charged propagators
in \eqs\nr{G_T}--\nr{hatG_E} are reproduced. 

\vspace*{3mm}

Turning finally to fermions (in practice we only consider quarks, 
which have the largest Yukawa couplings), the action has a similar
matrix structure as in \eq\nr{S_2}. With the example of top quarks, and 
denoting $t^{ }_\rmii{L,R} \equiv a^{ }_\rmii{L,R} t$, where
$
 a^{ }_\rmii{L,R} \equiv (\mathbbm{1} \mp \gamma^{ }_5)/2
$ 
are the chiral projectors, the quadratic part of the HTL 
effective action reads
\be
 S \; \supset \; \Tint{\{P\}}
 \bigl(
  \begin{array}{cc} 
    \bar{t}^{ }_{\rmii{L}}
      \; 
    \bar{t}^{ }_{\rmii{R}}
  \end{array}
 \bigr)
  \biggl(
  \begin{array}{cc} 
     \bsl{L}^{ }_{\!\P} & m^{ }_t 
     \\
     m^{ }_t & \bsl{R}^{ }_{\!\P}  
  \end{array}
 \biggr)
  \biggl(
  \begin{array}{c} 
    t^{ }_{\rmii{L}}
     \\ 
    t^{ }_{\rmii{R}}
  \end{array}
 \biggr)
 \;, \la{S_2_top}
\ee
where $\Tinti{\{P\}}$ denotes a fermionic Matsubara sum-integral. 
The four-vectors $L$ and $R$, introduced in ref.~\cite{hw}, have the 
forms 
\ba
 \bsl{L}^{ }_{\!\P} & = & L^{ }_{\P;\mu} \gamma^{\mu}_{ }
 \;, \la{L_def} \\[2mm]
 L^{ }_{\P;\mu} & \equiv & 
 \bigl\{ 
 p^{ }_0 
 \, \bigl( 1 + c^\rmii{W}_{\P;{t}^{ }_\rmiii{L}}\bigr)
 \,,\, p^{ }_i 
 \, \bigl( 1 + c^\rmii{P}_{\P;{t}^{ }_\rmiii{L}}\bigr)
 \bigr\} 
 \;, \la{L_vec}
\ea
and similarly for $L\to R$, 
where we have already gone over to 
Minkowskian Dirac matrices and carried out a Wick rotation
(in Minkowskian spacetime we use the metric signature
$+$$-$$-$$-$).
Here the HTL self-energies, the analogues of \eq\nr{PiT} 
and \nr{hatPiE}, read
\ba
 c^\rmii{W}_{\P;{t}^{ }_\rmiii{L}} & \equiv & 
  - \frac{m_{{t}^{ }_\rmiii{L}}^2}{2 p p^0_{ }}
    \ln \frac{p^0_{ } + p + i 0^+_{ }}{p^0_{ } - p + i 0^+_{ }} 
 \;, \la{cW} \\[2mm] 
 c^\rmii{P}_{\P;{t}^{ }_\rmiii{L}} & \equiv & 
  \frac{m_{{t}^{ }_\rmiii{L}}^2}{p^2_{ }}
  \biggl[\, 
  1  - \frac{p^0_{ }}{2 p}
    \ln \frac{p^0_{ } + p + i 0^+_{ }}{p^0_{ } - p + i 0^+_{ }} 
  \,\biggr]
 \;,  \la{cP}
\ea
and similarly with $L\to R$. 
For top and bottom quarks, 
the thermal mass parameters
read (cf.,\ e.g.,\ ref.~\cite{masses})
\ba
 m^2_{t^{ }_\rmiii{L}} & = & 
 m^2_{b^{ }_\rmiii{L}} \; \approx \; 
 T^2_{ }\biggl( 
   \frac{g_1^2}{288} + \frac{3 g_2^2}{32} + \frac{g_3^2}{6}
   + \frac{h_t^2 + h_b^2}{16}
 \biggr)
 \;, \la{mtL} \\[2mm]
 m^2_{t^{ }_\rmiii{R}} & \approx & 
 T^2_{ }\biggl( 
   \frac{g_1^2}{18} + \frac{g_3^2}{6}
   + \frac{h_t^2}{8}
 \biggr)
 \;, \quad  
 m^2_{b^{ }_\rmiii{R}} \; \approx \; 
 T^2_{ }\biggl( 
   \frac{g_1^2}{72} + \frac{g_3^2}{6}
   + \frac{h_b^2}{8}
 \biggr)
 \;. \la{mbR}
\ea

The inversion of the matrix from \eq\nr{S_2_top} was worked out 
in ref.~\cite{hw}. Showing the chiral projectors explicitly, 
the result can be written as 
\ba
 \bsl\Delta^{-1}_{\P;t} & = & 
 \F^{t}_{\P}\, 
 \biggl\{ 
   \aL \bigl(\, R^2_{\P}\, \bsl{L}^{ }_{\!\P}
             - m_t^2 \bsl{R}^{ }_{\!\P}   \,\bigr)\, \aR 
 \; - \; 
 \aL m^{ }_t \biggl(
   L^{ }_{\P}\cdot R^{ }_{\P} - m_t^2 
   + 
   \frac{
        [\bsl{L}^{ }_{\!\P},\bsl{R}^{ }_{\!\P}]
        }{2}
 \biggr) 
 \aL 
 \nn[2mm]
 & & \; + \,  
   \aR \bigl(\, L^2_{\P}\, \bsl{R}^{ }_{\!\P}
             - m_t^2 \bsl{L}^{ }_{\!\P}   \,\bigr)\, \aL
 \; - \; 
 \aR m^{ }_t \biggl(
   L^{ }_{\P}\cdot R^{ }_{\P} - m_t^2 
   - 
   \frac{
        [\bsl{L}^{ }_{\!\P},\bsl{R}^{ }_{\!\P}]
        }{2}
 \biggr) 
 \aR 
 \;\biggr\}
 \;, \hspace*{6mm} \la{t_prop_1} \\[2mm]
 \F^{t}_{\P} & \equiv & 
 \frac{1}{
  L^2_\P\, R^2_\P - 2 m_t^2 L^{ }_{\P}\cdot R^{ }_{\P} + m_t^4 
 }
 \;. \la{t_prop_2}
\ea

We note that if we only keep the contribution from the SU(3) 
coupling $g_3^2$ in \eqs\nr{mtL}--\nr{mbR}, which is the largest
individual term, then the thermal masses of the chiral states
agree with each other. In this situation, the quarks are vectorlike, 
with $L = R$. Consequently the propagator from \eqs\nr{t_prop_1}
and \nr{t_prop_2} can be greatly simplified, 
\ba
 \bsl\Delta^{-1}_{\P;t} & \stackrel{L\to R}{=} & 
 \frac{   
 m^{ }_t \,\mathbbm{1}
 - p^{ }_0 \gamma^0_{ } 
 \, \bigl( 1 + c^\rmii{W}_{\P}\bigr)
 - p^{ }_i \gamma^i_{ }
 \, \bigl( 1 + c^\rmii{P}_{\P}\bigr)
 }{ \Delta^{ }_{\P;t}  } 
 \;, \la{ferm_prop} \\[3mm] 
 \Delta^{ }_{\P;t}
 & \equiv &  
 m_t^2 
 + p^2_{ }\, \bigl(\, 1 + c^\rmii{P}_\P \,\bigr)^2_{ }
 - (p^0_{ })^2_{ }\, \bigl(\, 1 + c^\rmii{W}_\P \,\bigr)^2_{ }
 \;, \la{Delta_t}
\ea
where the thermal masses appearing in 
$ c^\rmii{W}_{\P} $ and
$ c^\rmii{P}_{\P} $ (cf.\ \eqs\nr{cW} and \nr{cP})
are now defined by retaining only the SU(3) 
parts of \eqs\nr{mtL} and \nr{mbR},  
\be
 m^2_{t^{ }_\rmiii{L=R}} 
 \; \equiv \; 
 \frac{g_3^2 T^2_{ }}{6}
 \;. \la{mtLR}
\ee

\vspace*{3mm}

\begin{figure}[t]

 \hspace*{-0.1cm}
 \centerline{%
  \epsfysize=7.2cm\epsfbox{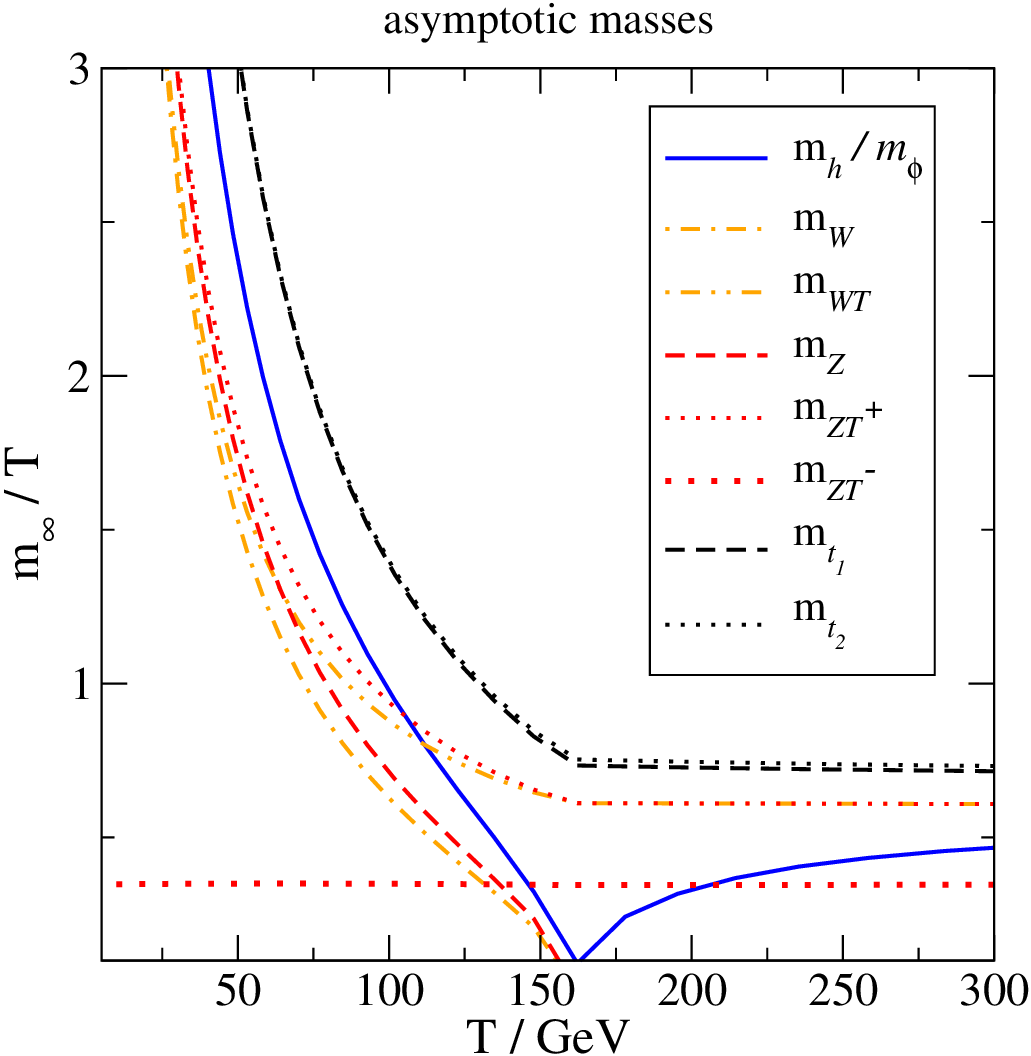}%
 }
 
 \caption[a]{\small
   An illustration of the asymptotic masses in various channels: 
   $m^{ }_h$ and $m^{ }_\phi$, 
   from \eqs\nr{m_h} and \nr{m_phi}; 
   $m_{\W}^{ }$ and $m_{\W\T}^{ }$, 
   from \eqs\nr{GT_inv} and \nr{GE_inv}; 
   $m_{\Z}^{ }$, $m^{ }_{\Z\T^+}$ and $m^{ }_{\Z\T^-}$,  
   from \eqs\nr{rhoE_Z} and \nr{poles_Z};
   and  
   $m^{ }_{ t_1}$ and $m^{ }_{ t_2}$, 
   from \eq\nr{t_masses}.
   We remark that $m_{\W\T}^{ }$ and $m^{ }_{\Z\T^+}$ become
   degenerate at $T > \Tc^{ }$, reflecting the absence of
   a Higgs mechanism, whereas $m^{ }_{\Z\T^-}$ represents the
   thermal asymptotic mass of a photon or hypercharge excitation. 
   }
 
 \la{fig:asy_masses}
\end{figure}

We end this subsection by stressing that 
when HTL corrections are included, the spin-$\frac{1}{2}$ 
and spin-1 propagators 
that we are faced with 
(\eqs\nr{G_T}--\nr{hatG_E} for $W$, 
\eqs\nr{F_T}--\nr{hatF_E} for $Z$, 
\eqs\nr{t_prop_1}--\nr{Delta_t} for $t$)
are non-polynomial functions of $p^0_{ }$ and $p$. 
In this situation there is no unambiguous notion of a ``mass''; 
rather, a dispersion relation, if defined as a location of a pole, 
depends on the spatial momentum in a non-covariant way. As  
will become apparent later on (cf.\ discussion around \eq\nr{power_counting}), 
for our problem relevant are hard momenta, $p \gg \pi T$. Then 
we can parametrize dispersion relations by what are called 
{\em asymptotic masses}, 
as
$ 
 p^0_{ } \approx p + m^2_{\infty}/(2 p ) + \rmO(1/p^3)
$.
The asymptotic masses can be defined in all cases, but there 
are more channels than in vacuum, with the $W$ splitting into
two channels (with asymptotic masses $m_{\W}^2$ and $m_{\W\T}^2$, 
cf.\ \eqs\nr{GT_inv} and \nr{GE_inv}); the $Z$ splitting into
three channels (with the asymptotic masses
$m_{\Z}^2$ and $m^2_{\Z\T^\pm}$,  
cf.\ \eqs\nr{rhoE_Z} and \nr{poles_Z}); and the $t$ splitting
into two channels (with the asymptotic masses
$m^2_{ t_1}$ and $m^2_{ t_2}$, 
cf.\ \eq\nr{t_masses}). We illustrate the numerical magnitudes 
of these asymptotic masses
in \fig\ref{fig:asy_masses}, 
together with the effective Higgs mass,
$m_h^{ }$ at $T < \Tc{ }$ from \eq\nr{m_h}
and 
$m^{ }_\phi$ at $T > \Tc^{ }$ from \eq\nr{m_phi}.

%
\subsection{Diagrams for a retarded singlet scalar self-energy}
\la{ss:diagrams}

Making use of the notation introduced in \se\ref{ss:notation}, 
and suppressing sum-integration symbols, the 
contributions of the various diagrams to the self-energy 
defined in \eq\nr{Pi_def} read
\ba
 \AmplHa    \hspace*{+0.3cm} 
 \Pi^{ }_{K;\varphi} & \supset &
 \frac{ \kappa^2_{ } }{2}
 \,\Delta^{-1}_{-K-P;\widetilde\varphi}
 \,\Bigl\{
 \,\Delta^{-1}_{Q;h}
 \,\Delta^{-1}_{P-Q;h}
 \nn[2mm]
 & & 
 + 
 \,\Delta^{-1}_{Q;G^{ }_\iZ}
 \,\Delta^{-1}_{P-Q;G^{ }_\iZ}
 +2
 \,\Delta^{-1}_{Q;G^{ }_\iW}
 \,\Delta^{-1}_{P-Q;G^{ }_\iW}   
 \,\Bigr\}
 \;, \la{AmplHa}
 \\[2mm]
 \AmplHb    \hspace*{+0.3cm} 
 \Pi^{ }_{K;\varphi} & \supset &
 - \kappa^2_{ } m_h^2 
 \,\Delta^{-1}_{-K-P;\widetilde\varphi}
 \,\Delta^{-1}_{P;h}
 \,\Bigl\{
 3 \,\Delta^{-1}_{Q;h}
 \,\Delta^{-1}_{P-Q;h}
 \nn[2mm]
 & & 
 + 
 \,\Delta^{-1}_{Q;G^{ }_\iZ}
 \,\Delta^{-1}_{P-Q;G^{ }_\iZ}
 +2
 \,\Delta^{-1}_{Q;G^{ }_\iW}
 \,\Delta^{-1}_{P-Q;G^{ }_\iW}   
 \,\Bigr\}
 \;, \la{AmplHb}
 \\[2mm]
 \AmplHf    \hspace*{+0.3cm} 
 \Pi^{ }_{K;\varphi} & \supset &
 \frac{ \kappa^2_{ } m^4_h }{2}
 \,\Delta^{-1}_{-K-P;\widetilde\varphi}
 \,\Delta^{-2}_{P;h}
 \Bigl\{
 9
 \,\Delta^{-1}_{Q;h}
 \,\Delta^{-1}_{P-Q;h}
 \nn[2mm]
 & & 
 +
 \,\Delta^{-1}_{Q;G^{ }_\iZ}
 \,\Delta^{-1}_{P-Q;G^{ }_\iZ}
 + 
 2 
 \,\Delta^{-1}_{Q;G^{ }_\iW}
 \,\Delta^{-1}_{P-Q;G^{ }_\iW}
 \, \Bigr\}
 \;, \la{AmplHf}
 \\[2mm]
 \AmplHc    \hspace*{+0.3cm} 
 \Pi^{ }_{K;\varphi} & \supset &
 - 2 \kappa^3_{ } v^2_{ }
 \,\Delta^{-1}_{-K-P;\widetilde\varphi}
 \,\Delta^{-1}_{-K-Q;\widetilde\varphi}
 \,\Delta^{-1}_{P;h}
 \,\Delta^{-1}_{P-Q;h}
 \;, \la{AmplHc}
 \\[2mm]
 \AmplHd    \hspace*{+0.3cm} 
 \Pi^{ }_{K;\varphi} & \supset &
 3 \kappa^3_{ } v^2_{ } m_h^2 
 \,\Delta^{-1}_{-K-P;\widetilde\varphi}
 \,\Delta^{-1}_{-K-Q;\widetilde\varphi}
 \,\Delta^{-1}_{P;h}
 \,\Delta^{-1}_{Q;h}
 \,\Delta^{-1}_{P-Q;h}
 \;, \la{AmplHd}
 \\[2mm]
 \AmplHe    \hspace*{+0.3cm} 
 \Pi^{ }_{K;\varphi} & \supset &
 \kappa^4_{ } v^4_{ }  
 \,\Delta^{-1}_{-K-P;\widetilde\varphi}
 \,\Delta^{-1}_{Q;\widetilde\varphi}
 \,\Delta^{-1}_{P-Q;\widetilde\varphi}
 \,\Delta^{-1}_{-K-Q;h}
 \,\Delta^{-1}_{P;h}
 \;, \la{AmplHe}
 \\[2mm]
 \AmplHg    \hspace*{+0.3cm} 
 \Pi^{ }_{K;\varphi} & \supset &
 \kappa^4_{ } v^4_{ }  
 \,\Delta^{-2}_{-K-P;\widetilde\varphi}
 \,\Delta^{-1}_{P-Q;\widetilde\varphi}
 \,\Delta^{-1}_{-K-Q;h}
 \,\Delta^{-1}_{P;h}
 \;, \la{AmplHg}
 \\[2mm]
 \AmplHi    \hspace*{+0.3cm} 
 \Pi^{ }_{K;\varphi} & \supset &
 \kappa^2_{ }
 \,\Delta^{-1}_{-K-P;\widetilde\varphi}
 \,\Delta^{-2}_{P;h}
 (P^{ }_\mu + Q^{ }_\mu)
 (P^{ }_\nu + Q^{ }_\nu)
 \Bigl\{ 
 \nn[2mm]
 & & 
 + 
 \,\Delta^{-1}_{Q;G^{ }_\iW}
 \,[\Delta^{-1}_{P-Q;W}]^{ }_{\mu\nu}
 [\, 2 m^2_{\W} \,]
 \nn[2mm]
 & & 
 + 
 \,\Delta^{-1}_{Q;G^{ }_\iZ}
 \,[\Delta^{-1}_{P-Q;Z}]^{ }_{\mu\nu}
 [\, m^2_{\Z} \,]
  \, \Bigr\} 
 \;, \la{AmplHi}
 \\[2mm]
 \AmplHj    \hspace*{+0.3cm} 
 \Pi^{ }_{K;\varphi} & \supset &
 2\, \kappa^2_{ }
 \,\Delta^{-1}_{-K-P;\widetilde\varphi}
 \,\Delta^{-2}_{P;h}
 \Bigl\{ 
 \,[\Delta^{-1}_{Q;W}]^{ }_{\mu\nu}
 \,[\Delta^{-1}_{P-Q;W}]^{ }_{\nu\mu}
 [\, 2 m^4_{\W} \,]
 \nn[2mm]
 & & 
  + 
 \,[\Delta^{-1}_{Q;Z}]^{ }_{\mu\nu}
 \,[\Delta^{-1}_{P-Q;Z}]^{ }_{\nu\mu}
 [\, m^4_{\Z} \,]
 \, \Bigr\} 
 \;, \la{AmplHj}
 \\[2mm]
 \AmplHk    \hspace*{+0.3cm} 
 \Pi^{ }_{K;\varphi} & \supset &
 -\,\kappa^2_{ }
 \,\Delta^{-1}_{-K-P;\widetilde\varphi}
 \,\Delta^{-2}_{P;h}
 \Bigl\{ 
 \,\Delta^{-1}_{Q;c^{ }_\iW}
 \,\Delta^{-1}_{P-Q;\bar{c}^{ }_\iW}
 [\, 2 {m'}^4_{\!\!\W} \,] 
 \nn[2mm]
 & & 
  + 
 \,\Delta^{-1}_{Q;c^{ }_\iZ}
 \,\Delta^{-1}_{P-Q;\bar{c}^{ }_\iZ}
 [\, {m'}^4_{\!\!\Z} \,] 
 \, \Bigr\} 
 \;, \la{AmplHk}
 \\[2mm]
 \AmplHl    \hspace*{+0.3cm} 
 \Pi^{ }_{K;\varphi} & \supset &
 -\,\kappa^2_{ } \Nc^{ }
 \,\Delta^{-1}_{-K-P;\widetilde\varphi}
 \,\Delta^{-2}_{P;h}
 \,\Bigl\{\,  
 m_t^2\,
 \tr\bigl[
 \,\bsl\Delta^{-1}_{Q;t}
 \,\bsl\Delta^{-1}_{Q-P;t}
 \,\bigr]
 \nn[2mm]
 & & + 
 m_b^2\,
 \tr\bigl[
 \,\bsl\Delta^{-1}_{Q;b}
 \,\bsl\Delta^{-1}_{Q-P;b}
 \,\bigr]
 \,\Bigr\}
 \;. \la{AmplHl}
\ea

%
\subsection{Cancellation of gauge parts}

An important consistency check of the computation is that all gauge 
dependence cancels. In terms of \eqs\nr{AmplHa}--\nr{AmplHl}, 
this is equivalent to the disappearance of the masses 
${m'}^{ }_{\!\!\W}$ and ${m'}^{ }_{\!\!\Z}$. Such effects originate
from the Goldstone propagators in 
\eqs\nr{AmplHa}, \nr{AmplHb}, \nr{AmplHf} and \nr{AmplHi}; 
the longitudinal parts of the gauge propagators in 
\eqs\nr{AmplHi} and \nr{AmplHj}; 
and the ghost propagators in \eq\nr{AmplHk}. Having written the
gauge propagators in analogous forms in 
\eqs\nr{htl_prop_w_pre}--\nr{htl_prop_w} and
\nr{htl_prop_z_pre}--\nr{hatF_E}, respectively, the cancellation
takes place in the same way in the charged and neutral sectors. 
For instance, 
collecting together the gauge terms of the charged sector, we find
\ba
 \Pi^{ }_{K;\varphi} & \supset &
 \kappa^2_{ }
   \Delta^{-1}_{-K-P;\widetilde\varphi}
 \,\Delta^{-2}_{P;h}
 \,\Delta^{-1}_{Q;G^{ }_\iW}
 \,\Delta^{-1}_{P-Q;G^{ }_\iW}
 \Bigl\{ 
 \nn[2mm] 
 & & 
 + (P^2_{ } + m_h^2)^2{ } 
 - 2 m_h^2 (P^2_{ } + m_h^2)
 + m_h^4
 - 2 {m'}^4_{\!\!\W}
 \nn[2mm]
 & & 
 - 2 (P^{ }_\mu + Q^{ }_\mu)(P^{ }_\nu + Q^{ }_\nu)
     (P^{ }_\mu - Q^{ }_\mu)(P^{ }_\nu - Q^{ }_\nu)
 \nn[2mm]
 & &  
 + 4 Q^{ }_\mu Q^{ }_\nu (P^{ }_\mu - Q^{ }_\mu)(P^{ }_\nu - Q^{ }_\nu)
 \Bigr\}
 \nn[2mm] 
 & + & 
 2 \kappa^2_{ } m^2_\W
   \Delta^{-1}_{-K-P;\widetilde\varphi}
 \,\Delta^{-2}_{P;h}
 \,\Delta^{-1}_{Q;G_\iW}
 \,[\Delta^{-1}_{P-Q;W}]^\rmi{phys}_{\mu\nu}
 \Bigl\{ 
 \nn[2mm] 
 & & 
 + (P^{ }_\mu  + Q^{ }_\mu)
   (P^{ }_\nu  + Q^{ }_\nu )
 - 4 Q^{ }_\mu Q^{ }_\nu
 \Bigr\}
 \;. \la{gauge_W}
\ea
By completing squares in the first part, {\it viz.}
\ba
 && \hspace*{-0.5cm} 
 -\, 2 (P^2_{ } - Q^2_{ })^2_{ }
 + 4(Q\cdot P - Q^2_{ })^2_{ }
 \; = \; 
  2 {m'}^4_{\!\!\W} - P^4_{ } 
 \la{gauge_1st} \\ 
 & + & 
   \Delta^2_{P-Q;G^{ }_\iW}
 + 2 \Delta^{ }_{P-Q;G^{ }_\iW} \Delta^{ }_{Q;G^{ }_\iW}
 - \Delta^2_{Q;G^{ }_\iW}
 - 2 \Delta^{ }_{P-Q;G^{ }_\iW} [P^2_{ }+ 2 {m'}^2_{\!\!\W} ]
 + 2 \Delta^{ }_{Q;G^{ }_\iW} P^2_{ }
 \;, \nonumber 
\ea
we note that 
the terms on the first line of \eq\nr{gauge_1st}
cancel against the second line
of \eq\nr{gauge_W}, whereas the terms of the second
line of \eq\nr{gauge_1st} remove at least one of the 
propagators, leaving over terms in 
which one loop is a 
tadpole, i.e.\ factorized from the rest. Such terms yield no
cut corresponding to a $2\to 2$ process. 
In the second part of \eq\nr{gauge_W}, 
after writing 
$
 (P^{ }_\mu + Q^{ }_\mu) = 
 (P^{ }_\mu - Q^{ }_\mu) + 2 Q^{ }_\mu
$,
the term 
$
 [\Delta^{-1}_{P-Q;W}]^\rmi{phys}_{\mu\nu} (P^{ }_\mu - Q^{ }_\mu)
$
yields a structure containing no poles, whereas 
$4 Q^{ }_\mu Q^{ }_\nu$ cancels. In total, then, 
gauge dependence indeed cancels from the $2\to 2$ cut of \eq\nr{gauge_W}.
An analogous cancellation takes place in the neutral sector. 

%
\subsection{Physical parts}
\la{ss:phys_parts}

Moving over to the physical part, we just write down some examples
for now. From \eq\nr{AmplHj}, 
the physical part of the neutral gauge boson propagator yields
\ba
 \AmplHj    \hspace*{+0.3cm} 
 \Pi^{ }_{K;\varphi} & \supset &
 \frac{ \kappa^2_{ } }{2}
 \,\Delta^{-1}_{-K-P;\widetilde\varphi}
 \,\Delta^{-2}_{P;h}
 \Bigl\{\, 
 \nn[2mm]
 & & 
 \;
 \overset{\rmii{(a)}}{+}
 \;
 \bigl[\, P^2_{ } - Q^2_{ }- (P-Q)^2_{ } \,\bigr]^2_{ }
 \widehat{\F}^\rmii{E}_{Q}
 \,\widehat{\F}^\rmii{E}_{P-Q}
 \nn[2mm]
 & & 
  \;
 \overset{\rmii{(b)}}{+}
  \;
  8 m^2_\Z Q^{ }_\mu Q^{ }_\nu\, 
    (\mathbbm{P}-\mathbbm{Q})^\rmii{T}_{\mu\nu}
    \,\widehat{\F}^\rmii{E}_Q 
    \,\bigl( \F^\rmii{T}_{P-Q} - \F^\rmii{E}_{P-Q} \bigr)
 \nn[2mm]
 & & 
  \;
 \overset{\rmii{(c)}}{+}
 \;
    4 m^4_\Z  
    \,\bigl[\,  
     \mathbbm{Q}^\rmii{T}_{\mu\nu}
    (\mathbbm{P}-\mathbbm{Q})^\rmii{T}_{\mu\nu}
    - 2
    \,\bigr]\,
    \bigl( \F^\rmii{T}_{Q} - \F^\rmii{E}_{Q} \bigr)\,
    \bigl( \F^\rmii{T}_{P-Q} - \F^\rmii{E}_{P-Q} \bigr)\,
 \nn[2mm]
 & & 
  \;
 \overset{\rmii{(d)}}{+}
  \;
  8 m^4_\Z \, \F^\rmii{T}_{Q} \, \F^\rmii{T}_{P-Q}
 \, \Bigr\} 
 \;, \la{AmplHj_phys}
\ea
where we have again left out terms that do not yield 
a $2\to 2$ cut. The labelling ``(a)--(d)'' will be employed later on, 
when discussing the various terms.
Charged gauge bosons give 
an analogous expression, once we replace 
$\F^\rmii{T,E}_{ } \to \G^\rmii{T,E}_{ }$ 
and multiply the result by~2.

For quarks, 
inserting \eq\nr{t_prop_1} into
\eq\nr{AmplHl} yields
\ba
 \AmplHl    \hspace*{+0.3cm} 
 \Pi^{ }_{K;\varphi} & \supset &
 - 2 \kappa^2_{ } m_t^2 \Nc^{ }
 \,\Delta^{-1}_{-K-P;\widetilde\varphi}
 \,\Delta^{-2}_{P;h}
 \,\F^{t}_{\Q}
 \,\F^{t}_{\Q-\P}
 \,\Bigl\{\,  
 \nn[2mm]
 & & \hspace*{-2cm} 
 +\,
  L^{ }_{\Q}\cdot R^{ }_{\Q-\P} 
  \bigl( R^2_{\Q} L^2_{\Q-\P} + m_t^4 \bigr)
 \,+\, 
  R^{ }_{\Q}\cdot L^{ }_{\Q-\P} 
  \bigl( L^2_{\Q} R^2_{\Q-\P} + m_t^4 \bigr)
  \nn[2mm]
 & & \hspace*{-2cm}
 +\, m_t^2 \,\Bigl[\,
  2 \bigl( L^{ }_{\Q}\cdot R^{ }_{\Q} - m_t^2 \bigr)
  \bigl( L^{ }_{\Q-\P}\cdot R^{ }_{\Q-\P} - m_t^2 \bigr)
  \nn[2mm]
 & & \hspace*{-2cm}
 +\, 
  2 \,\bigl(\, 
   L^{ }_{\Q}\cdot R^{ }_{\Q-\P} \, 
   R^{ }_{\Q}\cdot L^{ }_{\Q-\P}
  - 
   L^{ }_{\Q}\cdot L^{ }_{\Q-\P} \,
   R^{ }_{\Q}\cdot R^{ }_{\Q-\P}
   \,\bigr)
  \nn[2mm]
 & & \hspace*{-2cm}
 -\,
  L^{ }_{\Q}\cdot L^{ }_{\Q-\P} \,
  \bigl(  
    R^{2}_{\Q}
  + R^{2}_{\Q-\P}
  \bigr)
 \,-\,
  R^{ }_{\Q}\cdot R^{ }_{\Q-\P} \,
  \bigl(  
    L^{2}_{\Q}
  + L^{2}_{\Q-\P}
  \bigr)
  \,\Bigr]
 \,\Bigr\}
 \;, \hspace*{6mm} \la{AmplHl_phys}
\ea
with an analogous expression for the bottom quarks. Here $L$ and $R$
are four-vectors, defined via 
\eqs\nr{L_def}--\nr{mbR}.
In the chirally
invariant limit, $L\to R$, 
\eq\nr{AmplHl_phys} reduces to the much simpler expression
that can be obtained from \eq\nr{ferm_prop}, {\it viz.}
\ba
 \AmplHl    \hspace*{+0.3cm} 
 \Pi^{ }_{K;\varphi} & \stackrel{L\to R}{\supset} &
 4 \kappa^2_{ } m_t^2 \Nc^{ }
 \,\Delta^{-1}_{-K-P;\widetilde\varphi}
 \,\Delta^{-2}_{P;h}
 \,\Delta^{-1}_{Q;t}
 \,\Delta^{-1}_{Q-P;t}
 \,\Bigl\{\,  
 \nn[2mm]
 & & 
 +\, 
 \vec{q}\cdot(\vec{q-p})
 \bigl( 1 + c^\rmii{P}_{\Q} \bigr)
 \bigl( 1 + c^\rmii{P}_{\Q-\P} \bigr)
 \nn[2mm]
 & & 
 -\, 
 q^0_{ }(q^0_{ } -p^0_{ })
 \bigl( 1 + c^\rmii{W}_{\Q} \bigr)
 \bigl( 1 + c^\rmii{W}_{\Q-\P} \bigr)
 - m^2_t
 \,\Bigr\}
 \;, \la{AmplHl_phys_appro}
\ea
where the thermal mass appearing in 
$ c^\rmii{W}_{\Q} $ and
$ c^\rmii{P}_{\Q} $ 
is defined via \eq\nr{mtLR}. 

%
\section{Cuts of self-energy diagrams}
\la{se:cuts}

Having determined the self-energy 
$
  \Pi^{ }_{K;\varphi} 
$
(cf.\ \se\ref{ss:phys_parts}), 
the next task is to analytically continue it to Minkowskian
spacetime and then extract its imaginary part, or ``cut'', 
$
  \im \Pi^{ }_{\K;\varphi}
$, 
which is what is needed in \eq\nr{Gamma_def} and then  
\eq\nr{sigmav}.
We illustrate here how this can be done, if we do not know
{\it a priori} whether the final-state particles can be 
represented as on-shell excitations. This means that they will
appear as general ``spectral functions'', $\varrho^{ }_\Q$
(cf.\ \eq\nr{varrho}). 

%
\subsection{A prototypical example}

Let us consider a structure similar to what appears in the gauge 
contribution, \eq\nr{AmplHj_phys},  
but simplified to the core. Denoting by 
$\F^{ }_{ }$ a resummed propagator, 
by $\Phi$ terms that will not be cut, 
and by $\;\deltabar(P)$ a Dirac-delta normalized as 
$\Tinti{P}\;\deltabar(P)=1$, we focus on
\ba
 I^{ }_{K} & \equiv & 
 \Delta^{-1}_{-K-P;\tilde\varphi}\,
 \F^{ }_{Q}\, \F^{ }_{P-Q}\,\Phi
 \la{def_I}
 \\[2mm]
 & \equiv &
 \Tint{P,Q} 
 \frac{  \F^{ }_{Q}\, \F^{ }_{P-Q}\,\Phi }
      {(K+P)^2_{ } + m_\varphi^2}\, 
 \nn[2mm]
 & \stackrel{P\,\equiv\, -H-K}{=}& 
 \Tint{H,Q}
 \frac{ \F^{ }_{Q}\, \F^{ }_{-K-H-Q}\,\Phi }
 {H^2_{ } + m_\varphi^2}\, 
 \; = \; 
 \Tint{H,Q,R} 
 \;\deltabar(K+H+Q+R) 
 \frac{  \F^{ }_{Q}\, \F^{ }_{R}\,\Phi  }
 {H^2_{ } + m_\varphi^2}\, 
 \nn[2mm]
 & \stackrel{\epsilon_h^2 \;\equiv\; h^2_{ } + m_\varphi^2 }{=} &  
 \Tint{H,Q,R} 
 \int_{-\infty}^{\infty} \! \frac{{\rm d} q^{ }_0}{\pi}
 \int_{-\infty}^{\infty} \! \frac{{\rm d} r^{ }_0}{\pi} \;
 \frac{\deltabar(K+H+Q+R)}{h_n^2 + \epsilon_h^2}
 \frac{\varrho^{ }_\Q}{q^{ }_0 - i q^{ }_n}
 \frac{\varrho^{ }_\R}{r^{ }_0 - i r^{ }_n} \,\Phi
 \;, \hspace*{7mm}
\ea
where we represented $\F^{ }_Q$ in a spectral representation, 
\be
 \F^{ }_Q 
 \; = \; 
 \int_{-\infty}^{\infty} \! \frac{{\rm d} q^{ }_0}{\pi}
 \frac{\varrho^{ }_\Q}{q^{ }_0 - i q^{ }_n}
 \;, \quad
 \varrho^{ }_\Q \equiv \im \F^{ }_Q 
 |^{ }_{q^{ }_n \to -i (q^{ }_0 + i 0^+_{ })}
 \;. \la{varrho}
\ee
For free particles, spectral functions amount to localized
``poles'' (cf.\ \eq\nr{rho_free}), but more generally they
are continuous ``cuts''. 

We now express the temporal part of $\;\deltabar(K+H+Q+R)$ as a Fourier
integral, and carry out the Matsubara sums, 
\ba
 I^{ }_K 
 & = & 
 \int_{\vec{h,q,r}} \deltabar(\vec{k+h+q+r})
 \int_{-\infty}^{\infty} \! \frac{{\rm d} q^{ }_0}{\pi}
 \int_{-\infty}^{\infty} \! \frac{{\rm d} r^{ }_0}{\pi}
 \;
 \varrho^{ }_\Q
 \;
 \varrho^{ }_\R
 \,\Phi
 \nn[2mm]
 & \times &  
 \int_0^\beta \! {\rm d}\tau \, e^{i k^{ }_n\tau}_{ }
 \biggl\{ 
   \underbrace{ 
   T\sum_{h_n} \frac{e^{i h^{ }_n\tau}_{ }}{h_n^2 + \epsilon_h^2}
   }_{
   \frac{\niB^{ }(\epsilon^{ }_h)}{2\epsilon^{ }_h}
   \bigl[ e^{(\beta-\tau)\epsilon^{ }_h}_{ } + e^{\tau\epsilon^{ }_h}_{ }\bigr]
   }
 \biggr\}
 \biggl\{
    \underbrace{ 
    T\sum_{q_n} \frac{e^{i q^{ }_n\tau}_{ }}{q^{ }_0 - i q^{ }_n} 
    }_{ 
     \niB(q^{ }_0)\, e^{\tau q^{ }_0}_{ }
    }
 \biggr\}
 \biggl\{
    \underbrace{ 
    T\sum_{r_n} \frac{e^{i r^{ }_n\tau}_{ }}{r^{ }_0 - i r^{ }_n} 
    }_{ 
     \niB(r^{ }_0)\, e^{\tau r^{ }_0}_{ }
    }
 \biggr\}
 \nn[2mm]
 & = & 
 \int_{\vec{h,q,r}} \deltabar(\vec{k+h+q+r})
 \int_{-\infty}^{\infty} \! \frac{{\rm d} q^{ }_0}{\pi}
 \int_{-\infty}^{\infty} \! \frac{{\rm d} r^{ }_0}{\pi} 
 \;
 \varrho^{ }_\Q
 \;
 \varrho^{ }_\R
 \,\Phi
 \, 
 \frac{\niB^{ }(\epsilon^{ }_h)\niB^{ }(q^{ }_0)\niB^{ }(r^{ }_0)}
 {2\epsilon^{ }_h}
 \nn[2mm]
 & \times & 
 \biggl\{
   \frac{e^{\beta(q^{ }_0 + r^{ }_0)}_{ } - e^{\beta\epsilon^{ }_h}_{ } }
        {i k^{ }_n - \epsilon^{ }_h + q^{ }_0 + r^{ }_0}
 + 
   \frac{e^{\beta(\epsilon^{ }_h + q^{ }_0 + r^{ }_0)}_{ } - 1 }
        {i k^{ }_n + \epsilon^{ }_h + q^{ }_0 + r^{ }_0}
 \biggr\}
 \;. 
\ea
We then set $k^{ }_n \to -i (\omega + i 0^+_{ })$ 
and take the imaginary part, by making use of 
\be
 \im \frac{1}{\Delta + i 0^+_{ }} = -\pi \delta(\Delta)
 \;. 
\ee
This yields
\ba
 \im I^{ }_{\K} 
 & = &
 \frac{1}{2}
 \int_{\vec{h,q,r}} 
 \frac{
  \deltabar(\vec{k+h+q+r}) }
 {2\epsilon^{ }_h}
 \int_{-\infty}^{\infty} \! \frac{{\rm d} q^{ }_0}{\pi}
 \int_{-\infty}^{\infty} \! \frac{{\rm d} r^{ }_0}{\pi} 
 \;
 \varrho^{ }_\Q
 \;
 \varrho^{ }_\R
 \,\Phi
 \, 
 \nn[2mm]
 & \times & 
 \Bigl\{\,
 2\pi \delta(\omega - \epsilon^{ }_h + q^{ }_0 + r^{ }_0) 
 \, \bigl[\, 
    \bar\nB^{ }(\epsilon^{ }_h)
    \nB^{ }(q^{ }_0)
    \nB^{ }(r^{ }_0)
    - 
    \nB^{ }(\epsilon^{ }_h)
    \bar\nB^{ }(q^{ }_0)
    \bar\nB^{ }(r^{ }_0)
 \,\bigr]
 \nn[2mm]
 & &  + \, 
 2\pi \delta(\omega + \epsilon^{ }_h + q^{ }_0 + r^{ }_0) 
 \, \bigl[\, 
    \nB^{ }(\epsilon^{ }_h)
    \nB^{ }(q^{ }_0)
    \nB^{ }(r^{ }_0)
    - 
    \bar\nB^{ }(\epsilon^{ }_h)
    \bar\nB^{ }(q^{ }_0)
    \bar\nB^{ }(r^{ }_0)
 \,\bigr]
 \,\Bigr\}
 \nn[2mm]
 & 
  \underset{\R\to-\R}{
  \overset{\Q\to-\Q}{=}} 
 &
 \frac{1}{2}
 \int_{\vec{h,q,r}} 
 \frac{
  \deltabar(\vec{k+h-q-r}) }
 {2\epsilon^{ }_h}
 \int_{-\infty}^{\infty} \! \frac{{\rm d} q^{ }_0}{\pi}
 \int_{-\infty}^{\infty} \! \frac{{\rm d} r^{ }_0}{\pi} 
 \;
 \varrho^{ }_\Q
 \;
 \varrho^{ }_\R
 \,\Phi
 \, 
 \la{cut_I} \\[2mm]
 & \times & 
 \Bigl\{\,
 2\pi \delta(\omega - \epsilon^{ }_h - q^{ }_0 - r^{ }_0) 
 \, \bigl[\, 
    \bar\nB^{ }(\epsilon^{ }_h)
    \bar\nB^{ }(q^{ }_0)
    \bar\nB^{ }(r^{ }_0)
    - 
    \nB^{ }(\epsilon^{ }_h)
    \nB^{ }(q^{ }_0)
    \nB^{ }(r^{ }_0)
 \,\bigr]
 \nn[2mm]
 & &  + \, 
 2\pi \delta(\omega + \epsilon^{ }_h - q^{ }_0 - r^{ }_0) 
 \, \bigl[\, 
    \nB^{ }(\epsilon^{ }_h)
    \bar\nB^{ }(q^{ }_0)
    \bar\nB^{ }(r^{ }_0)
    - 
    \bar\nB^{ }(\epsilon^{ }_h)
    \nB^{ }(q^{ }_0)
    \nB^{ }(r^{ }_0)
 \,\bigr]
 \,\Bigr\}
 \;, \hspace*{6mm} \nonumber 
\ea
where in the second step we made use of the (bosonic) properties 
\be
 \varrho^{ }_{-\Q} \; = \; - \varrho^{ }_\Q
 \;, \quad
 \nB^{ }(-q^{ }_0) \; = \; - \bar\nB^{ }(q^{ }_0)
 \;, \quad
 \bar\nB^{ }(q^{ }_0)
 \; \equiv \; 
 1 + \nB^{ }(q^{ }_0)
 \;. 
\ee

In order to understand which part of the large integration domain
of \eq\nr{cut_I} is important, let us for a moment 
consider an unresummed propagator, with 
\be
 \F^{ }_Q
 \; = \;
 \frac{1}{q_n^2 + \epsilon_q^2}
 \;
 \stackrel{\rmii{\nr{varrho}}}{\Rightarrow} 
 \;
 \varrho^{ }_\Q 
 \; = \; 
 \frac{\pi}{2 \epsilon^{ }_q}
 \,\bigl[\,
  \delta(q^{ }_0 - \epsilon^{ }_q) - \delta(q^{ }_0 + \epsilon^{ }_q) 
 \,\bigr]
 \;. \la{rho_free}
\ee
If we insert this in \eq\nr{cut_I}, the two channels there 
split into four different channels each, yielding in total eight
possibilities, corresponding to 
$1\to 3$, $2\to 2$, $3\to 1$, and a hypothetical $4\to 0$ process. 
Among these, only the $2\to 2$ ones are kinematically realized 
in the non-relativistic limit. Among the $2\to 2$ channels, 
only one is relevant for the inelastic rate, namely that with 
$\varphi$ and $\tilde\varphi$ in the initial state. 
This originates from
the second row of \eq\nr{cut_I}, with $q^{ }_0$ and $r^{ }_0$ taken 
positive. Furthermore, we note that 
\ba
 \nB^{ }(\epsilon^{ }_h) 
 \bar\nB^{ }(q^{ }_0)
 \bar\nB^{ }(r^{ }_0)
 & = & 
 \nB^{ }(\epsilon^{ }_h) 
 \nB^{ }(q^{ }_0) 
 \nB^{ }(r^{ }_0)
 \, e^{(q^{ }_0 + r^{ }_0)/T}_{ } 
 \;,  \\[2mm]
 \bar\nB^{ }(\epsilon^{ }_h) 
 \nB^{ }(q^{ }_0)
 \nB^{ }(r^{ }_0)
 & = & 
 \nB^{ }(\epsilon^{ }_h) 
 \nB^{ }(q^{ }_0) 
 \nB^{ }(r^{ }_0)
 \, e^{\epsilon^{ }_h/T}_{ } 
 \;. \la{kinematics}
\ea
Given that 
$
 q^{ }_0 + r^{ }_0
 = \omega + \epsilon^{ }_h
 \gg \epsilon^{ }_h
$, 
the first factor, which we may refer to as a loss term, 
dominates by an exponentially large amount over the gain term
($\omega/T \ge m^{ }_\varphi/T \gg 1$). 
Moreover, since $q^{ }_0$ and $r^{ }_0$ carry the energy 
released from the co-annihilation of $\varphi$ and $\widetilde\varphi$, 
each of them is large. Therefore, 
the final-state Bose enhancements, 
$ \nB^{ }(q^{ }_0) $ 
and 
$ \nB^{ }(r^{ }_0) $,
are exponentially small, 
and can be omitted in practice, 
implying that 
$
  \bar \nB^{ }(q^{ }_0) 
  \approx 1 
  \approx 
  \bar \nB^{ }(r^{ }_0) 
$.
Similarly, the initial-state factor 
$
 \nB^{ }(\epsilon^{ }_h)
$, 
as well as the corresponding 
$
 \nB^{ }(\omega)
$
from \eq\nr{sigmav}, can be replaced by the leading terms
in a dilute expansion, 
\be
 \nB^{ }(\epsilon^{ }_h) \to e^{-\beta \epsilon^{ }_h }_{}
 \;, \quad
 \nB^{ }(\omega) \to e^{-\beta \omega}_{ }
 \;. \la{dilute}
\ee 
Note that as long as we do not replace the energies
by their non-relativistic limits, the dilute expansion
is rapidly convergent. 
With these simplifications, \eq\nr{cut_I}, 
with $\varrho^{ }_\Q$ and $\varrho^{ }_\R$ restored to their
general forms, serves as the starting point for the next 
steps (cf.\ \eq\nr{cut_I_2}).

%
\subsection{Choice of kinematic variables}

Before we insert the spectral functions 
$\varrho^{ }_\Q$ and $\varrho^{ }_\R$
into \eq\nr{cut_I}, 
it is helpful to rewrite the integration measure in terms of 
new variables. 
Starting from \eqs\nr{sigmav}
and \nr{cut_I}; making use of the simplifications explained between
\eqs\nr{kinematics} and \nr{dilute}; and noting from 
\eqs\nr{AmplHj_phys}--\nr{AmplHl_phys_appro} 
that the four-momenta appearing in the weight
of the integrand can normally be expressed in terms of 
${Q}$ and
$R \equiv P - {Q} $, which become 
$\Q$ and $\R$ after going over
to the spectral representation, we consider an integral of the
form 
\ba
 \langle \sigma \vrel \rangle 
 &
 \underset{\rmii{\nr{cut_I},\nr{dilute}}}{
 \overset{\rmii{\nr{sigmav}}}{\supset}} 
 &
 \frac{1}{n^{2}_\rmi{eq}}   
 \int_{\vec{k,h,q,r}} 
 \int_{-\infty}^{\infty} \! \frac{{\rm d} q^{ }_0}{\pi}
 \int_{-\infty}^{\infty} \! \frac{{\rm d} r^{ }_0}{\pi} 
 \frac{(2\pi)^4_{ }
  \delta^{(4)}_{ }(\K + \H - \Q - \R) }
 {4 \,\omega\, \epsilon^{ }_h}
 \nn[2mm]
 & & \; \times
 \;
 \varrho^{ }_\Q
 \;
 \varrho^{ }_\R
 \, 
 \, e^{ -\beta ( \omega + \epsilon^{ }_h )}_{ }
 \,
 \Phi(\Q,\R)
 \;. \hspace*{6mm} \la{cut_I_2} 
\ea
We now re-introduce the center-of-mass four-momentum 
$\P = (p^{ }_0,\vec{p})$ and write 
\be
 \delta^{(4)}_{ }(\K + \H - \Q - \R) 
 = 
 \int \! {\rm d}^4_{ }\P \, 
 \delta^{(4)}_{ }(\K + \H - \P) 
 \, 
 \delta^{(4)}_{ }(\P - \Q - \R) 
 \;. \la{def_P}
\ee
Then we change the order of integrations, and 
consider the outer integral 
\be
 I \; \equiv \; 
 \int_{\vec{k,h}}
 \frac{ 
 (2\pi)^4_{ } \delta(\omega + \epsilon^{ }_h - p^{ }_0)
              \delta^{(3)}_{ }(\vec{k+h-p})
 }{ 4 \,\omega\, \epsilon^{ }_h }
 \; = \; 
 \int_{\vec{k}}
 \frac{2\pi \delta( \omega + \epsilon^{ }_{pk} - p^{ }_0 )}
 {4 \, \omega \, \epsilon^{ }_{pk}}
 \;, \la{factorize}
\ee
where
$
 \omega \equiv \sqrt{k^2_{ } + m_\varphi^2 }
$
and  
$
 \epsilon^{ }_{pk} \equiv \sqrt{(\vec{p-k})^2_{ }  + m_\varphi^2}
$.
The Dirac-$\delta$ can give a contribution
if $p^{ }_0 > \sqrt{p^2_{ } + 4 m_\varphi^2}$.
Rather than $k$, it is helpful to take $\omega$
as an integration variable, and then 
the integration boundaries can be established as 
\be
 k^{ }_{\pm} 
 \; 
 \overset{\P^2_{ }\,>\,4m_\varphi^2}{=} 
 \;
 \frac{p}{2} \pm 
 \frac{p^{ }_0}{2} 
 \sqrt{ 1 - \frac{4m_\varphi^2}{\P^2_{ }} }
 \; \Rightarrow \;
 \omega^{ }_{\pm} 
 \; 
 \overset{\P^2_{ }\,>\,4m_\varphi^2}{=} 
 \;
 \frac{p^{ }_0}{2} \pm 
 \frac{p}{2} 
 \sqrt{ 1 - \frac{4m_\varphi^2}{\P^2_{ }} }
 \;, \la{boundaries_1}
\ee
where
\be
 \P^2_{ }\; \equiv \; p_0^2 - p^2_{ }
 \;. 
\ee 
The integral over the 
angle between $\vec{k}$ and $\vec{p}$ can be carried out, 
with the Jacobian yielding $\epsilon^{ }_{pk}/(p k)$.
Combining these steps, we find 
\be
 I
 \; 
 \underset{\rmii{\nr{boundaries_1}}}{
 \overset{\rmii{\nr{factorize}}}{=}} 
 \;
 \theta\bigl( p^{ }_0 - \sqrt{p^2_{ } + 4 m_\varphi^2} \bigr)
 \int_{\omega_-}^{\omega_+} \! \frac{ {\rm d}\omega }{8\pi p}
 \;. \la{I_result}
\ee
Furthermore we note that in \eq\nr{cut_I_2}, 
the exponential factor contains $\omega$, 
but according to \eq\nr{factorize}, we can replace
$\omega + \epsilon^{ }_h\to p^{ }_0$. Then the integrand
is independent of $\omega$, and the $\omega$-integral can be trivially 
carried out. This finally yields a somewhat simpler representation
for \eq\nr{cut_I_2}, 
\ba
 \langle \sigma \vrel \rangle 
 & 
 \underset{\rmii{\nr{I_result},\nr{boundaries_1}}}{
 \overset{\rmii{\nr{cut_I_2}}}{\supset}}
 &  
 \frac{1}{n^{2}_\rmi{eq}}   
 \overbrace{
 \int \! {\rm d}^3_{ }\vec{p} 
 }^{4\pi\int_0^\infty \! {\rm d}p\, p^2_{ } }
 \int_{\sqrt{p^2_{ } + 4 m_\varphi^2}}^\infty \! {\rm d}p^{ }_0
 \, \frac{
    e^{ -\beta p^{ }_0 }_{ }
    \,
    \sqrt{1 - \frac{ 4 m_\varphi^2 }{ \P^2_{ }}}
 }{8\pi}
 \nn[2mm] 
 & \times & 
 \int_{\vec{q,r}} 
 \int_{-\infty}^{\infty} \! \frac{{\rm d} q^{ }_0}{\pi}
 \int_{-\infty}^{\infty} \! \frac{{\rm d} r^{ }_0}{\pi} 
 \,\delta^{(4)}_{ }(\P - \Q - \R)
 \varrho^{ }_\Q
 \,
 \varrho^{ }_\R
 \,
 \Phi(\Q,\R)
 \la{cut_I_3} 
 \;.
\ea

In \eq\nr{cut_I_3}, 
$\P$ is effectively the four-momentum of an off-shell
Higgs boson. The formula indicates that we need to evaluate
its $1\to 2$ decays into Standard Model particles, with thermal 
effects contained in 
$
 \varrho^{ }_\Q
$
and
$
 \varrho^{ }_\R
$
and in the ``matrix element squared'' 
$
 \Phi(\Q,\R)
$.
The Higgs is {\em not} at rest with respect to the medium but, as follows
from the thermal weight $e^{-\beta p^{ }_0}_{ }$, carries
a typical momentum $p \sim \sqrt{m^{ }_\varphi T} \gg \pi T$.

We remark in passing that if we move towards the relativistic regime, i.e.\
$\pi T \gsim m^{ }_\varphi $, then it can happen that because of the thermal
mass corrections experienced by Standard Model particles 
(cf.\ \fig\ref{fig:asy_masses}), 
$m^{ }_\varphi \sim m^{ }_\phi$. In this situation the kinematics 
changes from what we have assumed above, most simply because 
it is not sufficient to have $p_0^2 - p^2 > 4 m_\varphi^2$, 
but $\P^2_{ }$ also needs to be large enough 
to set the final-state particles on-shell
(provided that their spectral functions are of the type in \eq\nr{rho_free}).

%
\subsection{Results for the high-temperature phase ($T > \Tc^{ }$)}

Let us first apply the formalism to a simple example. 
For this we go to temperatures $T > \Tc^{ }$, 
where most of the contributions drop out. 
The remaining result reads 
\ba
 \AmplHa    \hspace*{+0.3cm} 
 \lim_{v\to 0}
 \Pi^{ }_{K;\varphi} & \supset &
 \Delta^{-1}_{-P-K;\widetilde\varphi}
 \,\Delta^{-1}_{Q;\phi}
 \,\Delta^{-1}_{P-Q;\phi}
 \,
   \overbrace{ 
   \bigl\{\, 
   2 \kappa^2_{ }
 \,\bigr\} }^{\,\equiv\,\Phi^{ }_{\phi\phi} }
 \;, \la{AmplHsym}
\ea
where $\phi$ denotes scalar excitations
(with 4 real degrees of freedom).

Inserting unresummed spectral functions from \eq\nr{rho_free} 
into \eq\nr{cut_I_2}, we obtain
\ba
  \langle \sigma \vrel \rangle 
 &
 \underset{\rmii{\nr{rho_free}},\rmii{\nr{cut_I_2}}}{
 \overset{\rmii{\nr{AmplHsym}}}{\approx}}
 &  
 \frac{1}{n^{2}_\rmi{eq}}
 \int \! 
 \frac{1}{(2\pi)^{12}_{ }}
 \frac{{\rm d}^3_{ }\vec{k}}{2 \omega}
 \frac{{\rm d}^3_{ }\vec{h}^{ }_{\tilde\varphi}}
   {2 \epsilon^{ }_{\tilde\varphi}}
 \frac{{\rm d}^3_{ }\vec{q}^{ }_{\phi_1}}
   {2 \epsilon^{ }_{\phi_1}}
 \frac{{\rm d}^3_{ }\vec{r}^{ }_{\phi_2}}
   {2 \epsilon^{ }_{\phi_2}}
 \,(2\pi)^4 \delta^{(4)}(\K + \H^{ }_{\tilde\varphi} 
                       - \Q^{ }_{\phi_1} - \R^{ }_{\phi_2} )
 \nn[2mm] 
 & \times & 
  e^{-\beta(\omega + \epsilon^{ }_{\tilde\varphi})}_{ }
 \bigl\{ \, 2 \kappa^2_{ } \,\bigr\} 
 \;. \la{sigma_conf}
\ea 
{}From here we see that $\Phi^{ }_{\phi\phi} = 2\kappa^2_{ }$ corresponds
to the usual matrix element squared, summed over final-state
degeneracies, and with a prefactor $1/2!$ to cancel the
overcounting for the integration over the momenta of identical
final-state particles, {\it viz.}  
\be
 2\kappa^2_{ } = \frac{1}{2!} \,4\, |\mathcal{M}|^2_{ }
 \;, \quad
 |\mathcal{M}| = \kappa
 \;.
\ee

For a practical integration, it is beneficial to proceed
with \eq\nr{cut_I_3}. The integrals over $q^{ }_0$ and $r^{ }_0$
can be carried out with the help of \eq\nr{rho_free}. The phase-space
integral over $\vec{q}$ and $\vec{r}$ can also be performed, as we 
recall in appendix~\ref{app:A}. Then we are left over with a three-dimensional
integral, similar to \eq\nr{cut_I_4}, which is rapidly convergent
and easy to evaluate numerically. 

%
\subsection{Results for the HTL-resummed $WW$ channel 
in the Higgs phase ($T < \Tc^{ }$)}
\la{ss:higgs_WW}

Our main challenge is to generalize the results to the Higgs phase, 
including contributions from the non-polynomial HTL self-energies. 
In the language of \eqs\nr{cut_I_2} and \nr{def_P}, the large
initial-state energy $p^{ }_0 = \omega+\epsilon^{ }_h \ge 2 m^{ }_\varphi$ 
is transmitted to 
the Standard Model energies, $q^{ }_0$ and $r^{ }_0$. Moreover, 
because of the Boltzmann weight 
\be
 e^{-\beta p^{ }_0 }_{ }
 \; \le \; 
 e^{-\beta \sqrt{p^2_{ } + 4 m_\varphi^2 } }_{ }
 \; \approx \;
 e^{-\beta \,[\,2 m^{ }_\varphi + p^2_{ } / (4 m^{ }_\varphi)\,] }_{ }
 \;, \la{boltzmann}
\ee
the center-of-mass 
spatial momentum cannot be larger than $p^2 \sim 4 m^{ }_\varphi T$. 
The momenta of the decay products must sum up as 
$\vec q + \vec r = \vec p$. Then at least one of them must be 
of the same order as $p$, say $q\sim 2 \sqrt{ m^{ }_\varphi T }$. 
But given that $m^{ }_\varphi \gg \pi T$, this implies that 
$q \gg \pi T \gg m^{ }_\W, m^{ }_\rmii{E$2$}$. In this domain, 
the Standard Model particles are pole-like (see below), 
with energies 
$q^{ }_0 \approx q + m^2_{\infty}/(2q) \approx q$, 
where $m^{ }_{\infty}$ corresponds to the  
asymptotic mass of each channel 
(cf.\ \fig\ref{fig:asy_masses} on p.~\pageref{fig:asy_masses}). 
However, having 
$q^{ }_0 \approx q \sim 2 \sqrt{ m^{ }_\varphi T }$ is not enough: 
the two energies must sum into the much larger
$2m^{ }_\varphi \gg  2 \sqrt{ m^{ }_\varphi T }$. 
From this conflict we conclude that {\em both} 
momenta must be much {\em larger}
than $\sim 2 \sqrt{ m^{ }_\varphi T }$, and back-to-back, 
so that there is a cancellation between them. All in all, 
these arguments indicate that 
\be
 q \; \sim \; m^{ }_\varphi \;\gg\; \pi T 
 \;, \quad
 q^{ }_0 - q \; \sim \; \frac{m_\infty^2}{ m^{ }_\varphi}
 \;\ll\; q
 \;. \la{power_counting}
\ee

Adopting the power counting from \eq\nr{power_counting}, 
let us first consider the contribution from 
charged gauge bosons ($WW$). The starting point is  
\eq\nr{AmplHj_phys}, but multiplied with a factor~2, and with
$\F^\rmii{T,E}_{ } \to \G^\rmii{T,E}_{ }$,  
where the resummed propagators are from \eqs\nr{G_T}--\nr{hatG_E}.
Denoting 
\be 
 R \; \equiv  \; P-Q
 \;, \la{R_def}
\ee 
the result can be written in the form 
\ba
 \AmplHj    \hspace*{+0.3cm} 
 \Pi^{ }_{K;\varphi} 
 & 
 \underset{Z\to W}{
 \overset{\rmii{\nr{AmplHj_phys}}}{\supset}}
 &
 \kappa^2_{ }
 \,\Delta^{-1}_{-K-P;\widetilde\varphi}
 \,\Delta^{-2}_{P;h}
 \Bigl\{\, 
 \nn[2mm]
 & & 
  \;
 \overset{\rmii{(a)}}{+}
  \; 
 \bigl(\, P^2_{ } - Q^2_{ }- R^2_{ } \,\bigr)^2_{ }
 \widehat{\G}\hspace*{0.3mm}{}^\rmii{E}_{Q}
 \,\widehat{\G}\hspace*{0.3mm}{}^\rmii{E}_{R}
 \nn[2mm]
 & & 
  \;
 \overset{\rmii{(b)}}{+}
  \;  
    8 m^2_\W Q^{ }_\mu Q^{ }_\nu 
    \,\mathbbm{R}^\rmii{T}_{\mu\nu}
    \,\widehat{\G}\hspace*{0.3mm}{}^\rmii{E}_Q 
    \,\bigl( \G^\rmii{T}_{R} - \G^\rmii{E}_{R} \bigr)
 \nn[2mm]
 & & 
  \;
 \overset{\rmii{(c)}}{+}
  \;  
    4 m^4_\W 
    \,\bigl(\,  
     \mathbbm{Q}^\rmii{T}_{\mu\nu}
     \mathbbm{R}^\rmii{T}_{\mu\nu}
    - 2
    \,\bigr)\,
    \bigl( \G^\rmii{T}_{Q} - \G^\rmii{E}_{Q} \bigr)\,
    \bigl( \G^\rmii{T}_{R} - \G^\rmii{E}_{R} \bigr)\,
 \nn[2mm]
 & & 
  \;
 \overset{\rmii{(d)}}{+}
  \;
    8 m^4_\W \, \G^\rmii{T}_{Q} \, \G^\rmii{T}_{R}
 \, \Bigr\} 
 \;. \la{AmplHj_WW}
\ea
This can be readily used in \eq\nr{cut_I_3}, once we identify 
the spectral functions $\varrho^{ }_{\Q,\R}$.

In order to find $\varrho^{ }_{\Q,\R}$, we need to inspect the
propagators in \eqs\nr{G_T}--\nr{hatG_E} in the kinematic 
domain of \eq\nr{power_counting}. For the self-energies from 
\eqs\nr{PiT} and \nr{hatPiE}, we find
\ba
 \Pi^\rmii{T$2$}_{\Q}
 & 
  \overset{\rmii{\nr{PiT}}}{
  \underset{q^{ }_0 - q \,\ll\, q}{\approx}} 
 & 
 \frac{  m_\rmii{E$2$}^2 }{2}
 \;, \la{PiT_appro}
 \\[2mm]
 \widehat\Pi^\rmii{E$2$}_{\Q}
 & 
  \overset{\rmii{\nr{hatPiE}}}{
  \underset{q^{ }_0 - q \,\ll\, q}{\approx}} 
 & 
 \frac{  m_\rmii{E$2$}^2 }{q^2_{ }}
 \biggl( 1 - \frac{1}{2} \ln \frac{2q}{q^{ }_0 - q} \biggr)
 \;. \la{hatPiE_appro}  
\ea
This means that 
\ba
 \bigl\{\, \G^\rmii{T}_{\Q} \,\bigr\}^{-1}_{ }
 & 
  \overset{q^{ }_0 - q \,\ll\, q}{\approx} 
 & 
 - q_0^2 + q^2_{ } 
 + \overbrace{ 
 m_\W^2 +  \frac{  m_\rmii{E$2$}^2 }{2}
 }^{ 
 \; \equiv \; 
  m_{\W\T}^2 
 }
 \;, \la{GT_inv} \\[2mm]
 \bigl\{\, \G^\rmii{E}_{\Q} \,\bigr\}^{-1}_{ }
 & 
  \overset{q^{ }_0 - q \,\ll\, q}{\approx} 
 & 
 \bigl(\, 
 - q_0^2 + q^2_{ } 
 \,\bigr)
 \biggl[\, 
  1 + 
   \frac{  m_\rmii{E$2$}^2 }{q^2_{ }}
   \biggl( 1 - \frac{1}{2} \ln \frac{2q}{q^{ }_0 - q} \biggr)
 \,\biggr]
 + m_\W^2 
 \nn[2mm]
 & 
  \overset{q \,\gg\, m^{ }_\rmiii{E2}}{\approx}
 & 
 - q_0^2 + q^2_{ } + m_\W^2
 \;. \la{GE_inv}
\ea
This in turn implies that at leading order, 
\ba
 \varrho^\rmii{T}_{\Q}
 \; \equiv \; 
 \im \G^\rmii{T}_{\Q}
 & 
   \underset{q^{ }_0 - q \;\ll\; q}
  {\overset{\rmii{\nr{rho_free},\nr{GT_inv}}}{\approx}}
 &  
 \frac{\pi \,\delta(q^{ }_0 - \epsilon^q_{\W\T})}{2 \epsilon^q_{\W\T}}
 \;, \quad
 \epsilon^q_{\W\T} 
 \; \equiv \; 
 \sqrt{q^2_{ } + m^2_{\W\T}}
 \;, 
 \la{rhoT_Q} \\[2mm]
 \varrho^\rmii{E}_{\Q}
 \; \equiv \; 
 \im \G^\rmii{E}_{\Q}
 & 
   \underset{q^{ }_0 - q \;\ll\; q}
  {\overset{\rmii{\nr{rho_free},\nr{GE_inv}}}{\approx}}
 &  
 \frac{\pi \,\delta(q^{ }_0 - \epsilon^q_{\W})}{2 \epsilon^q_{\W}}
 \;, \quad
 \epsilon^q_{\W} 
 \; \equiv \; 
 \sqrt{q^2_{ } + m^2_{\W}}
 \;. \la{rhoE_Q}
\ea
The difference between \eqs\nr{rhoT_Q} and \nr{rhoE_Q}, and thus the absence
of thermal corrections in the longitudinal channel, is a consequence
of the fact that the HTL correction appears as a multiplicative 
rather than additive factor in \eq\nr{GE_inv}.
For the same reason, up to corrections suppressed by 
$\rmO(m_\rmii{E$2$}^2/q^2_{ })$, 
\be
 {\widehat{\varrho}}\hspace*{0.4mm}{}^\rmii{E}_{\Q}
 \; \equiv \; 
 \im {\widehat{\G}}\hspace*{0.3mm}{}^\rmii{E}_{\Q}
 \; 
   \underset{\rmii{\nr{hatPiE_appro}}}{\overset{\rmii{\nr{hatG_E}}}{\approx}}
 \; 
 \varrho^\rmii{E}_{\Q}
 \;. \la{hat_rhoE_Q}
\ee

A first message conveyed by \eqs\nr{rhoT_Q} and \nr{rhoE_Q}
is that the poles in the transverse and electric channels are {\em not}
the same. Therefore 
the 2nd and 3rd structures in \eq\nr{AmplHj_WW}, which contain
differences between the two channels, do not cancel exactly. 
However, their difference is small, 
effectively of $\rmO(m_\rmii{E$2$}^2/q^2_{ })$. 
We return to this below. Moreover, the pole in the electric channel, 
which is sometimes also called the longitudinal channel (with respect 
to spatial momentum), does {\em not} experience any thermal 
corrections at this order (cf.\ \eq\nr{rhoE_Q}). 

\vspace*{3mm}

As a next step, we need the integral from the 2nd line
of \eq\nr{cut_I_3}. This standard computation 
is summarized in appendix~\ref{app:A}. In the domain where the spectral
functions are dominated by poles,   
and assuming $m^{ }_1 + m^{ }_2 < 2 m^{ }_\varphi$,
the result reads
\be
 \langle \sigma \vrel \rangle 
 \;
 \underset{\rmii{\nr{J_result}}}{
 \overset{\rmii{\nr{cut_I_3}}}{\supset}}
 \;  
 \frac{1}{n^{2}_\rmi{eq}} \frac{1}{\pi (4\pi)^4_{ }}  
 \int_0^\infty \! {\rm d}p\, p \,
 \int_{\sqrt{p^2_{ } + 4 m_\varphi^2}}^\infty \! {\rm d}p^{ }_0
 \, 
    e^{ -\beta p^{ }_0 }_{ }
    \,
    \sqrt{1 - \frac{ 4 m_\varphi^2 }{ \P^2_{ }}}
 \int_{\epsilon_1^-}^{\epsilon_1^+} 
 \! {\rm d}\epsilon_1^q \; 
 \Phi^{ }_{ }|^{ }_{\epsilon_2^r \,=\, p^{ }_0 - \epsilon_1^q}
 \la{cut_I_4}
 \;, 
\ee
where $\epsilon_1^\pm$ are given by \eq\nr{boundaries_2}, 
and the angle is fixed through 
\be
  \vec{q}\cdot\vec{r} 
 \;
  \overset{\vec r \,=\, \vec p - \vec q}{=}
 \; 
  \vec{q}\cdot\vec{p} - q^2_{ }
 \;=\; 
  \frac{p^2_{ } 
 \overbrace{
 - (\vec{p-q})^2 - m_2^2 }^{ -(\epsilon_2^r)^2_{ }}
 \overbrace{
 - q^2_{ } - m_1^2 }^{ -(\epsilon_1^q)^2_{ }}
 + m_1^2  + m_2^2}{2}
 \;. \la{angles}
\ee
For reference, we remark that if $\Phi\approx 1$ and 
$2 m^{ }_\varphi \gg T$, 
then 
$
  \langle \sigma \vrel \rangle \approx 1/(32\pi m_\varphi^2)
$.

We apply \eq\nr{cut_I_4} to \eq\nr{AmplHj_WW}.
Let us estimate the parametric magnitudes 
of the various contributions, in the domain of \eq\nr{power_counting}. 
In particular, given the large values of $q^{ }_0$ and $q$, we may
ask what happens at leading order in an expansion in $m_1^2/m^2_{\varphi}$ 
and $m_2^2/m^2_{\varphi}$, where $ p^{ }_0 \sim q\sim r \sim m^{ }_\varphi$
and $m^{ }_{1,2} \ll p \sim 2\sqrt{m^{ }_\varphi T} \ll p^{ }_0$.

We note that the projectors in
the 2nd and 3rd structures in \eq\nr{AmplHj_WW}
can be written as 
\ba
 Q^{ }_\mu Q^{ }_\nu 
    \,\mathbbm{R}^\rmii{T}_{\mu\nu} 
 & 
  \overset{\rmii{\nr{projectors}}}{=}
 & 
 \frac{q^2_{ }r^2_{ } - (\vec{q}\cdot\vec{r})^2_{ }}{r^2_{ }}
 \;, \la{transverse_1} \\[2mm]
     \mathbbm{Q}^\rmii{T}_{\mu\nu}
     \mathbbm{R}^\rmii{T}_{\mu\nu}
    - 2
 & 
  \overset{\rmii{\nr{projectors}}}{=}
 & 
 \frac{q^2_{ }\delta^{ }_{ij} - q^{ }_i q^{ }_j}{q^2_{ }}
 \frac{r^2_{ }\delta^{ }_{ij} - r^{ }_i r^{ }_j}{r^2_{ }}
 - 2
 \; = \; 
 - \frac{q^2_{ }r^2_{ } - (\vec{q}\cdot\vec{r})^2_{ }}{q^2_{ }r^2_{ }}
 \;. \la{transverse_2}
\ea
For vanishing masses, \eq\nr{angles} implies that 
\be
 \vec{q}\cdot\vec{r}
 \; 
  \underset{\epsilon_1^q\,=\,q\;,\;\epsilon_2^r\,=\,p^{ }_0-q}
 {\overset{\rmii{\nr{angles}~with~$m^{ }_{1,2}\to 0$}}
 {\longrightarrow}} 
 \; 
 q \, r - \frac{\P^2_{ }}{2}
 \;, \quad
 q^2_{ }r^2_{ } - (\vec{q}\cdot\vec{r})^2_{ }
 \;\longrightarrow\;
 \P^2_{ }\, 
 \biggl(
  q\,r - \frac{\P^2_{ }}{4} 
 \biggr)
 \;. \la{transverse_4}
\ee
If we represent \eq\nr{range_appro} as 
\be
 q \;=\; \frac{p^{ }_0 + x \, p}{2}
 \;, \quad
 x \;\in\; (-1,+1)
 \;, \la{def_x}
\ee
then 
\be
 q^2_{ }r^2_{ } - (\vec{q}\cdot\vec{r})^2_{ }
 \; 
   \underset{\rmii{\nr{transverse_4}\,,\,\nr{def_x}}}
  {\overset{s\,\equiv\, \P^2_{ }}{\longrightarrow}} 
 \;
 \frac{p^2_{ } s (1-x^2_{ })}{4} 
 \;. \la{transverse_3}
\ee
We recall once more that
\be
 s 
 \;
 \overset{\rmii{\nr{cut_I_4}}}{\ge}
 \;
 4 m_\varphi^2
 \;, \quad 
 q^2_{ } \;\sim\; r^2_{ } 
 \;
 \overset{\rmii{\nr{power_counting}}}{\sim}
 \;
 s
 \;, \quad 
 p^2_{ } 
 \;
 \overset{\rmii{\nr{boltzmann}}}{\sim}
 \;
 4 m^{ }_\phi T 
 \;
 \overset{\pi T \ll m^{ }_\varphi}{\ll}
 \;
 s
 \;. \la{power_counting_2} 
\ee

We can now estimate the magnitudes of the various terms
in \eq\nr{AmplHj_WW}. The first one yields
\be
 \Phi^\rmi{(a)}_{\W\W}
 \; = \; 
 \kappa^2_{ }
 \, 
 \frac{( s  - 2 m_\W^2)^2_{ }}{(s - m_h^2)^2_{ }}
 \; 
 \underset{m^{ }_h,m^{ }_\W \,\ll\, m^{ }_\varphi}{
 \overset{\rmii{\nr{power_counting_2}}}{\approx}}
 \;  
 \kappa^2_{ }
 \;. \la{Phi_WW_a}
\ee
The second gives
\be
 \Phi^\rmi{(b)}_{\W\W}
 \; 
   \underset{\rmii{\nr{transverse_3}}}
  {\overset{\rmii{\nr{transverse_1}}}{\sim}} 
 \; 
 \kappa^2_{ }
 \frac{m_\W^2 p^2_{ } s}{(s - m_h^2)^2_{ } r^2_{ }}
 \frac{m_\rmii{E2}^2}{r^2_{ }}
 \; 
 \underset{m^{ }_h \,\ll\, m^{ }_\varphi}{
 \overset{\rmii{\nr{power_counting_2}}}{\sim}}
 \; 
 \kappa^2_{ } \frac{ m_\W^2 p^2_{ }m_\rmii{E2}^2}{s^3_{ }}
 \;. \la{Phi_WW_b}
\ee
The third produces
\be
 \Phi^\rmi{(c)}_{\W\W}
 \; 
   \underset{\rmii{\nr{transverse_3}}}
  {\overset{\rmii{\nr{transverse_2}}}{\sim}} 
 \; 
 \kappa^2_{ }
 \frac{m_\W^4 p^2_{ } s}{(s - m_h^2)^2_{ } q^2_{ } r^2_{ }}
 \frac{m_\rmii{E2}^4}{ q^2_{ } r^2_{ }}
 \; 
 \underset{m^{ }_h \,\ll\, m^{ }_\varphi}{
 \overset{\rmii{\nr{power_counting_2}}}{\sim}}
 \; 
 \kappa^2_{ } \frac{ m_\W^4 p^2_{ }m_\rmii{E2}^4}{s^5_{ }}
 \;.  \la{Phi_WW_c}
\ee
The fourth evaluates to 
\be
 \Phi^\rmi{(d)}_{\W\W}
 \; = \; 
 \kappa^2_{ }
 \, 
 \frac{8 m_\W^4}{(s - m_h^2)^2_{ }}
 \; 
 \underset{m^{ }_h \,\ll\, m^{ }_\varphi}{
 \overset{\rmii{\nr{power_counting_2}}}{ \approx }}
 \;  
 \kappa^2_{ } \frac{8 m_\W^4}{ s^2_{ }}
 \;. \la{Phi_WW_d}
\ee
So we see that \eq\nr{Phi_WW_a} dominates by a large amount, 
and \eq\nr{Phi_WW_d} is the second most important contribution. 
In other words, the vacuum matrix element, 
given in \eq\nr{Phi_WW_0}, gives a good approximation. 
These estimates can be confirmed numerically, 
as illustrated in \fig\ref{fig:sigma_thermal_WW}.
The corresponding 
unresummed results for all channels can be found 
in appendix~\ref{app:B}, whereas resummed results for the 
$ZZ$ and $t\bar{t}$ channels 
are collected in appendices~\ref{app:C} and \ref{app:D}, respectively, 
with numerical plots shown in 
\figs\ref{fig:sigma_thermal_ZZ} (on p.~\pageref{fig:sigma_thermal_ZZ})
and \ref{fig:sigma_thermal_tt} (on p.~\pageref{fig:sigma_thermal_tt}). 

\begin{figure}[t]

 \hspace*{-0.1cm}
 \centerline{%
  \epsfysize=7.2cm\epsfbox{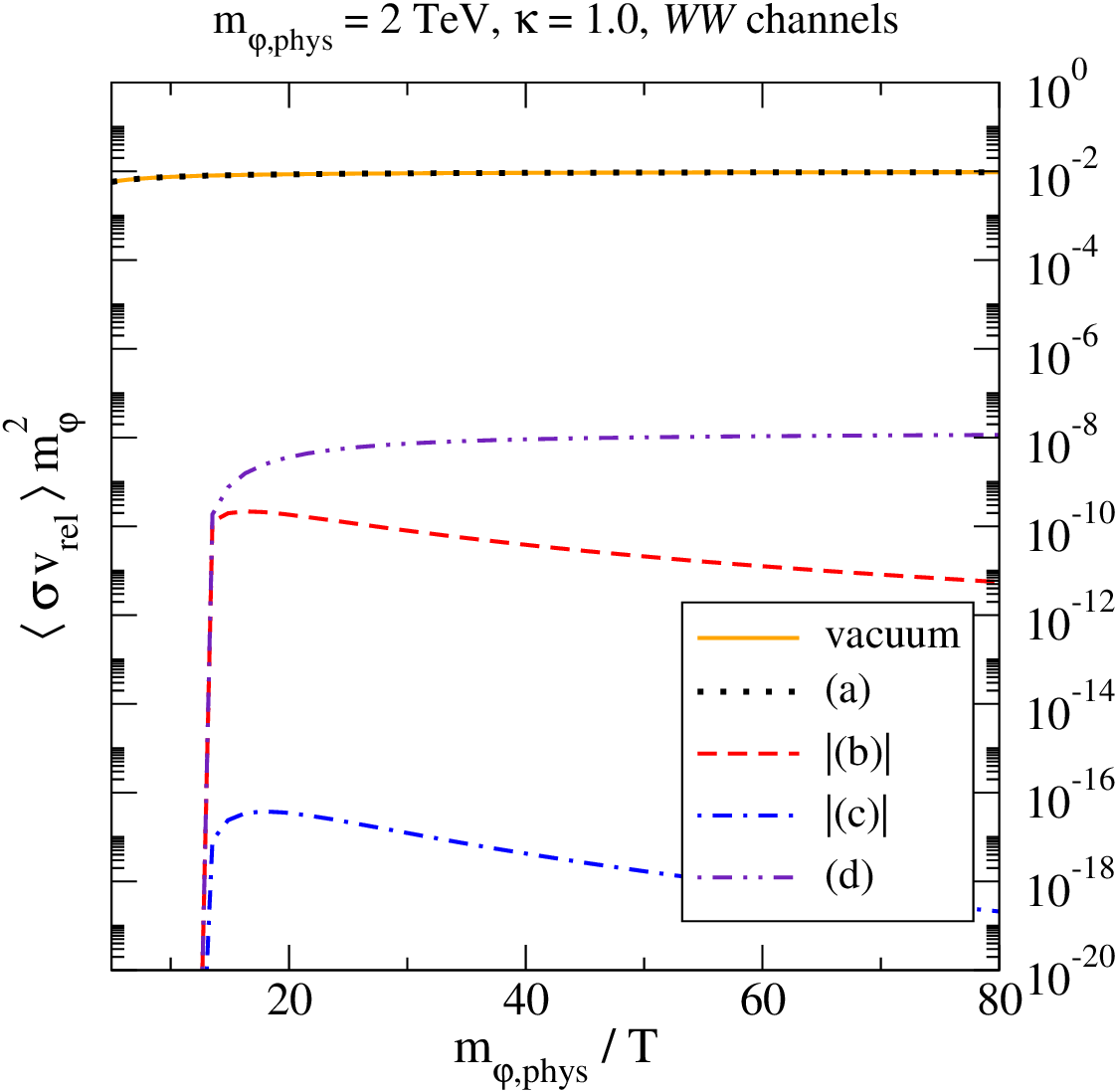}%
 }
 
 \caption[a]{\small
   An illustration of the cross sections originating from the 
   various terms in \eq\nr{AmplHj_WW}, 
   for the benchmark values 
   $m^{ }_{\varphi,\rmii{phys}} = 2$~TeV, 
   $\kappa = 1.0$. The notation ``$|$(b)$|$'' indicates that 
   this channel gives a negative contribution, and we show
   the absolute value. The channels (b)--(d) give
   a vanishing contribution as we 
   go to $T > \Tc^{ }$, where $m_\W^2\to 0$.
   }
 
 \la{fig:sigma_thermal_WW}
\end{figure}

%
\section{Phenomenological determination of dark matter abundance}
\la{se:pheno}

\begin{figure}[t]
 
 \hspace*{-0.1cm}
 \centerline{%
  \epsfysize=7.4cm\epsfbox{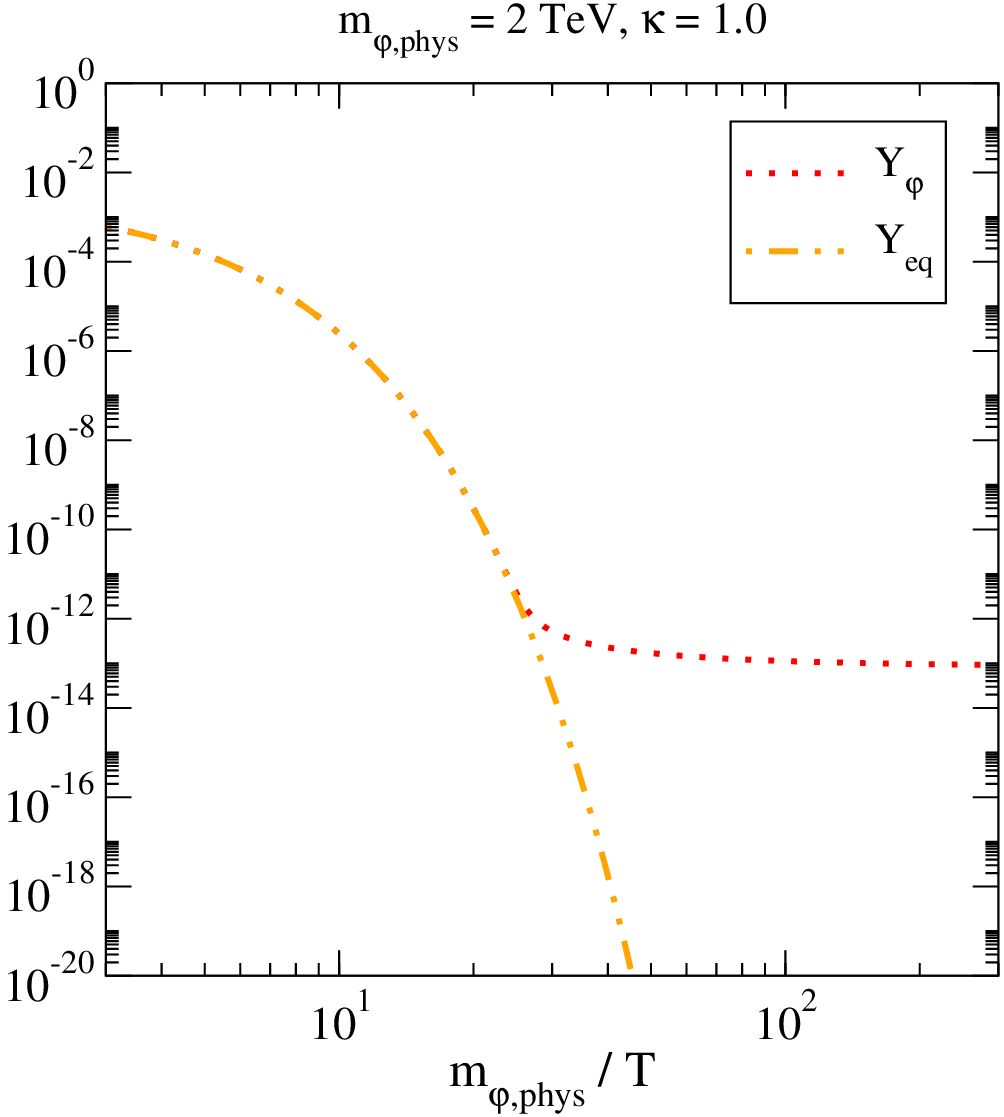}%
  \hspace{0.8cm}%
  \epsfysize=7.4cm\epsfbox{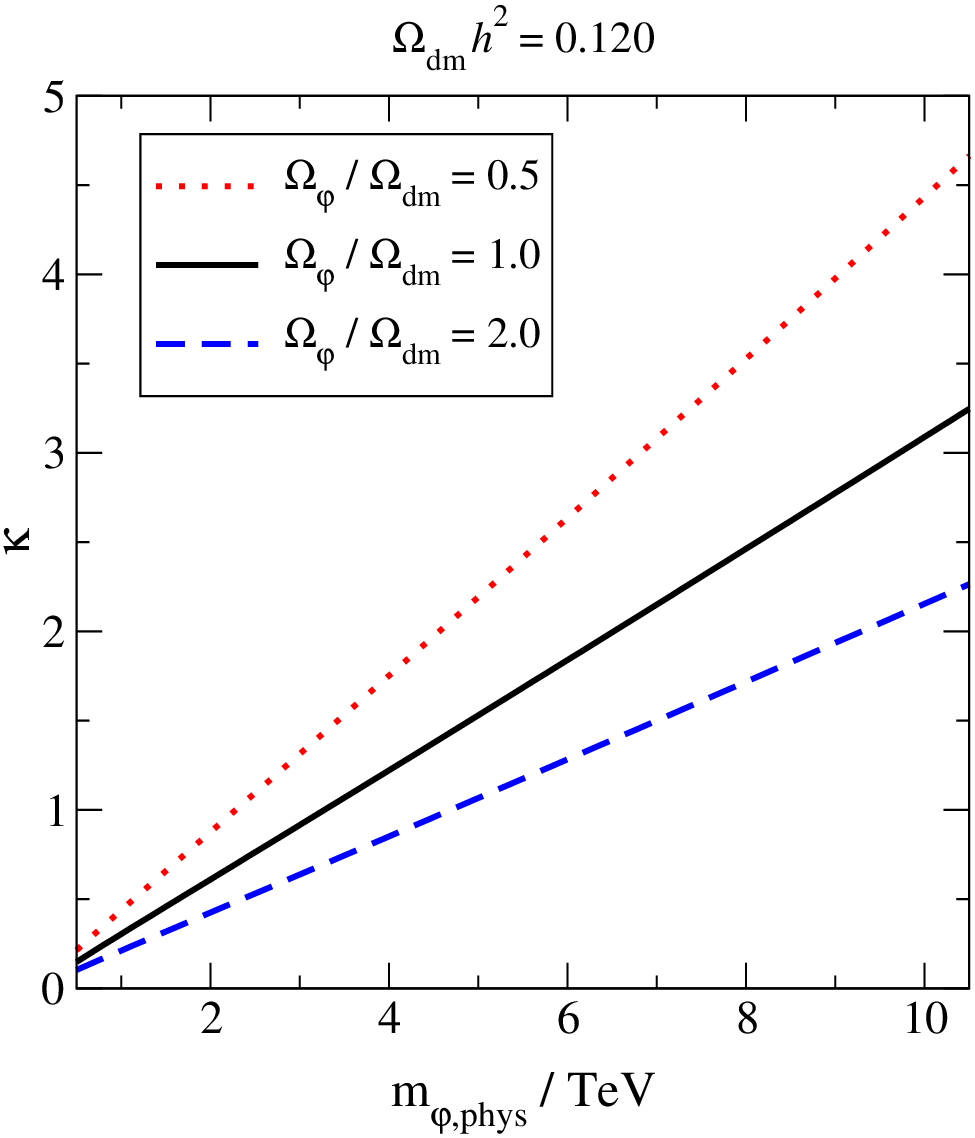}%
 }
 
 \caption[a]{\small
   Left: The solution of \eq\nr{lw2} for the benchmark values
   $m^{ }_{\varphi,\rmii{phys}} = 2$~TeV, 
   $\kappa = 1.0$.
   Right: $\Omega^{ }_\varphi/\Omega^{ }_\rmi{dm}$ in the plane
   of $m^{ }_{\varphi,\rmii{phys}}$ and $\kappa$. 
   Within the plot resolution, the results agree with 
   refs.~\cite{singlet4,singlet6}.
   }
 
 \la{fig:solution}
\end{figure}

The averaged annihilation cross section that we have determined for
the $WW$ channel in \se\ref{se:cuts}, and for the other channels in 
appendices~B--D, 
yields the dark matter relic density according
to \eq\nr{lw}. For a practical integration, it is helpful to replace
the physical time as an integration variable
through $x \equiv \ln (T^{ }_\rmi{max}/T)$, 
where the choice of $T^{ }_\rmi{max}$ is
a matter of convention (it has no influence).\footnote{%
 In the literature, the variable 
 $z = m^{ }_{\varphi} / T$ is frequently employed, but  
 it is less convenient than our $x$, given that $ m^{ }_{\varphi} $ depends 
 on temperature, via $v$ and an explicit thermal 
 correction (cf.\ \eq\nr{m_varphi}). To use it 
 properly would require the determination of an associated Jacobian.   
 } 
The Jacobian between
$t$ and $x$ follows from the Friedmann equations, yielding 
$ {\rm d}x/{\rm d}t = 3 c_s^2 H $,  where 
$c_s^{2} = \partial p / \partial e$ is 
the speed of sound squared. The number densities are 
conveniently normalized to the entropy density, $s$, 
and we denote 
$Y^{ }_\varphi \equiv n^{ }_\varphi / s$
and $Y^{ }_\rmi{eq} \equiv n^{ }_\rmi{eq}/s$.
Then \eq\nr{lw} is converted into 
\be
 \partial^{ }_x Y^{ }_\varphi
 \;
  \overset{\rmii{\nr{lw}}}{\approx}
 \; 
 - \frac{\langle \sigma v^{ }_\rmii{rel} \rangle s}
 {3 c_s^2 H}
 \, 
 \bigl( Y_\varphi^2 - Y_\rmi{eq}^2 \bigr)
 \;. \la{lw2} 
\ee
The current energy fraction in the singlet scalar particles, 
compared with the observational value, 
$
 \Omega^{ }_\rmi{dm} h^2_{ } \approx 0.120
$, 
can be expressed as 
$
 \Omega^{ }_\varphi / \Omega^{ }_\rmi{dm} 
 \approx
 2.29\, 
 (m^{ }_{\varphi,\rmii{phys}} / \mbox{eV})
 \, Y^{ }_\varphi (x^{ }_\rmii{0})
$, 
where $x^{ }_\rmii{0}$ refers to the present universe. 
For the thermodynamic functions $s$, $c_s^2$, and $e$ 
(the last one determines the Hubble rate via 
$H^2 = 8\pi G e/3 $), we employ
the tabulations in ref.~\cite{eos15}, 
and for the running couplings and masses of the Standard Model, 
the multiloop values obtained in ref.~\cite{mDebye}, 
see in particular the explanation in a paragraph below its eq.~(4.12). 

Let us briefly comment on the running of the coupling $\kappa$
that connects the singlet scalar to the Standard Model
(cf.\ \eq\nr{L}). Assuming that $\kappa$ is much larger
than $\lambda^{ }_\varphi$ and the Standard Model quartic 
coupling $\lambda^{ }_h$, the $\msbar$ coupling in principle
runs as
\be
 \bmu \frac{{\rm d}\kappa(\bmu)}{{\rm d}\bmu}
 \; \supset \; \frac{\kappa^2_{ }(\bmu)}{4\pi^2_{ }}
 \; \Rightarrow \; 
 \kappa(\bmu)
 \; \simeq \; 
 \frac{4\pi^2_{ }}{\ln(\Lambda/\bmu)}
 \;, 
\ee
where $\Lambda$ is an integration constant (``Landau pole''). 
However, our physics takes place
in the non-relativistic regime, 
$\pi T,\sqrt{m^{ }_\varphi T} \ll m^{ }_\varphi$.
Therefore the UV energy scale that fixes $\bmu$, 
stays put around 
the rest mass, $\bmu \sim m^{ }_\varphi$, and we do 
not need to run $\kappa$ in practice. 

\vspace*{3mm}

As an example of a solution, our benchmark displayed
in \figs\ref{fig:sigma_thermal_WW}, 
\ref{fig:sigma_vac_all}, 
\ref{fig:sigma_thermal_ZZ} and 
\ref{fig:sigma_thermal_tt}, 
namely $m^{ }_{\varphi,\rmii{phys}} = 2$~TeV and
$\kappa = 1.0$, yields 
$Y^{ }_\varphi(x^{ }_\rmii{0}) \approx 8.48 \times 10^{-14}_{ }$ and
$\Omega^{ }_\varphi / \Omega^{ }_\rmi{dm} \approx 0.388$,
if we integrate all the way to the QCD scale 
($T\sim 0.2$~GeV, $m^{ }_{\varphi,\rmii{phys}}/T\sim 10^4_{ }$).
The early part of the corresponding solution 
is illustrated in \fig\ref{fig:solution}(left).

Subsequently, we scan the singlet
scalar mass between 0.5 and 10.5~TeV.
The dependence of the final results on the dark matter mass
and the coupling $\kappa$ is shown in 
\fig\ref{fig:solution}(right).
Even though several technical details of the analysis
are different, the results agree numerically with 
refs.~\cite{singlet4,singlet6}. To be concrete, 
as a specific benchmark, 
ref.~\cite{singlet6} cites 
$\Omega^{ }_{\varphi}\hspace*{0.3mm} h^2_{ } = 0.1131 $
for $m^{ }_{\varphi,\rmii{phys}} = 9.79$~TeV and $\kappa = 3.1$.
We obtain $\Omega^{ }_{\varphi}\hspace*{0.3mm} h^2_{ } \approx 0.114 $
for the same $\kappa$, and $\kappa \approx 3.12 $ for the same
$\Omega^{ }_{\varphi}\hspace*{0.3mm} h^2_{ } $, even if it has to 
be remarked that unfortunately the freeze-out dynamics takes 
place mostly in the ``easy'' $T > \Tc^{ }$ domain 
for this benchmark.

If, in contrast, we go towards singlet scalar masses below 0.5~TeV, it 
is clear that thermal corrections become increasingly important. Most
simply, this can be seen by considering temperatures 
$T > 2 m^{ }_{\varphi,{\rm phys}} \sim 1$~TeV.
In the freeze-out scenario, we would think that such temperatures are
``not important'', as dark matter should be well equilibrated. However, 
as seen from \fig\ref{fig:asy_masses}, 
many asymptotic masses are larger than 0.5~TeV
in this situation. Therefore, the $2\to 2$ channel gets partly closed; 
more generally, the kinematics that we have assumed gets modified. 
To carry out a proper computation, including the initial equilibration, 
requires a more complicated analysis than here, with thermal mass
corrections playing an $O(1)$ role. 

%
\section{Conclusions}
\la{se:concl}

When we dream about finding dark matter through 
direct or indirect detection, or collider experiments, 
we necessarily focus on the Standard Model particles that 
participate in these reactions, because they are what we have 
access to. In contrast, in cosmology, it is rather the inclusive 
annihilation cross section of the initial-state dark-matter particles that 
is important: in principle it does not matter in which precise 
way the energy released is distributed. 
Nevertheless, it may still be interesting
to {\em ask} what actually happens, and this is the angle of 
approach that 
we have pursued in the current paper. In particular, 
we have determined the thermally averaged singlet scalar
annihilation cross section by taking into account
the full structures predicted by Hard Thermal Loop effective theories
on the Standard Model side. 

On the qualitative level, the results of our analysis can be 
understood from the cross sections in the $WW$-channel, 
sketched in \eqs\nr{Phi_WW_a}--\nr{Phi_WW_d}. The dominant 
process is represented by \eq\nr{Phi_WW_a}, which is independent
of both the temperature and the Higgs mechanism. The largest
correction is the vacuum one in \eq\nr{Phi_WW_d}, induced by 
the Higgs mechanism. Thermal effects appear as ``power corrections'': 
according to \eqs\nr{Phi_WW_b} and \nr{Phi_WW_c}, they are proportional
to the 2nd or 4th power of the Debye mass, which in turn is proportional
to the gauge coupling times the temperature (cf.\ \eq\nr{mE2}). However, 
these corrections are very 
much suppressed, as is visible in the numerical
results in \fig\ref{fig:sigma_thermal_WW}. Consequently, the
final phenomenological results, displayed in \fig\ref{fig:solution}, 
agree well with the literature, in which no Hard Thermal Loop 
resummation was implemented. 
We remark in passing that if one would nevertheless like to 
pursue higher-order corrections, it would be 
recommendable to return to the unifying theoretical framework
mentioned in the introduction, as otherwise the analysis
becomes exceedingly complicated.  

Even if Hard Thermal Loop resummation had little influence on the
final results of our
computation, it may play a more substantial role for physics in which the large
external scale of our problem, $m^{ }_\varphi$, is absent. 
Within the singlet scalar model itself, this could happen if 
we reduce $m^{ }_{\varphi,\rmii{phys}}$ below $0.5$~TeV, whereby
the coupling $\kappa$ is becoming weak 
(cf.\ \fig\ref{fig:solution}(right), and the discussion 
at the end of \se\ref{se:pheno}).
For very small $\kappa$, 
the question of the initial equilibration of $\varphi$ needs to 
be raised, meaning that we could eventually
enter the domain of the ``freeze-in''
mechanism (cf.,\ e.g.,\ ref.~\cite{tb}). 
The diagrams that we considered 
in \se\ref{se:feynman} are a subset of those
relevant for such scenarios, 
however the kinematics of \se\ref{se:cuts}
cannot be simplified in the same way as in the current study, and 
in addition $1\leftrightarrow 2$ processes and their LPM resummation 
needs to be included. In this situation thermal effects are
expected to dominate the physics. 
Other general problems in which soft-scale physics 
around the electroweak crossover
can be important, are low-scale leptogenesis, taking place 
close to when the sphaleron rate switches off~\cite{sphaleron}, or the  
real-time physics that is associated with a first-order electroweak
phase transition, present in many extensions of the 
Standard Model. We hope that the full set of HTL-resummed propagators
and mixings that we 
have assembled together (cf.\ \se\ref{ss:notation}) may facilitate
studies in such contexts.

%
%
\appendix
\renewcommand{\thesection}{\Alph{section}} 
\renewcommand{\thesubsection}{\Alph{section}.\arabic{subsection}}
\renewcommand{\theequation}{\Alph{section}.\arabic{equation}}

\newpage

%
\section{Two-particle phase space integral}
\la{app:A}

In this appendix, we recall the evaluation 
of the integral from the 2nd line
of \eq\nr{cut_I_3}. 
To simplify the notation, we do this for
two free spectral functions with different masses, 
\ba
 J^{ }_{12} 
 & \equiv & 
 \int_{\vec{q,r}} 
 \int_{-\infty}^{\infty} \! \frac{{\rm d} q^{ }_0}{\pi}
 \int_{-\infty}^{\infty} \! \frac{{\rm d} r^{ }_0}{\pi} 
 \,\delta^{(4)}_{ }(\P - \Q - \R)
 \frac{\pi \delta(q^{ }_0 - \epsilon^q_1)}{2 \epsilon^q_{1}}
 \,
 \frac{\pi \delta(r^{ }_0 - \epsilon^r_2)}{2 \epsilon^r_{2}}
 \nn[2mm] 
 & = & 
 \int_{\vec{q,r}} 
 \frac{
      \delta(p^{ }_0 - \epsilon^q_1 - \epsilon^r_2)
      \,\delta^{(3)}_{ }(\vec{p-q-r})
      }{4 \epsilon^q_1 \epsilon^r_2 }
 \; = \; 
 \frac{1}{(2\pi)^3_{ }}
 \int_\vec{q} 
 \frac{\delta(p^{ }_0 - \epsilon^q_1 - \epsilon^{pq}_2)
      }{4 \epsilon^q_1 \epsilon^{pq}_2 }
 \;. \la{def_J} 
\ea
This is closely related to the integral 
in \eq\nr{factorize}, except that the masses
are non-degenerate, and the result is smaller by
$1/(2\pi)^4_{ }$.
 
Extremizing the argument of the Dirac-$\delta$ 
with respect to $q$, we find that 
\be
 \min (\epsilon^q_1 + \epsilon^{pq}_2)
 = \sqrt{p^2_{ } + (m^{ }_1 + m^{ }_2)^2}
 \;. 
\ee
The energy $p^{ }_0$ has to be larger than this, in order for the
energy constraint to be realized. To determine the 
range of $q$ in which this happens, 
we vary the angle between $\vec{q}$ and $\vec{p}$, requiring
\be
 \sqrt{q^2_{ }+ m^2_1} + \sqrt{(q-p)^2_{ } + m_2^2 }
 \;<\; p^{ }_0 \;<\; 
 \sqrt{q^2_{ }+ m^2_1} + \sqrt{(q+p)^2_{ } + m_2^2 }
 \;. 
\ee
This leads to a minimal and maximal $q$, denoted by 
$q^{ }_\pm$, in analogy with \eq\nr{boundaries_1}. It is
helpful to express the outcome in terms 
of $\epsilon_1^q$, denoting 
$
 \epsilon_1^\pm \equiv \epsilon_1^{q_\pm}
$.
We find 
\be
 \epsilon_1^\pm 
 \; = \; 
 \frac{p^{ }_0\,(\P^2_{ }+ m_1^2 - m_2^2)
 \pm p \kallen(\P^2_{ },m_1^2,m_2^2)}{2\P^2_{ }}
 \;, \la{boundaries_2}
\ee
where the K\"all\'en function is defined as 
\be
 \kallen^{ } ( x,m_1^2,m_2^2 )
  \; \equiv \; 
 \sqrt{     x^2+m_1^4+m_2^4
              - 2 x ( m_1^2  + m_2^2 )
              - 2 m_1^2 m_2^2 
      }
 \;. \la{kallen}
\ee
For equal masses, \eq\nr{boundaries_2} reduces to 
\eq\nr{boundaries_1}, and for 
$m_1^2,m_2^2 \ll \P^2_{ }$, to 
\be 
 \epsilon_1^\pm 
 \; \approx \; 
 \frac{p^{ }_0 \pm p}{2}
 \;. \la{range_appro}
\ee
Carrying out the angular integral, 
the Jacobian yields $\epsilon^{pq}_{2}/(p q)$.
Combining these steps, we find 
\be
 J^{ }_{12} = 
 \frac{
 \theta\bigl( p^{ }_0 - \sqrt{p^2_{ } + (m_1^{ }+ m_2^{ })^2} \bigr)
 }{(2\pi)^4_{ }}
 \int_{\epsilon_1^-}^{\epsilon_1^+} 
 \! \frac{ {\rm d}\epsilon_1^q }{8\pi p}
 \;, \la{J_result}
\ee
generalizing on \eq\nr{I_result}. 

\newpage

%
\section{Leading-order matrix elements squared in the unresummed limit}
\la{app:B}

For completeness, we tabulate here the matrix elements squared
pertinent to the unresummed limit in the Higgs phase
($T < \Tc^{ }$), using a normalization that can 
be directly inserted in \eq\nr{cut_I_4}.
The corresponding result for the high-temperature phase
($T > \Tc^{ }$) can be found in \eq\nr{AmplHsym},
and reads $\Phi^{ }_{\phi\phi} = 2 \kappa^2_{ }$.
The results are illustrated numerically 
in \fig\ref{fig:sigma_vac_all}.

\begin{figure}[t]

 \hspace*{-0.1cm}
 \centerline{%
  \epsfysize=7.2cm\epsfbox{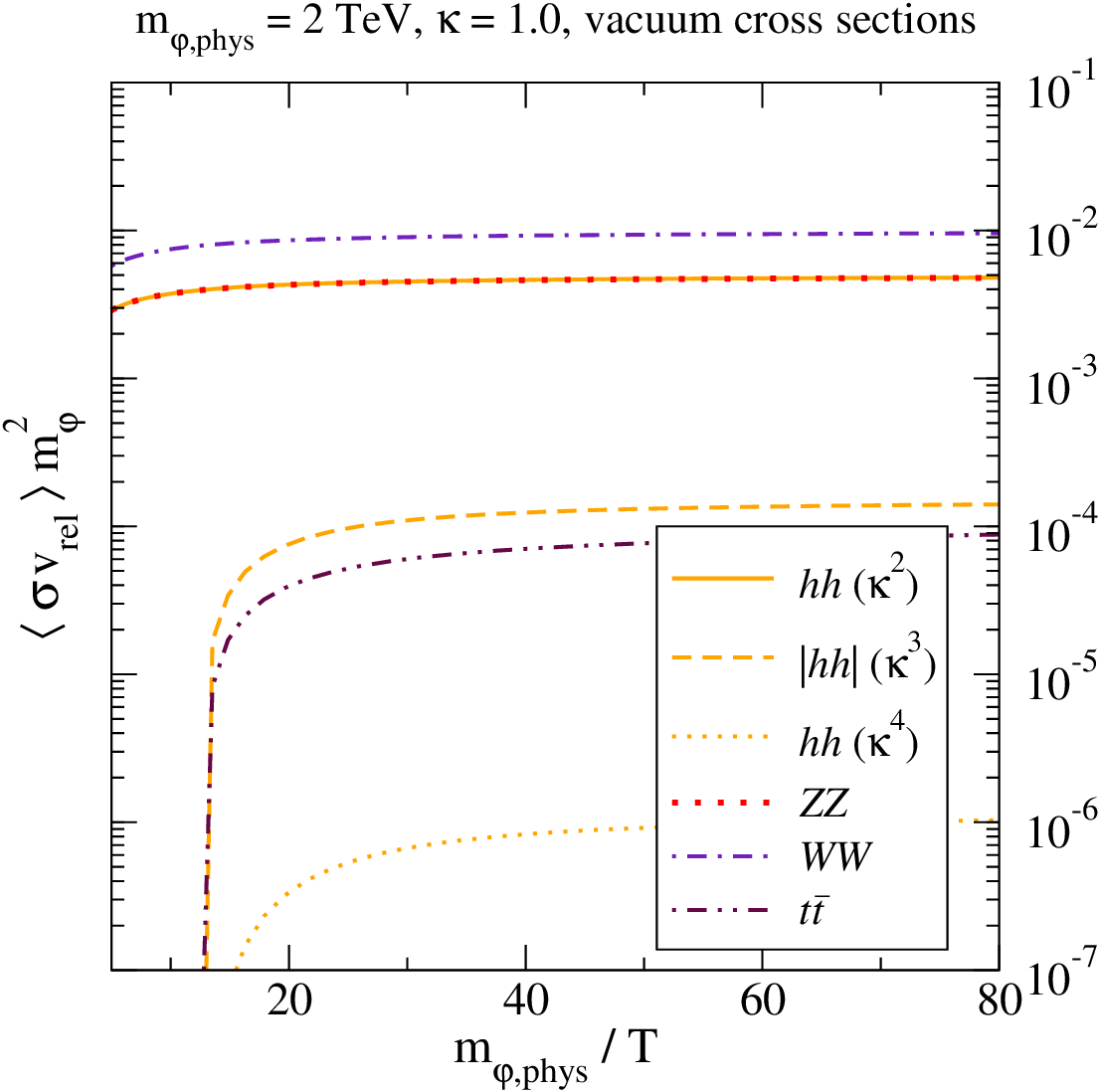}%
 }
 
 \caption[a]{\small
   An illustration of the cross sections following from 
   \eqs\nr{k2}, \nr{k3}, \nr{k4}, \nr{Phi_ZZ_0}, \nr{Phi_WW_0} 
   and \nr{Phi_tt_0}, for the benchmark values 
   $m^{ }_{\varphi,\rmii{phys}} = 2$~TeV, 
   $\kappa = 1.0$.
   The notation ``$|hh|$'' indicates that 
   this channel gives a negative contribution, and we show
   the absolute value.  
   The $b\bar{b}$ channel is
   obtained from \eq\nr{Phi_tt_0} with 
   $m^{ }_t \to m^{ }_b$ but is too small to be visible. 
   }
 
 \la{fig:sigma_vac_all}
\end{figure}

Employing the Mandelstam variables $s$, $t$, $u$, we consider first
the Higgs contributions from \eqs\nr{AmplHa}--\nr{AmplHf}. They yield
\ba
 \Phi^{ }_{hh}
 & \supset &  
 \frac{\kappa^2_{ }}{2} 
 \biggl\{\, 
   1 + \frac{6 m_h^2}{s-m_h^2 } + \frac{9 m_h^4}{(s-m_h^2)^2_{ }}
  \,\biggr\}
 \; = \; 
 \frac{\kappa^2_{ }}{2} 
 \, 
 \frac{( s  + 2 m_h^2)^2_{ }}{(s - m_h^2)^2_{ }}
 \;. \la{k2}
\ea
For the Higgs effects of $\rmO(\kappa^3_{ })$ from 
\eqs\nr{AmplHc}--\nr{AmplHd}, we find
\ba
 \Phi^{ }_{hh}
 & \supset &  
 \kappa^3_{ } v^2_{ } 
 \biggl\{\, 
   \frac{2}{t-m_\varphi^2 } 
 + \frac{6 m_h^2}{ (t - m_\varphi^2) ( s - m_h^2 )}
  \,\biggr\}
 \; = \; 
 2 \kappa^3_{ } v^2_{ } 
 \, 
 \frac{ s  + 2 m_h^2 }{ (t - m_\varphi^2) (s - m_h^2) }
 \;. \la{k3}
\ea
By a renaming of integration variables, this could
be symmetrized under $t\leftrightarrow u$, however this does
not simplify the expression.  
The Higgs effects of $\rmO(\kappa^4_{ })$ from 
\eqs\nr{AmplHe}--\nr{AmplHg} produce 
\be
 \Phi^{ }_{hh}
 \;\supset\;  
 \kappa^4_{ } v^4_{ } 
 \biggl\{\, 
   \frac{1}{(t-m_\varphi^2)^2_{ } } 
 + \frac{1}{ (t - m_\varphi^2) ( u - m_\varphi^2 )}
  \,\biggr\}
 \;, \la{k4}
\ee
which can be written in many equivalent forms, 
by making use of 
$
 s + t  +u = 2(m_h^2 + m_\varphi^2)
$,
as well as the symmetry $t\leftrightarrow u$.

For a practical integration, we note that in \eq\nr{cut_I_4}, 
only the center-of-mass momentum, $\P$, and a final-state energy, 
$\epsilon_1^q$, appear as integration variables. To turn 
\eqs\nr{k3} and \nr{k4} into a form in which we can use the same
integration measure, we have to integrate over the initial-state
momenta $\vec{k}$ and $\vec{h}^{ }_{\tilde\varphi}$
(cf.\ \eq\nr{sigma_conf}), 
appearing in $t$ or $u$. The Boltzmann weight, 
which breaks Lorentz invariance, only contains $p^{ }_0$ in the
non-relativistic regime, not the initial-state energies separately. 
Therefore the average over $\vec{k}$ and $\vec{h}^{ }_{\tilde\varphi}$
can be expressed as a Lorentz-invariant
integral over $\K$ and $\H$, depending on the invariants 
$\P^2_{ }$, $\P\cdot\Q$, and $\Q^2_{ }$. This integral can  
be conveniently carried out in the frame $\vec{q} = \vec{0}$, 
and subsequently re-expressed in a covariant form. 
After a tedious computation, 
this implies that we can effectively substitute~\cite{cas}
\ba
 \Phi^{ }_{hh}
 & \to &
 \frac{\kappa^2_{ }}{2} 
 \, 
 \biggl\{\, 
 \frac{( s  + 2 m_h^2)^2_{ }}{(s - m_h^2)^2_{ }}
 - \frac{ 8 \kappa v^2_{ } ( s  + 2 m_h^2) \, F(\xi) }
        { (s - m_h^2) (s - 2 m_h^2)}
 + \frac{8 \kappa^2_{ }v^4_{ }}{ (s - 2 m_h^2)^2_{ }}
   \, \biggl[
   \frac{1}{1 - \xi^2_{ }} + F(\xi)
   \biggr] 
  \,\biggr\}
  \;, \hspace*{8mm} \\[2mm]
 \xi & \equiv & 
 \frac{s}{s-2 m_h^2}
 \sqrt{1 - \frac{4 m_h^2}{s}}
 \sqrt{1 - \frac{4 m_\varphi^2}{s}}
 \;, \quad 
 F(\xi) \; \equiv \; 
 \frac{\mbox{artanh}\,\xi}{\xi}
 \;. 
\ea

Turning to gauge effects from \eq\nr{AmplHj_phys}, we note that if 
we omit resummation, i.e.\ set the self-energies to zero in 
\eqs\nr{F_T}--\nr{hatF_E}, then 
\be
 \F^\rmii{T}_{P} 
 \stackrel{\Pi^\rmiii{Ti}_{P}\to 0}{\longrightarrow}
 \Delta^{-1}_{P;Z}
 \;, \quad
 \F^\rmii{E}_{P} 
 \stackrel{\Pi^\rmiii{Ei}_{P}\to 0}{\longrightarrow}
 \Delta^{-1}_{P;Z}
 \;, \quad
 \widehat{\F}^\rmii{E}_{P}
 \stackrel{\widehat\Pi^\rmiii{Ei}_{P}\to 0}{\longrightarrow}
 \Delta^{-1}_{P;Z}
 \;. 
\ee
Then only the first and last term of \eq\nr{AmplHj_phys} remain. 
Adding the contribution from the charged gauge bosons, this yields
\ba
 \Phi^{ }_{\Z\Z}
 & = & 
 \frac{\kappa^2_{ }}{2} 
 \, 
 \frac{( s  - 2 m_\Z^2)^2_{ } + 8 m_\Z^4 }{(s - m_h^2)^2_{ }}
 \;, \la{Phi_ZZ_0}
 \\[2mm]
 \Phi^{ }_{\W\W}
 & = & 
 \kappa^2_{ }
 \, 
 \frac{( s  - 2 m_\W^2)^2_{ } + 8 m_\W^4 }{(s - m_h^2)^2_{ }}
 \;. \la{Phi_WW_0}
\ea
Finally, the quark contribution from \eq\nr{AmplHl_phys} turns into
\ba
 \Phi^{ }_{t\bar{t}}
 & = &  
 2 \kappa^2_{ } m_t^2 \Nc^{ }
 \, 
 \frac{ s  - 4 m_t^2  }{(s - m_h^2)^2_{ }}
 \;. \la{Phi_tt_0}
\ea

\newpage

%
\section{HTL-resummed $ZZ$ channel}
\la{app:C}

We repeat here the analysis of \se\ref{ss:higgs_WW} for the 
$ZZ$ channel, whose matrix element squared was given in 
\eq\nr{AmplHj_phys}.
The results are illustrated numerically 
in \fig\ref{fig:sigma_thermal_ZZ}.

\begin{figure}[t]

 \hspace*{-0.1cm}
 \centerline{%
  \epsfysize=7.2cm\epsfbox{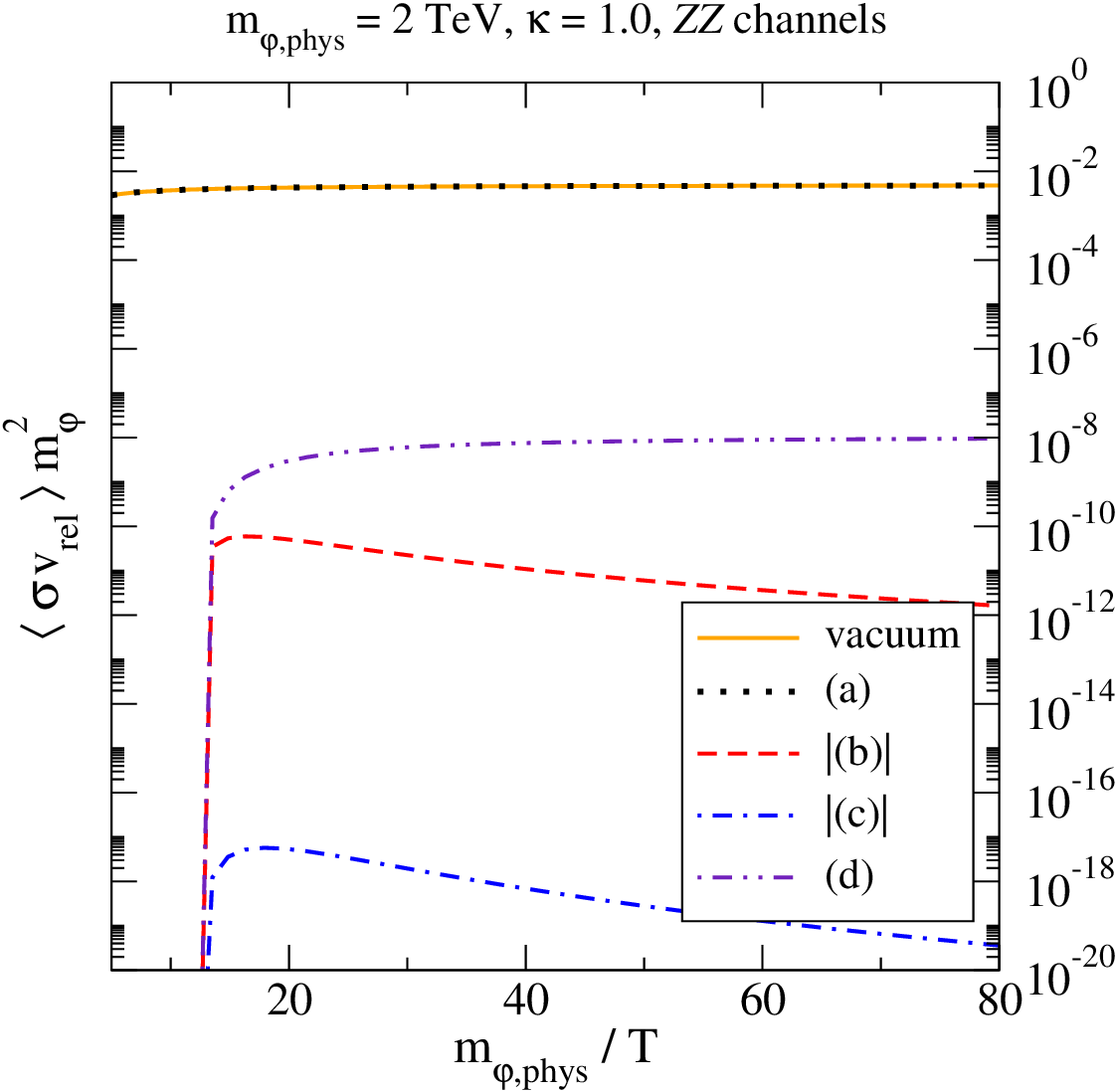}%
 }
 
 \caption[a]{\small
   An illustration of the cross sections following from the 
   various terms in \eq\nr{AmplHj_phys}, 
   for the benchmark values 
   $m^{ }_{\varphi,\rmii{phys}} = 2$~TeV, 
   $\kappa = 1.0$. The notation ``$|$(b)$|$'' indicates that 
   this channel gives a negative contribution, and we show
   the absolute value. 
   }
 
 \la{fig:sigma_thermal_ZZ}
\end{figure}

The starting point is to generalize the propagators 
$ \G^\rmii{T}_{\Q} $ and 
$ \G^\rmii{E}_{\Q} $, 
appearing in \eqs\nr{GT_inv} and \nr{GE_inv},
into the corresponding 
$ \F^\rmii{T}_{\Q} $ and 
$ \F^\rmii{E}_{\Q} $, defined according to  
\eqs\nr{F_T} and \nr{F_E}, respectively. 
For  $ \F^\rmii{E}_{\Q} $
(and $ \widehat{\F}^\rmii{E}_{\Q} $), 
the task is simple, because according to 
\eq\nr{hatPiE_appro}
the self-energies
$\widehat{\Pi}^\rmii{E$i$}_{\Q}$ are small in 
the kinematic domain of interest. Therefore, 
following \eqs\nr{rhoE_Q} and \nr{hat_rhoE_Q},
we find vacuum-like spectral functions in the E channel,  
\be
 \im {\widehat{\F}}\hspace*{0.3mm}{}^\rmii{E}_{\Q}
 \;
  \overset{\rmii{\nr{hatF_E}}}
 {\underset{\rmii{\nr{hatPiE_appro}}}{\approx}}
 \;
 \im {\F}\hspace*{0.3mm}{}^\rmii{E}_{\Q}
 \;
  \overset{\rmii{\nr{F_E}}}
 {\underset{\rmii{\nr{hatPiE_appro}}}{\approx}}
 \;
 \frac{\pi \,\delta(q^{ }_0 - \epsilon^q_{\Z})}{2 \epsilon^q_{\Z}}
 \;, \quad
 \epsilon^q_{\Z} 
 \; \equiv \; 
 \sqrt{q^2_{ } + m^2_{\Z}}
 \;. \la{rhoE_Z}
\ee

The situation is more complicated in the T channel, where the 
thermal self-energy correction is of the same order as the vacuum
mass, cf.\ \eq\nr{PiT_appro}. 
Employing Euclidean conventions
for notational simplicity, the propagator is 
\be
 {\F}\hspace*{0.3mm}{}^\rmii{T}_{Q}
 \; 
  \underset{\rmii{\nr{PiT_appro}}}{\overset{\rmii{\nr{F_T}}}{\approx}}
 \; 
  \frac{Q^2_{ } + \frac{1}{2} 
 ( c^2_{ } m^2_\rmiii{E1} + s^2_{ } m^2_\rmiii{E2} ) }
 {
 (Q^2_{ } + \frac{1}{2} m^2_\rmiii{E1} )
 (Q^2_{ } + \frac{1}{2} m^2_\rmiii{E2} )
 + m^2_\Z\,
 [ Q^2_{ } + \frac{1}{2} 
 ( c^2_{ } m^2_\rmiii{E1} + s^2_{ } m^2_\rmiii{E2} ) ]
 }
 \;.
 \la{F_T_appro_1}
\ee
Searching for the poles, which we parametrize through
\ba
 m^2_{\Z\T^\pm}
 & \equiv & 
 \frac{1}{2}\,
 \biggl[\,
  m_\Z^2 + \frac{1}{2} \bigl( m^2_\rmiii{E1} + m^2_\rmiii{E2} \bigr) 
 \nn[2mm]
 & \pm &
 \sqrt{
  m_\Z^4 + m_\Z^2 (c^2_{ } - s^2_{ })
 \bigl( m^2_\rmiii{E2} - m^2_\rmiii{E1} \bigr)
 + 
 \frac{1}{4} 
 \bigl( m^2_\rmiii{E2} - m^2_\rmiii{E1} \bigr)^2_{ }
 } 
 \,\biggr]
 \;, \la{poles_Z}
\ea
the propagator may be partially fractioned, 
\ba
 {\F}\hspace*{0.3mm}{}^\rmii{T}_{Q}
 \; 
  \underset{\rmii{\nr{poles_Z}}}
 {\overset{\rmii{\nr{F_T_appro_1}}}{=}}
 \; 
   \frac{
         c^2_{\Z\T}
        }{Q^2_{ } + m^2_{\Z\T^+} }
 \; + \; 
   \frac{
         s^2_{\Z\T}
        }{Q^2_{ } + m^2_{\Z\T^-} }
 \;. \hspace*{2mm} \la{F_T_appro_2}
\ea
Here we have defined
\ba
 c^2_{\Z\T}
 & \equiv &
  \frac{1}{ m^2_{\Z\T^+} - m^2_{\Z\T^-} }
 \Bigl[\; 
        m^2_{\Z\T^+}
     -
        \frac{1}{2} ( c^2_{ } m^2_\rmiii{E1} + s^2_{ } m^2_\rmiii{E2} )
 \Bigr]
 \;, \\[2mm]
 s^2_{\Z\T}
 & \equiv &
  \frac{1}{ m^2_{\Z\T^+} - m^2_{\Z\T^-} }
 \Bigl[\; 
        \frac{1}{2} ( c^2_{ } m^2_\rmiii{E1} + s^2_{ } m^2_\rmiii{E2} )
     -
        m^2_{\Z\T^-}
 \Bigr]
 \;.
\ea
These ``residues'' sum to unity, 
\ba
 && \hspace*{-1.5cm}
 c^2_{\Z\T}
 + 
 s^2_{\Z\T}
 \nn[2mm] 
 & = & 
 \frac{1}{ m^2_{\Z\T^+} - m^2_{\Z\T^-} }
 \Bigl[\; 
        m^2_{\Z\T^+}
     -
        \frac{1}{2} ( c^2_{ } m^2_\rmiii{E1} + s^2_{ } m^2_\rmiii{E2} )
 \; + \; 
        \frac{1}{2} ( c^2_{ } m^2_\rmiii{E1} + s^2_{ } m^2_\rmiii{E2} )
     -  
         m^2_{\Z\T^-}
 \;\Bigr]
 \nn[2mm]
 & = & 1
 \;, 
\ea
so that they may be interpreted as 
thermally modified trigonometric weights squared.

For the spectral function, it follows from \eq\nr{F_T_appro_2} that 
\be
 \im {\F}\hspace*{0.3mm}{}^\rmii{T}_{\Q}
 \;
  \overset{\rmii{\nr{F_T_appro_2}}}{\approx}
 \;
 \frac{\pi}{2}\,
 \biggl[\,
 c^2_{\Z\T}
 \frac{\delta(q^{ }_0 - \epsilon^q_{\Z\T^+})}{\epsilon^q_{\Z\T^+}}
 + 
 s^2_{\Z\T}
 \frac{\delta(q^{ }_0 - \epsilon^q_{\Z\T^-})}{\epsilon^q_{\Z\T^-}}
 \,\biggr]
 \;, \quad
  \epsilon^q_{\Z\T^\pm} 
 \; \equiv \; 
 \sqrt{q^2_{ } + m^2_{\Z\T^\pm}}
 \;. \la{rhoT_Z}
\ee
When we insert this into \eq\nr{AmplHj_phys}, we obtain contributions
of the same types as analyzed in \eqs\nr{Phi_WW_a}--\nr{Phi_WW_d}, 
however with each transverse pole splitting into two separate
contributions, according to \eq\nr{rhoT_Z}. 
The power counting
employed in \eqs\nr{Phi_WW_a}--\nr{Phi_WW_d} remains valid, 
and therefore the transverse channel gives subdominant contributions
to \eq\nr{AmplHj_phys}.

\newpage

%
\section{HTL-resummed $t\bar{t}$ channel}
\la{app:D}

We repeat here the analysis of \se\ref{ss:higgs_WW} for the 
$t\bar{t}$ channel, whose matrix element squared was given in 
\eq\nr{AmplHl_phys}.
The results are illustrated numerically 
in \fig\ref{fig:sigma_thermal_tt}. The common prefactor
$
 - 2 \kappa^2_{ } m_t^2 \Nc^{ }
 \,\Delta^{-1}_{-K-P;\widetilde\varphi}
 \,\Delta^{-2}_{P;h}
$ 
is not shown, as we focus on the part
$
 \,\F^{t}_{\Q}
 \,\F^{t}_{\Q-\P}
 \,\{\,  
 ...
 \,\}.
$

\begin{figure}[t]

 \hspace*{-0.1cm}
 \centerline{%
  \epsfysize=7.2cm\epsfbox{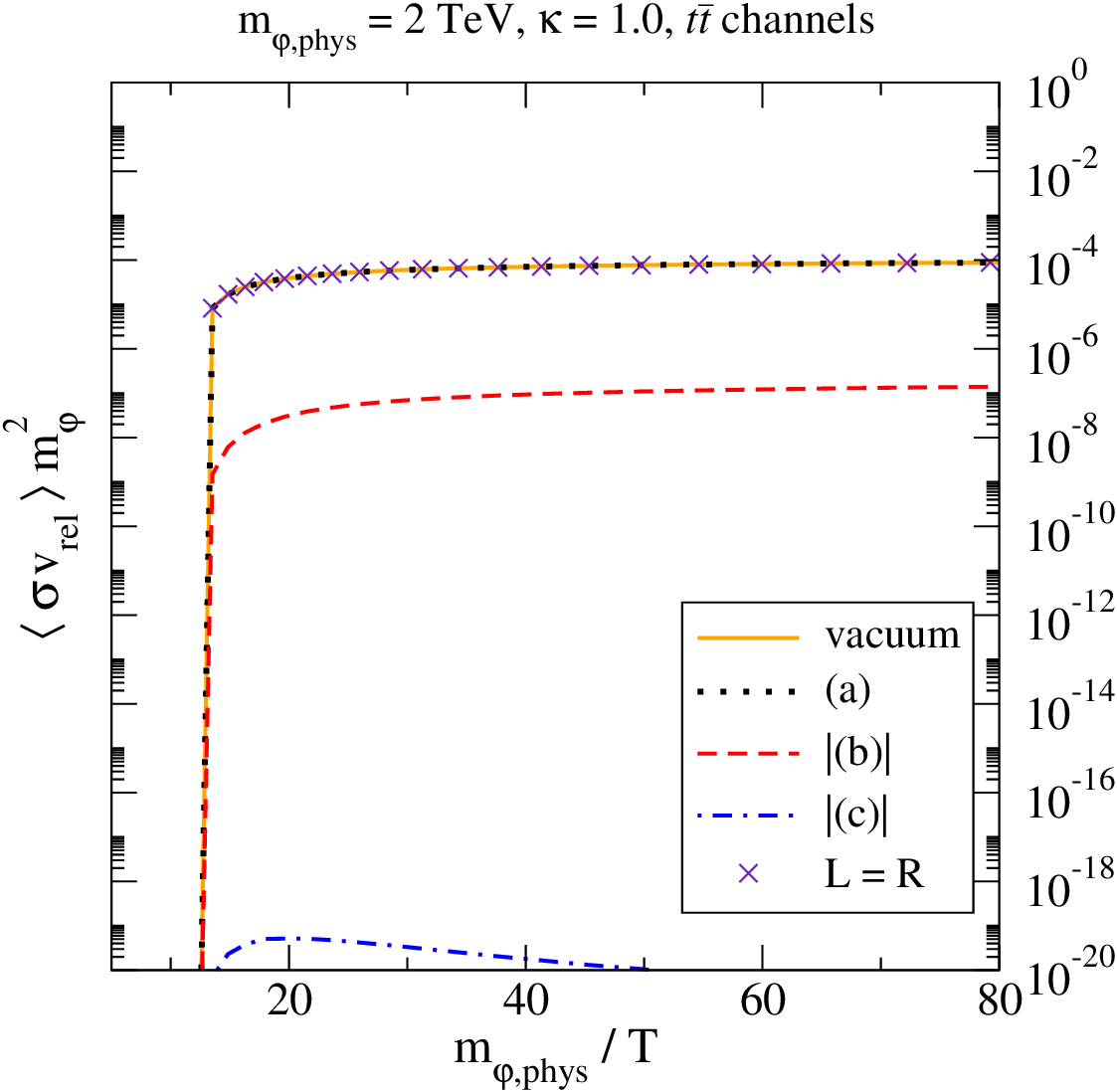}%
 }
 
 \caption[a]{\small
   An illustration of the cross sections following from the 
   various terms in \eq\nr{AmplHl_phys}, with the terms split
   up and labelled according to \eq\nr{tt_factorized}.  
   We have employed the benchmark values 
   $m^{ }_{\varphi,\rmii{phys}} = 2$~TeV, 
   $\kappa = 1.0$. The notation ``$|$(b)$|$'' indicates that 
   this channel gives a negative contribution, and we show
   the absolute value. The case (d) is not shown due to 
   its smallness, whereas $L=R$ illustrates the 
   outcome of the simplified expression in \eq\nr{AmplHl_phys_appro}.
   }
 
 \la{fig:sigma_thermal_tt}
\end{figure}

To get started, we need the self-energies from \eqs\nr{cW} and 
\nr{cP}. In the kinematic domain of \eq\nr{power_counting}, these can
be approximated as  
\ba
 c^\rmii{W}_{\Q;{t}^{ }_\rmiii{L}} & \approx & 
  - \frac{m_{{t}^{ }_\rmiii{L}}^2}{2 q^2_{ }}
    \ln \frac{2 q}{q^{ }_0 - q} 
 \;, \la{cW_appro} \\[2mm] 
 c^\rmii{P}_{\Q;{t}^{ }_\rmiii{L}} & \approx & 
  \frac{m_{{t}^{ }_\rmiii{L}}^2}{q^2_{ }}
  \biggl(\, 
  1  - \frac{1}{2}
    \ln \frac{2 q}{q^0_{ } - q} 
  \,\biggr)
 \;.  \la{cP_appro}
\ea
These play a role in four-vectors defined according 
to \eq\nr{L_vec}. Even if at first sight 
the corrections appear small compared with the tree-level
terms, the coefficients $\sim q^{ }_0,q$ 
in \eq\nr{L_vec} are large, 
so we need to be careful about when the corrections
can be neglected. 

Let us start by considering the propagator from \eq\nr{t_prop_2}. 
Recalling that we are using Minkowskian conventions here, the squares read
\ba
 L^2_\Q & = & 
  q_0^2 
 \, \bigl( 1 + c^\rmii{W}_{\Q;{t}^{ }_\rmiii{L}} \bigr)^2_{ }
  - q^2_{ } 
 \, \bigl( 1 + c^\rmii{P}_{\Q;{t}^{ }_\rmiii{L}} \bigr)^2_{ }
 \nn[2mm]
 & \overset{c^\rmiii{W,P}_{\Q;{t}^{ }_\rmiii{L}}\,\ll\, 1}{\approx} & 
 q_0^2 - q^2_{ } 
 + 2 q_0^2 c^\rmii{W}_{\Q;{t}^{ }_\rmiii{L}}
 - 2 q^2_{ } c^\rmii{P}_{\Q;{t}^{ }_\rmiii{L}}
 \nn[2mm]
 & \underset{q^{ }_0 \,\approx\, q}
  {\overset{\rmii{\nr{power_counting}}}{\approx}} & 
 q_0^2 - q^2_{ } 
 + 2 q^2_{ } \,\bigl(\, c^\rmii{W}_{\Q;{t}^{ }_\rmiii{L}}
 - c^\rmii{P}_{\Q;{t}^{ }_\rmiii{L}}
 \,\bigr)
 \; 
 \underset{\rmii{\nr{cP_appro}}}
 {\overset{\rmii{\nr{cW_appro}}}{\approx}}
 \; 
 q_0^2 - q^2_{ } - 2 m_{{t}^{ }_\rmiii{L}}^2
 \;, \la{L2}  \\[2mm]
 R^2_\Q & \approx & 
 q_0^2 - q^2_{ } - 2 m_{{t}^{ }_\rmiii{R}}^2
 \;, \la{R2} \\[2mm]
 L^{ }_\Q\cdot R^{ }_\Q
 & \approx & 
 \underbrace{
 q_0^2 - q^2_{ }
 }_{\;\equiv\;\Q^2_{ }}
 - m_{{t}^{ }_\rmiii{L}}^2 - m_{{t}^{ }_\rmiii{R}}^2
 \;. \la{LR}
\ea
Inserting in \eq\nr{t_prop_2} and factorizing the dependence
on $\Q^2_{ }$, we find 
\be
 \F^{t}_{\Q}
 \; \approx \; 
 \frac{1}{(\Q^2_{ } - m^2_{ t_1})(\Q^2_{ } - m^2_{ t_2})}
 \;, \quad
 m^2_{ t_1} \; \equiv \; 
 m_t^2 + 2 m_{{t}^{ }_\rmiii{L}}^2
 \;, \quad
 m^2_{ t_2} \; \equiv \; 
 m_t^2 + 2 m_{{t}^{ }_\rmiii{R}}^2
 \;. \la{t_masses}
\ee

Next, we need to tackle the numerator of \eq\nr{AmplHl_phys}. In order to 
simplify the notation, we denote 
\be
 \S \; \equiv \; \Q - \P
 \;. \la{S_def}
\ee
Also, even if there is a certain danger of confusion with the 
overall sign, we denote Minkowskian 
propagators similarly to \eq\nr{Delta_t}, {\it viz.} 
\be
 \Delta^{-1}_{\Q;t_i^{ }}
 \; \equiv \; 
 \frac{1}{m^2_{t_i} + q^2_{ } - q_0^2 }
 \; = \; 
 \frac{1}{m^2_{t_i} - \Q^2_{ }}
 \;, \quad i=1,2
 \;, 
\ee
where the masses are from \eq\nr{t_masses}.

Let us start with the first line ($\equiv\mbox{(i)}$) of \eq\nr{AmplHl_phys}. 
A straightforward computation yields 
\be
 R_\Q^2\, L_\S^2 + m_t^4 
 \;
 \underset{\rmii{\nr{R2}}}{\overset{\rmii{\nr{L2}}}{\approx}}
 \;
 (\Q^2 - m^2_{t^{ }_2})(\S^2 - m^2_{t^{ }_1})
 + 
 m_t^2 \, 
 \bigl( \Q^2 - m^2_{t^{ }_2} + \S^2 - m^2_{t^{ }_1} \bigr) 
 + 2 m^4_t
 \;. \la{for_i}
\ee
This then implies that
\ba
 \F^{t}_{\Q}
 \F^{t}_{\S}
 \{\mbox{(i)}\} 
 & \overset{\rmii{\nr{for_i}}}{\approx} & 
 L^{ }_\Q\cdot R^{ }_\S \,
 \bigl\{
    \Delta^{-1}_{\Q;t_1^{ }}
    \Delta^{-1}_{\S;t_2^{ }}
 - m_t^2 \, \bigl[\, 
    \Delta^{-1}_{\Q;t_1^{ }} \F^t_\S 
 +
    \Delta^{-1}_{\S;t_2^{ }} \F^t_\Q 
            \,\bigr]
 + 2 m^4_t\, \F^t_\Q \,\F^t_\S
 \bigr\}
 \nn[2mm]
 & + & 
 R^{ }_\Q\cdot L^{ }_\S \,
 \bigl\{
    \Delta^{-1}_{\Q;t_2^{ }}
    \Delta^{-1}_{\S;t_1^{ }}
 - m_t^2 \, \bigl[\, 
    \Delta^{-1}_{\Q;t_2^{ }} \F^t_\S 
 +
    \Delta^{-1}_{\S;t_1^{ }} \F^t_\Q 
            \,\bigr]
 + 2 m^4_t\, \F^t_\Q \,\F^t_\S
 \bigr\}
 \;. \hspace*{9mm} \la{res_i}
\ea
In a similar manner, 
the second line ($\equiv\mbox{(ii)}$) of \eq\nr{AmplHl_phys} yields
\be
 \F^{t}_{\Q}
 \F^{t}_{\S}
 \{\mbox{(ii)}\} 
 \;\approx\;
 \frac{m_t^2}{2}
 \bigl\{\,
    \Delta^{-1}_{\Q;t_1^{ }}
    \Delta^{-1}_{\S;t_1^{ }}
 + 
    \Delta^{-1}_{\Q;t_1^{ }}
    \Delta^{-1}_{\S;t_2^{ }}
 +
    \Delta^{-1}_{\Q;t_2^{ }}
    \Delta^{-1}_{\S;t_1^{ }}
 +
    \Delta^{-1}_{\Q;t_2^{ }}
    \Delta^{-1}_{\S;t_2^{ }} 
 \,\bigr\}
 \;. \la{res_ii}
\ee
For the third line ($\equiv\mbox{(iii)}$) of \eq\nr{AmplHl_phys}, we find
\ba
 \mbox{(iii)} & = & 
 2 m_t^2 \, \bigl(\, 
   L^{ }_{\Q}\cdot R^{ }_{\S} \, 
   R^{ }_{\Q}\cdot L^{ }_{\S}
  - 
   L^{ }_{\Q}\cdot L^{ }_{\S} \,
   R^{ }_{\Q}\cdot R^{ }_{\S}
   \,\bigr)
  \nn[2mm]
 & \overset{c^\rmiii{W,P}_{\Q;{t}^{ }_\rmiii{L,R}}\,\ll\, 1}{\approx} & 
 2 m_t^2 \bigl\{ \, 
 \bigl[\, 
 q^{ }_0 s^{ }_0   \bigl(\, 1 +  c^\rmii{W}_{\Q;{t}^{ }_\rmiii{L}}
                              +  c^\rmii{W}_{\S;{t}^{ }_\rmiii{R}}\,\bigr)
 - \vec{q}\cdot\vec{s}\, \bigl(\, 1 +  c^\rmii{P}_{\Q;{t}^{ }_\rmiii{L}}
                              +  c^\rmii{P}_{\S;{t}^{ }_\rmiii{R}}\,\bigr)
 \,\bigr]
  \nn[2mm]
 & & \hspace*{0.6cm} \times \, 
 \bigl[\, 
 q^{ }_0 s^{ }_0   \bigl(\, 1 +  c^\rmii{W}_{\Q;{t}^{ }_\rmiii{R}}
                              +  c^\rmii{W}_{\S;{t}^{ }_\rmiii{L}}\,\bigr)
 - \vec{q}\cdot\vec{s}\, \bigl(\, 1 +  c^\rmii{P}_{\Q;{t}^{ }_\rmiii{R}}
                              +  c^\rmii{P}_{\S;{t}^{ }_\rmiii{L}}\,\bigr)
 \,\bigr]
  \nn[2mm]
 & & \hspace*{0.6cm} - \, 
 \bigl[\, 
 q^{ }_0 s^{ }_0   \bigl(\, 1 +  c^\rmii{W}_{\Q;{t}^{ }_\rmiii{L}}
                              +  c^\rmii{W}_{\S;{t}^{ }_\rmiii{L}}\,\bigr)
 - \vec{q}\cdot\vec{s}\, \bigl(\, 1 +  c^\rmii{P}_{\Q;{t}^{ }_\rmiii{L}}
                              +  c^\rmii{P}_{\S;{t}^{ }_\rmiii{L}}\,\bigr)
 \,\bigr]
  \nn[2mm]
 & & \hspace*{0.6cm} \times \, 
 \bigl[\, 
 q^{ }_0 s^{ }_0   \bigl(\, 1 +  c^\rmii{W}_{\Q;{t}^{ }_\rmiii{R}}
                              +  c^\rmii{W}_{\S;{t}^{ }_\rmiii{R}}\,\bigr)
 - \vec{q}\cdot\vec{s}\, \bigl(\, 1 +  c^\rmii{P}_{\Q;{t}^{ }_\rmiii{R}}
                              +  c^\rmii{P}_{\S;{t}^{ }_\rmiii{R}}\,\bigr)
 \,\bigr]
 \,\bigr\}
 \;. \la{res_iii}
\ea
The correction factors appear in different places in the two terms 
in \eq\nr{res_iii}. However, expanding the result to linear order in them, 
all terms cancel. Finally, 
the fourth line ($\equiv\mbox{(iv)}$) of \eq\nr{AmplHl_phys} yields
\ba
 \F^{t}_{\Q}
 \F^{t}_{\S}
 \{\mbox{(iv)}\} 
 & \underset{\rmii{\nr{R2}}}{\overset{\rmii{\nr{L2}}}{\approx}} & 
 m_t^2\, L^{ }_\Q\cdot L^{ }_\S \,
 \bigl(\,
    \Delta^{-1}_{\Q;t_1^{ }} \F^t_\S 
 +
    \Delta^{-1}_{\S;t_1^{ }} \F^t_\Q 
 - 2 m^2_t\, \F^t_\Q \,\F^t_\S
 \,\bigr)
 \nn[2mm]
 & + & 
 m_t^2\, 
 R^{ }_\Q\cdot R^{ }_\S \,
 \bigl(\,
    \Delta^{-1}_{\Q;t_2^{ }} \F^t_\S 
 +
    \Delta^{-1}_{\S;t_2^{ }} \F^t_\Q 
 - 2 m^2_t\, \F^t_\Q \,\F^t_\S
 \,\bigr)
 \;. \hspace*{9mm} \la{res_iv}
\ea

Now, we note from \eqs\nr{res_i}--\nr{res_iv} that whenever 
$\F^t_\Q$ or $\F^t_\S$ appears, it is multiplied by $m^2_t$.
We then partial fraction
\ba
 m_t^2 \F^t_\Q 
 & \overset{\rmii{\nr{t_masses}}}{\approx} & 
 \frac{m_t^2}{(\Q^2_{ } - m^2_{ t_1})(\Q^2_{ } - m^2_{ t_2})}
 \; = \; 
 \frac{m_t^2}{m^2_{t^{ }_1} - m^2_{t^{ }_2}}
 \biggl(\, 
  \frac{1}{\Q^2_{ } - m^2_{t^{ }_1}}
 - 
  \frac{1}{\Q^2_{ } - m^2_{t^{ }_2}}
 \,\biggr)
 \nn[2mm]
 & = & 
 \underbrace{
 \frac{m^2_t}{2( m_{{t}^{ }_\rmiii{L}}^2 - m_{{t}^{ }_\rmiii{R}}^2)}
 }_{ 
 \;\equiv\;\chi^{ }_t
 }
 \bigl(\,
   \Delta^{-1}_{\Q;t_2^{ }}  
 - 
   \Delta^{-1}_{\Q;t_1^{ }}  
 \,\bigr)
 \;. \la{factor_t}
\ea
Thereby the expressions split into a large number of terms. 
Though it is a bit tedious, these can be regrouped into expressions
of 0th, 1st, or 2nd order in $\chi^{ }_t$. In the end, 
we also note that the result is symmetric in 
$
 \Q \leftrightarrow \S
$. 
By the substitution $\Q\to \P-\Q$, 
implying $\Q \to -\S$ and $\S \to -\Q$, 
terms can be combined. This produces
\ba
 \F^{t}_{\Q}
 \F^{t}_{\S}
 \{\mbox{(i--iv)}\} 
 &
 \overset{\rmii{(a)}}{\approx}
 &
 \Delta^{-1}_{\Q;t_1^{ }} \, \Delta^{-1}_{\S;t_2^{ }} \,
 \bigl(\,
  2 L^{ }_\Q\cdot R^{ }_\S + m_t^2 
 \,\bigr)
 \nn[2mm]
 & 
 \overset{\rmii{(b)}}{+}
 &  
 \bigl(\,
  \Delta^{-1}_{\Q;t_1^{ }} \, \Delta^{-1}_{\S;t_1^{ }}
 + 
   \Delta^{-1}_{\Q;t_2^{ }} \, \Delta^{-1}_{\S;t_2^{ }}
 \,\bigr) \,\frac{m_t^2}{2}
 \nn[2mm]
 & 
 \overset{\rmii{(c)}}{+}
 & 
 2\chi^{ }_t\, 
 \Delta^{-1}_{\Q;t_1^{ }}\, 
 \bigl(\, 
   \Delta^{-1}_{\S;t_2^{ }}
 -
   \Delta^{-1}_{\S;t_1^{ }} 
 \,\bigr)
 (L^{ }_\Q - R^{ }_\Q)\cdot L^{ }_\S
 \nn[2mm]
 &
 \overset{\rmii{(c)}}{-}
 & 
 2\chi^{ }_t\, 
 \Delta^{-1}_{\Q;t_2^{ }}\, 
 \bigl(\, 
   \Delta^{-1}_{\S;t_2^{ }}
 -
   \Delta^{-1}_{\S;t_1^{ }} 
 \,\bigr)
 (L^{ }_\Q - R^{ }_\Q)\cdot R^{ }_\S
 \nn[2mm]
 & 
 \overset{\rmii{(d)}}{+}
 & 
  2\chi^{2}_t\,
 \bigl(\, 
   \Delta^{-1}_{\Q;t_1^{ }}
 -
   \Delta^{-1}_{\Q;t_2^{ }} 
 \,\bigr)
 \bigl(\, 
   \Delta^{-1}_{\S;t_2^{ }}
 -
   \Delta^{-1}_{\S;t_1^{ }} 
 \,\bigr)
 (L^{ }_\Q - R^{ }_\Q)\cdot
 (L^{ }_\S - R^{ }_\S)
 \;. \hspace*{7mm} \la{tt_factorized} 
\ea
The labelling (a)--(d) corresponds to that shown 
in \fig\ref{fig:sigma_thermal_tt}. 

The terms in \eq\nr{tt_factorized} have a certain physical interpretation. 
As we see from \eq\nr{t_masses}, the state ``$t^{ }_1$'' corresponds to 
left-handed chirality, the state ``$t^{ }_2$'' to right-handed chirality. 
The term on the first line of \eq\nr{tt_factorized} is chirality conserving
(scalar splits into left and right). 
The term of the second line is not
(scalar into left-left or right-right), hence it is proportional to the 
chirality-breaking vacuum mass squared. On the third and 
fourth lines, the coefficient $\chi^{ }_t$ is inversely proportional
to chirality breaking, cf.\ \eq\nr{factor_t}. But there are {\em two}
differences multiplying $\chi^{ }_t$,
which are also proportional to chirality breaking. Therefore, 
the terms vanish if we envisage taking the chirally symmetric limit
(i.e.\ $g^{ }_1,g^{ }_2,h^{ }_t, h^{ }_b\to 0$, $g_3^{ }\neq 0$). 
Finally, on the fifth line, 
the coefficient $\chi^{2}_t$ is quadratically divergent
in chirality breaking, but there are {\em four} differences, 
more than compensating for the singular coefficient. 

Therefore, in the
would-be chirally symmetric limit, the three last lines drop out.
The first two lines can be combined
($m^{ }_{t^{ }_1} \to m^{ }_{t^{ }_2}$ if $L\to R$, 
cf.\ \eq\nr{t_masses}), and amount to  
\eq\nr{AmplHl_phys_appro}. 

\vspace*{3mm}

In order to take further steps, it is helpful to go over
to the variables in \eq\nr{transverse_4}. For this, we write
\be
 \S \; \underset{\rmii{\nr{S_def}}}{\overset{\rmii{\nr{R_def}}}{=}} \; -\R
 \;, \la{S_to_R}
\ee
which implies 
\ba
 L^{ }_{\Q}\cdot R^{ }_{\S}
 & \approx &
 + \, q^{ }_0 s^{ }_0   \bigl(\, 1 +  c^\rmii{W}_{\Q;{t}^{ }_\rmiii{L}}
                              +  c^\rmii{W}_{\S;{t}^{ }_\rmiii{R}}\,\bigr)
 - \vec{q}\cdot\vec{s}\, \bigl(\, 1 +  c^\rmii{P}_{\Q;{t}^{ }_\rmiii{L}}
                              +  c^\rmii{P}_{\S;{t}^{ }_\rmiii{R}}\,\bigr)
 \nn[2mm]
 & = & 
 -\, q^{ }_0 r^{ }_0   \bigl(\, 1 +  c^\rmii{W}_{\Q;{t}^{ }_\rmiii{L}}
                              +  c^\rmii{W}_{\R;{t}^{ }_\rmiii{R}}\,\bigr)
 + \vec{q}\cdot\vec{r}\, \bigl(\, 1 +  c^\rmii{P}_{\Q;{t}^{ }_\rmiii{L}}
                              +  c^\rmii{P}_{\R;{t}^{ }_\rmiii{R}}\,\bigr)
 \;. \la{LQ_RS}
\ea
Consider first the terms without the $c$-factors.  
We assume that the propagators have the masses
$m^{ }_i$ and $m^{ }_j$, and find 
\ba
 -\, q^{ }_0 r^{ }_0 + \vec{q}\cdot\vec{r} 
 & = & 
 -\epsilon^{q}_{i} \epsilon^{r}_j + \vec{q}\cdot\vec{r}
 \nn[2mm]
 & = & 
 \frac{1}{2} 
 \bigl[
  - ( \epsilon^{q}_{i} + \epsilon^{r}_j )^2_{ }
  + (\epsilon^{q}_{i})^2_{ } +  (\epsilon^{r}_j)^2_{ }
  + (\vec{q+r})^2_{ }- q^2_{ }- r^2_{ } 
 \bigr] 
 \nn[2mm]
 & = & 
 -\,\frac{1}{2} \,
 \bigl(\,
   s - m_i^2 - m_j^2
 \,\bigr)
 \;. 
\ea
Thereby
\be
 2 L^{ }_\Q\cdot R^{ }_\S + m_t^2 
 \; \supset \;
 -(s - m_i^2 - m_j^2 - m_t^2)
 \;. 
\ee
If we add the contribution from the second line of 
\eq\nr{tt_factorized} and send $m^{ }_i,m^{ }_j\to m^{ }_t$, 
the vacuum matrix element squared
in \eq\nr{Phi_tt_0} is recovered. 

In the correction terms in 
\eq\nr{LQ_RS}, which are already proportional to thermal masses
(cf.\ \eqs\nr{cW_appro} and \nr{cP_appro}), we can employ ultrarelativistic
kinematics from \eq\nr{transverse_4}. Hence,  
\ba
 L^{ }_{\Q}\cdot R^{ }_{\S}
 & \overset{\rmii{\nr{LQ_RS}}}{\supset} &
 -\, q^{ }_0 r^{ }_0   \bigl(\, c^\rmii{W}_{\Q;{t}^{ }_\rmiii{L}}
                              +  c^\rmii{W}_{\R;{t}^{ }_\rmiii{R}}\,\bigr)
 + \vec{q}\cdot\vec{r}\, \bigl(\, c^\rmii{P}_{\Q;{t}^{ }_\rmiii{L}}
                              +  c^\rmii{P}_{\R;{t}^{ }_\rmiii{R}}\,\bigr)
 \\[2mm]
 & \overset{\rmii{\nr{transverse_4}}}{\approx} & 
-\, q r   \bigl(\, c^\rmii{W}_{\Q;{t}^{ }_\rmiii{L}}
                              +  c^\rmii{W}_{\R;{t}^{ }_\rmiii{R}}\,\bigr)
 + \biggl( q r - \frac{s}{2} \biggr) 
         \, \bigl(\, c^\rmii{P}_{\Q;{t}^{ }_\rmiii{L}}
                              +  c^\rmii{P}_{\R;{t}^{ }_\rmiii{R}}\,\bigr)
 \nn[2mm]
 & = & 
 q r   \bigl(\, c^\rmii{P}_{\Q;{t}^{ }_\rmiii{L}}
               - c^\rmii{W}_{\Q;{t}^{ }_\rmiii{L}}
              +  c^\rmii{P}_{\R;{t}^{ }_\rmiii{R}}
                -  c^\rmii{W}_{\R;{t}^{ }_\rmiii{R}}\,\bigr)
 - \frac{s}{2} 
         \, \bigl(\, c^\rmii{P}_{\Q;{t}^{ }_\rmiii{L}}
                              +  c^\rmii{P}_{\R;{t}^{ }_\rmiii{R}}\,\bigr)
 \nn[2mm]
 & \approx & 
 \frac{r\, m_{{t}^{ }_\rmiii{L}}^2}{ q}
 + 
 \frac{q\, m_{{t}^{ }_\rmiii{R}}^2}{ r}
 - \frac{s}{2} 
         \, \biggl[\, 
  \frac{m_{{t}^{ }_\rmiii{L}}^2}{q^2_{ }}
  \biggl(\, 
  1  - \frac{1}{2}
    \ln \frac{2 q}{\epsilon^q_i - q} 
  \,\biggr)
 + 
  \frac{m_{{t}^{ }_\rmiii{R}}^2}{r^2_{ }}
  \biggl(\, 
  1  - \frac{1}{2}
    \ln \frac{2 r}{\epsilon^r_j - r} 
  \,\biggr)
 \,\biggr]
 \;. \hspace*{7mm} \nonumber
\ea
This does {\em not} simplify in any substantial way if we take
the chirally symmetric limit $L\to R$.

To summarize, in the fermionic channel,  
the first thermal corrections do {\em not} amount to 
simple mass modifications of vacuum cross sections, 
but have a more delicate structure. 
This remains true if we take the chirally symmetric limit $L\to R$.  
Nevertheless, in the domain of \eq\nr{power_counting}, the
mass corrections are small. This is also clearly 
visible from \fig\ref{fig:sigma_thermal_tt}.

\newpage

\small{
%

}

\end{document}